\documentclass[journal ]{new-aiaa}
\usepackage[utf8]{inputenc}
\usepackage{textcomp}

\usepackage{graphicx}
\usepackage{amsmath}
\usepackage[version=4]{mhchem}
\usepackage{siunitx}
\usepackage{longtable,tabularx}
\usepackage{subcaption}
\usepackage{float}
\title{Control of Reverse Flow on High-Speed Rotorcraft Using a Blade with a Sinusoidal Trailing Edge}
\author{Ranga Srinivas Gokul\footnote{Project Associate, Experimental Aerodynamics Group, Department of Aerospace Engineering, IIT Kanpur.}}
\author{Tufan Kumar Guha\footnote{Assistant Professor,  Department of Aerospace Engineering, Experimental Aerodynamics Group, IIT Kanpur; AIAA Member. \indent \indent Correspondence: tkguha@iitk.ac.in}} 
\affil{Indian Institute of Technology Kanpur, India}
\begin{document}
\maketitle
\begin{abstract}
One of the main challenges limiting the maximum forward speed of high-speed rotorcraft is the development of reverse flow on retreating blades. Flow separation at the sharp aerodynamic leading edge during reverse flow leads to negative lift, high drag, high pitching moment, and a pitching moment impulse. The study experimentally investigates the effectiveness of a sinusoidal trailing edge as a passive method for the control of reverse flow, under static pitching conditions. Three NACA 0015 two-dimensional blades were evaluated under both reverse and forward flow conditions. A baseline blade with a straight trailing edge and two modified blades with sinusoidal trailing edges, each having a spatial amplitude of 10\% of the chord and wavelengths of 5\% and 10\% of the chord, respectively. %Three NACA 0015 two-dimensional blades, a baseline blade with a straight trailing edge and two modified blades with sinusoidal trailing edges having spatial amplitudes of 10\% of the chord and wavelengths of 5\% and 10\% of the chord, were evaluated under both reverse and forward flow conditions. 
Experiments were conducted at chord-based Reynolds numbers of $\mathbf{0.7\times10^5}$ and $\mathbf{1.4\times10^5}$, consisting of planar Particle Image Velocimetry and aerodynamic load measurements. The study shows that a sinusoidal trailing edge significantly mitigates flow separation, leading to a large reduction in the associated problems over a wide range of angles of attack. The problem of pitching moment impulse is eliminated. The sinusoidal geometry with a lower wavelength of 5\% of the chord shows comparatively better performance. Additionally, both sinusoidal geometries minimally affect the forward flow characteristics.
\end{abstract}

\section*{Nomenclature}
{\renewcommand\arraystretch{1.0}
\noindent\begin{longtable*}{@{}l @{\quad=\quad} l@{}}
$A$  & sinusoidal amplitude \\
$c$  & model's chord \\
$C$ & sinusoid-crest \\
$C_D$ & drag coefficient \\
$C_L$ & lift coefficient \\
$C_M$ & moment coefficient \\
$F_A$ & aerodynamic-axial force (from load cell) \\
$F_N$ & aerodynamic-normal force (from load cell) \\
$M$ & sinusoid-mid \\
$Re_c$ & chord-based Reynolds number \\
$T$ & sinusoid-trough \\
$T_Z$ & pitching moment (from load cell)\\
$\bar{u}$ & mean streamwise velocity \\
$U_\infty$ & frestream velocity \\
$x$ & streamwise coordinate \\
$y$ & transverse coordinate \\
$z$ & spanwise coordinate \\
$\alpha$ & angle of attack \\
$\lambda$ & sinusoidal wavelength \\
\end{longtable*}}

\section{Introduction}
\lettrine[]{R}{otorcraft} have become integral to most civilian and military operations due to their operational flexibility and enhanced versatility. The widespread usage has spurred extensive research into increasing the forward speed of high-speed rotorcraft, which is currently limited to 250 knots \cite{sikorsky,problemsofdesign}. However, an increase in forward speed mandates a drop in rotor revolutions per minute (RPM) to avoid supersonic flow at the rotor tip and, as a consequence, shock-induced flow separation and wave drag. The coupled increment in forward speed and decrease in rotor RPM increases the advance ratio \cite{bramwell}, defined as the ratio between the freestream velocity and the rotor tip speed. However, at a high advance ratio, such as 0.75, a large portion of the retreating blade experiences reverse flow \cite{slowedrotor, Revflowexp, vorticallift, DeannaPassive, Cambermorphcomputation}. During reverse flow, the sharp geometric trailing edge becomes the aerodynamic leading edge. Therefore, the flow travels from a sharp leading edge to a blunt trailing edge of an airfoil.

The problems associated with reverse flow and the scope for flow control primarily originate from flow separation. The sharp aerodynamic leading edge forces the flow to separate over the suction side of the airfoil, even at small angles of attack, leading to the formation of a closed separation bubble \cite{DeannaPassive}. It should be noted that the pressure side of an airfoil during forward flow becomes the suction side during reverse flow, and vice versa \cite{Cambermorphcomputation}. The separation bubble gradually grows in size with increasing incidence, developing into a large stall vortex that is eventually shed into the wake \cite{Timeaveraged, DeannaPassive}. Since the flow separation is forced by the leading edge, the process is relatively insensitive to Reynolds number for thin airfoils such as NACA 0012 \cite{Reeffects}. The finite span of a blade and the continuous variation in its sweep angle during rotor revolution make the separation bubble highly three-dimensional and dynamic in nature \cite{vorticallift,nelson2024control}. The major drawbacks associated with reverse flow are the development of negative lift and an increase in drag and pitching moment, which are amplified when flow separation starts at the leading edge \cite{Cambermorphcomputation}. The suction from the low-pressure vortex core associated with flow separation leads to negative lift and contributes to the aerodynamic drag. The moment from the negative lift about the geometric quarter chord, where the blade is attached, induces a pitching moment impulse and unsteady torsional loads \cite{slowedrotor}. Cyclic variations in pitch, yaw, surge, and local freestream velocities over the rotor blades due to vehicle kinematics \cite{principles}, further add to the overall unsteadiness in loads. The cyclic formation and subsequent shedding of the separation vortex lead to dynamic stall and large hysteresis in aerodynamic forces and moments \cite{corkethomas, DeannaPassive}.  

Since the sharp aerodynamic leading edge causes flow separation during reverse flow, modification of rotor blade geometry to control the separation has been the subject of several studies. Investigations by Lind et al.\cite{Timeaveraged} experimentally compared the characteristics of a sharp (NACA 0012) and two blunt (ellipse, DBLN-526) geometric trailing edge airfoils at $Re = 1.1\times10^5$. Flow separation for the NACA 0012 airfoil in reverse flow was observed to start at the sharp leading edge at $\alpha=184^\circ$ ($4^\circ$ in reverse flow) as a closed bubble, and transition to a fully stalled flow at $189^\circ$ ($9^\circ$ in reverse flow). The blunt trailing edge airfoils showed a delay in deep stall to a higher angle of attack, with force measurements indicating lower drag values. However, flow attachment also led to higher negative lift and pitching moment coefficients. The study showed that there was no operating range in reverse flow configuration for which the blunt airfoils outperformed the standard NACA 0012 airfoil, when all three metrics of negative lift, drag, and pitching moment were taken together into consideration. Moreover, a blunt airfoil is expected to underperform during forward flow and hover. 
One other such modification is the implementation of a reflex camber at the geometric trailing edge, where the camber aligns the sharp edge with the freestream. Ko et al.\cite{DeannaPassive} experimentally investigated the effects of morphing the last quarter-chord of a finite-span NACA 63-218 blade over three static reflex angles of $5^\circ$,$10^\circ$ and $15^\circ$. SPIV measurements over both static and dynamic pitch conditions at $Re = 3.75\times10^5$ showed a mitigation of the leading-edge separation and significant reductions in wake width. Force measurements showed a 37.4\% reduction in drag values and a delay in stall angle during static pitching and a reduction in hysteresis during dynamic pitching. Effective implementation of this technique requires an active camber morphing mechanism.
Active flow-control methods investigated over rotor blades include boundary layer ingestion and plasma actuators. Studies by Crimi et al.\cite{TorsionalNASA} on boundary layer ingestion reported the overall mechanical complexity as the principal constraint. The suction rates affecting the overall control effectiveness showed a linear dependence on the rotor blade incidence. 
Experiments on the plasma actuators \cite{Plasmaactuator} showed that actuation, particularly near the aerodynamic leading edge, was more effective due to enhanced entrainment capabilities of coherent structures over the shear layer. The Sikorsky Aircraft Corporation also investigated the effects of modifying rotor blade geometry at the inboard regions to address the reverse flow problem \cite{X2D}. The modified blades showed a promising reduction in the adverse effects of reverse flow, albeit at the cost of performance during hover. Preliminary implementation of the Advancing Blade Concept, wherein negative lift on the retreating blade was offset using the opposite advancing blade, showed excessive vibration, control difficulties, and high fuel consumption \cite{sikorsky}. 
The above studies showcase the interest in addressing the reverse flow problem in both academia and industry. It also justifies the investigation of a simple passive flow control method that will require minimal modification of blade geometry. The current study proposes such a modification, which is based on two passive flow control techniques, leading edge tubercles and trailing edge serrations, which are discussed below. 

Non-straight leading and trailing edges play important roles in aerodynamic flow control research. Leading edge protuberances, such as those found on the flippers of humpback whales, help to delay stall and improve the lift-to-drag (L/D) characteristics \cite{Whaletubercles}. They do so by acting as leading-edge vortex generators, which energize the boundary layer. The vortices are generated at the protuberances along the streamwise direction, occurring as counter-rotating vortex pairs. The characteristics of these vortices are governed by two parameters, the spatial amplitude, A, and the spatial wavelength, $\lambda$. Investigations by Johari et al. \cite{LEPairfoil} reported the highest post-stall performance for protuberances with maximum spatial amplitudes and minimum spatial wavelengths; an observation that matches with the results from the present study. Combined experimental and numerical investigations by Hansen et al. \cite{LEPHansen} on a NACA 0021 airfoil with leading edge tubercles showed a gradual increase in size of the vortices along the streamwise direction, accompanied by a decrease in peak primary vorticity due to gradual diffusion. The vortices maintained greater flow attachment behind the tubercle peaks and increased surface shear stresses, consequently generating higher lift, but also higher drag values. Optimization studies by Favier et al. \cite{favier2012control} reported that protuberances with spatial amplitudes amounting to 7\% of the chord successfully suppressed wake vortex shedding within the deep stall regime. Larger spatial amplitudes, up to 10\% of the chord, resulted in a more gradual stall. 

Trailing edge serrations, widely seen in nocturnal birds of prey \cite{jaworski2020aeroacoustics}, have been the focus of recent studies for their noise attenuation capabilities. Acoustic studies have reported significantly lower sound pressure levels at frequencies below 1.6 kHz, with no detectable noise beyond 6.3 kHz \cite{SilentFlight}. The underlying noise reduction mechanisms can be categorized into two principal features, namely, the flow control devices located on the wing surface and the trailing edge, the latter of which is relevant to the present study. The key parameters influencing the trailing edge modifications, or serrations, are the profile, spatial amplitude, and spatial wavelength. Optimal noise reduction was observed for trailing edge serrations with an amplitude-to-chord ratio less than 15\% \cite{Optimumdesign}.
Experimental investigations on trailing-edge serrations over a flat plate by Moreau et al.\cite{FlatplateTE} reported a reduction of up to 13 dB in blunt vortex-shedding noise at high frequencies. The noise reduction was observed to be dependent on the Strouhal number and the serration wavelength, with a wavelength-to-amplitude ratio of 0.6 showing higher attenuation levels with no noise increase in the mid-frequency region. The noise reduction capability of the serrations was found to be tied to their influence on the hydrodynamic field at the source locations, particularly in modifying coherent structures in the wake and altering wake turbulence characteristics. Further studies on the extension of the trailing edge serrations by Sundeep et al. \cite{TEextension} showed lower attenuation losses due to flow misalignment, substantial reduction in root flow RMS fluctuations, and an increase in lift-to-drag ratios for mid-range angles of attack. 
\begin{figure}[H]
\centering
\includegraphics[scale=0.6]{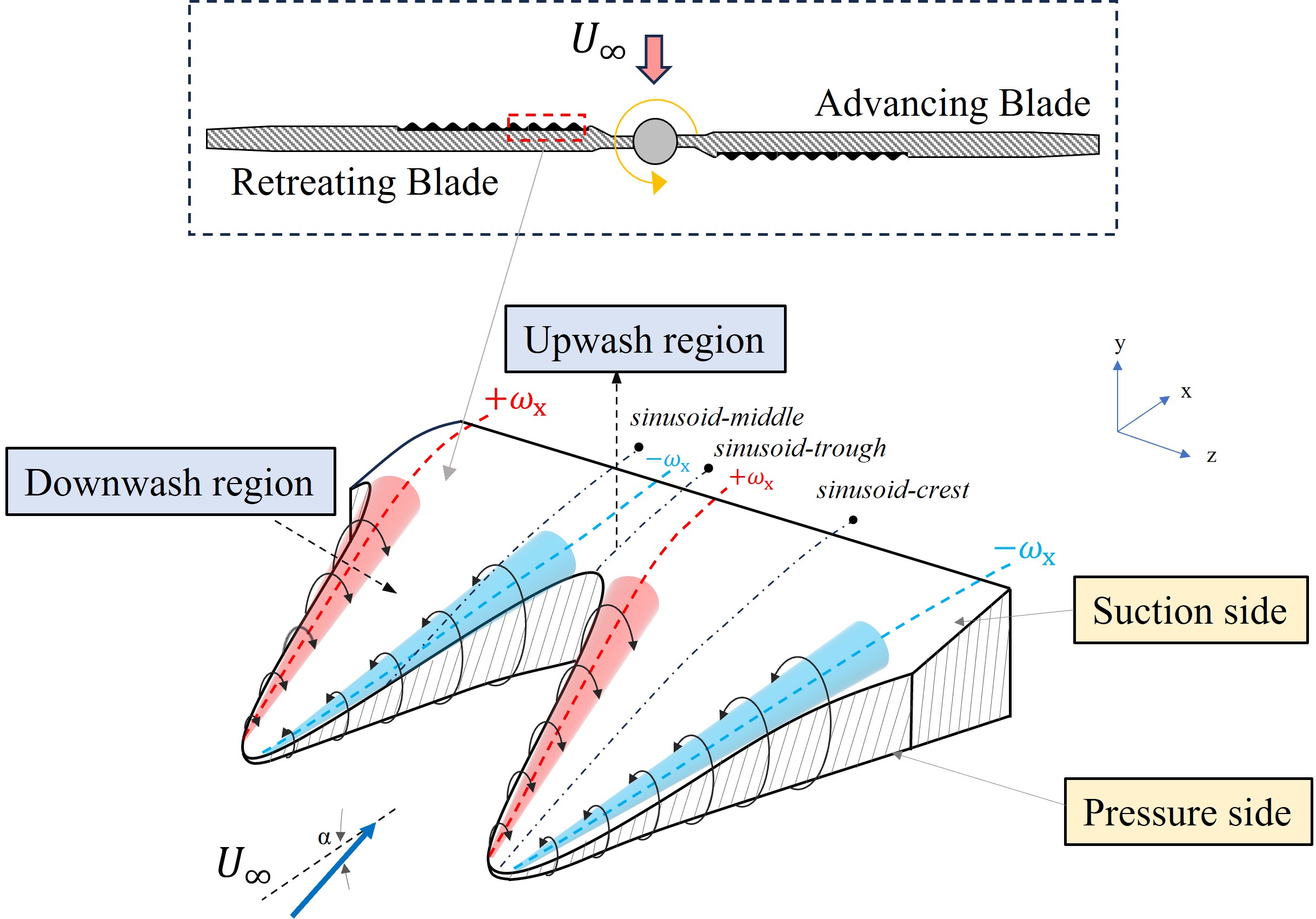}
\caption{Schematic of a two-bladed rotorcraft in forward flight with a proposed sinusoidal trailing edge near the root region of the blades, shown in the inset. Schematic of the expected vortical flowfield over the sinusoids during reverse flow configuration.}
\label{fig:serration vortices-schem}
\end{figure}

The present experimental study aims to investigate the effectiveness of a sinusoidal trailing edge to address the problem of flow separation during reverse flow. In real applications, this modification will only be implemented in the region of the blade that is expected to undergo reverse flow, based on the advance ratio. The sinusoids are expected to function similarly to leading-edge protuberances during reverse flow. The downwash from the sinusoid-generated vortices is expected to turn the incoming flow and align it with the surface of the blade. Thus mitigating flow separation at the sharp aerodynamic leading edge. During forward flow, as the sharp edge transitions to the aerodynamic trailing edge, the sinusoids may help in noise attenuation by acting as trailing edge serrations. The modification is thus effective over the entire revolution of the rotor blade.
A schematic explaining the implementation and the proposed flow control mechanism is shown in Fig. \ref{fig:serration vortices-schem}.  
Preliminary investigations by the authors \cite{Revflow_EAG} have shown the effectiveness of a sinusoidal trailing edge over a NACA 0015 airfoil in reverse flow. It consisted of preliminary planar Particle Image Velocimetry (PIV) measurements over two wall-to-wall, two-dimensional blades with NACA 0015 airfoil geometry. In one of the blades, a sinusoidal trailing edge was implemented, with sinusoidal amplitude and wavelength equal to 10\% of the chord length. Investigations at the midspan location showed that the sinusoidal trailing edge was highly effective in mitigating flow separation induced by reverse flow. However, important questions, such as the effect on the total aerodynamic loads experienced, remained unanswered. The present study consists of experiments on two-dimensional cantilevered NACA 0015 blades. Measurements consist of both planar PIV and aerodynamic forces and moments over a wide range of angles of attack. Aerodynamic loads were measured in both reverse and forward flow configurations. Forward flow measurements being important for a complete evaluation for rotorcraft applications.
Experiments were conducted at two chord-based Reynolds numbers, $Re_c=0.7\times10^5$ and $1.4\times10^5$. These Reynolds numbers can be expected in high-speed rotorcraft operating at an advance ratio of 0.77 \cite{Reeffects}, as well as in rotary-wing MAVs \cite{Timeaveraged}.  
Two sinusoidal trailing-edge blades were examined, both having the same amplitude as the previous study (0.1c) but different wavelengths (0.05c and 0.1c), to evaluate the effect of wavelength on the control authority. The results of this study establish a quantitative basis for evaluating sinusoidal trailing edges over a larger parametric space in reverse flow conditions, enabling a comprehensive characterization of their effects on full-scale rotors.

\section{Experimental Setup}
\subsection{Test Facility}
Experiments were conducted in the low-speed open-circuit wind tunnel facility at the Low-Speed Aerodynamics Laboratory at the Indian Institute of Technology, Kanpur. The test section is 1.1 m long with a rectangular cross-section of 0.5 m $\times$ 0.4 m, and a maximum freestream velocity of 50 m/s. The test section walls are made of clear glass, providing complete optical access from all sides. The inlet has an area contraction of 6.27 and is equipped with a series of flow conditioning devices to provide a streamlined flow in the test section. The freestream velocity in the test section was derived from the differential pressure between the inlet and a location in the contraction region. The differential pressure was measured using a Furness Control FCO560 pressure/flow calibrator having an uncertainty of $\pm$ 0.02 mbar. Experiments were conducted at freestream velocities of $U_{\infty} =$ 10 m/s and 20 m/s, corresponding to aerodynamic chord (= 0.1 m)-based Reynolds numbers of $Re_c = 0.7\times10^5$ and $1.4\times10^5$, respectively. Measurements of aerodynamic forces and moments were conducted at both velocities, while PIV was conducted at 20 m/s only. 
The freestream turbulence intensities in the test section for the two Reynolds numbers, measured using hot-wire anemometry, were found to be 0.3\% and 0.6\%, respectively. 

\subsection{Test Models}
Test models consisted of three wall-to-wall, two-dimensional NACA 0015 airfoil sections, also referred to as blades or airfoils in this study. All airfoils have a chord length of 0.1 m and a span of 0.4 m, giving them an aspect ratio of 4. One of the airfoils has a straight trailing edge and is referred to as the baseline airfoil. The other two airfoils have sinusoidal trailing edge geometries with spatial amplitudes equal to 10\% of the chord length and spatial wavelengths equal to 5\% and 10\% of the chord length, respectively. These modified models are referred to as A10$\lambda$10 and A10$\lambda$05, with A denoting the sinusoidal amplitude and $\lambda$ denoting the sinusoidal wavelength, respectively. It is important to note that the thickness of the NACA 0015 airfoil was not modified during the design of the sinusoids. Therefore, while the crest of the sinusoid was sharp, the trough was bluff, with a thickness equal to that of a NACA 0015 airfoil at 90\% of the chord (refer to the schematic in Fig. \ref{fig:serration vortices-schem}). The sharpness gradually reduced from the crest to the trough, with the bluff regions experiencing local flow separation, as shown later in the results. 
The effect of smoothing or tapering the geometry needs to be investigated in future studies, which will be aimed at optimizing the flow control technique. The test models were mounted vertically in a cantilevered configuration for both aerodynamic loads and PIV measurements. A clearance of approximately 1 mm was maintained between the model and the test section walls at the top and bottom, which is well within the acceptable range of 0.5\% span \cite{barlow1999low}. In the previous preliminary study by the group, which consisted solely of PIV measurements, the model was not cantilevered but instead mounted at both ends \cite{Revflow_EAG}.

The test models were manufactured as hollow shells using SLA (Stereolithography), utilizing ABS (Acrylonitrile Butadiene Styrene) with a print resolution of approximately 0.1 mm. Following fabrication, each blade was sanded and painted black to minimize surface reflections from the incident laser sheet during PIV. The models were mounted on a mild steel spar centered around 30\% of the chord from the geometric leading edge. 
The entire assembly, including the load cell, was mounted on a turntable, which was used to set static pitch angles, as shown in Fig. \ref{load_schm_a}. The static pitch angle for the present study was varied from $\alpha = 180^\circ$  to $\alpha = 200^\circ$ ($0^\circ$ to $20^\circ$ in reverse flow) for load measurements, and from $\alpha = 180^\circ$ to $\alpha = 194^\circ$ for PIV measurements. 
The test models at the maximum pitch angle produced a test section blockage of approximately 6.8\%. Since the blockage is low, no correction is applied to the data presented. 

\subsection{Measurements and Uncertainties}
\begin{figure}[H]
\centering
\begin{subfigure}[t]{0.45\textwidth}
    \vspace{0.1em}
    % \vspace{1em}
    \centering
    \includegraphics[width=\linewidth]{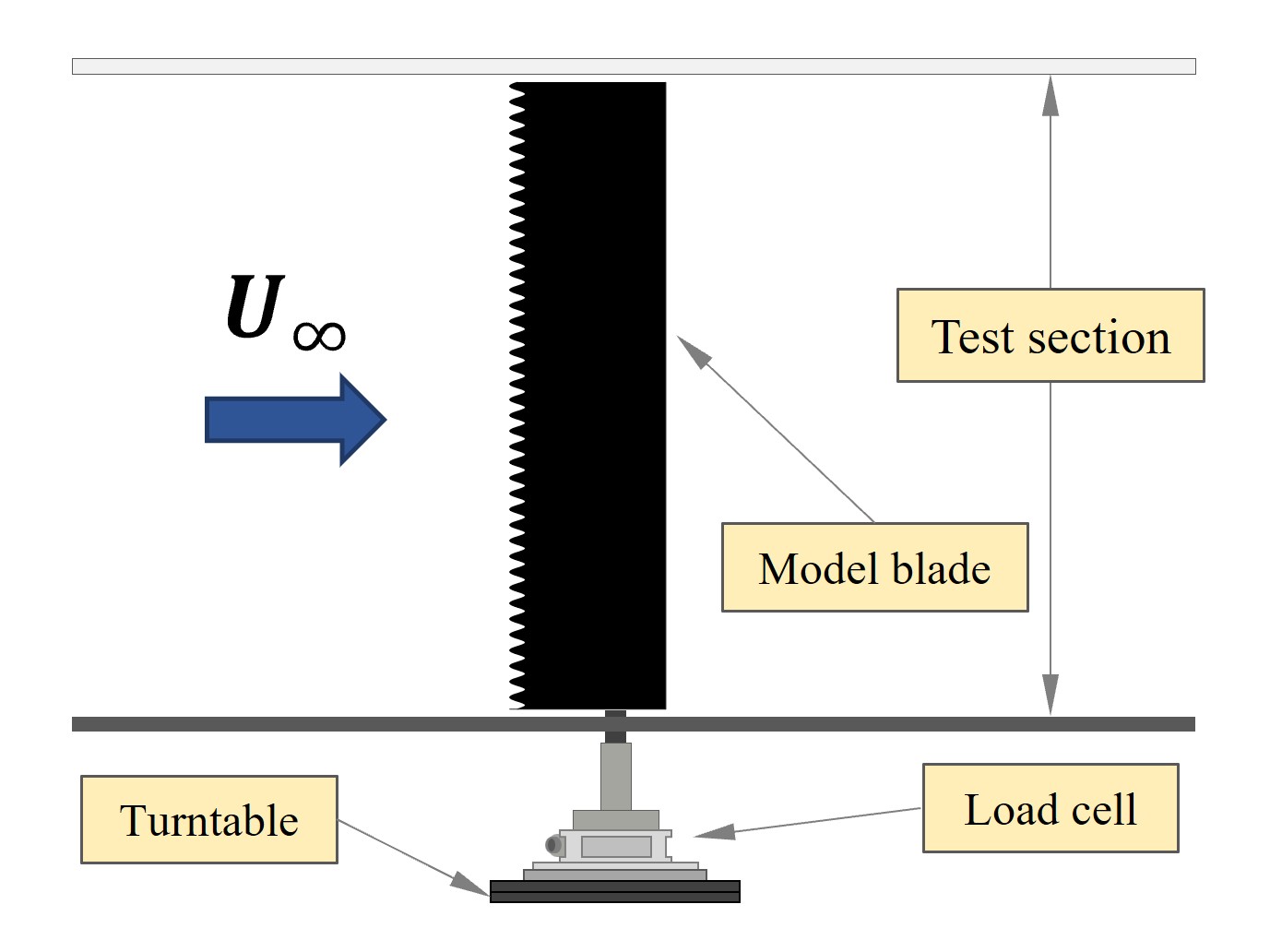}
    \caption{}
    \label{load_schm_a}
\end{subfigure}
% Right 
\begin{minipage}[t]{0.35\textwidth}
    \vspace{0.1em}
    \centering
    
    \begin{subfigure}[t]{0.8\textwidth}
        \centering
        \includegraphics[trim = 0 80 0 0,clip,width=\linewidth]{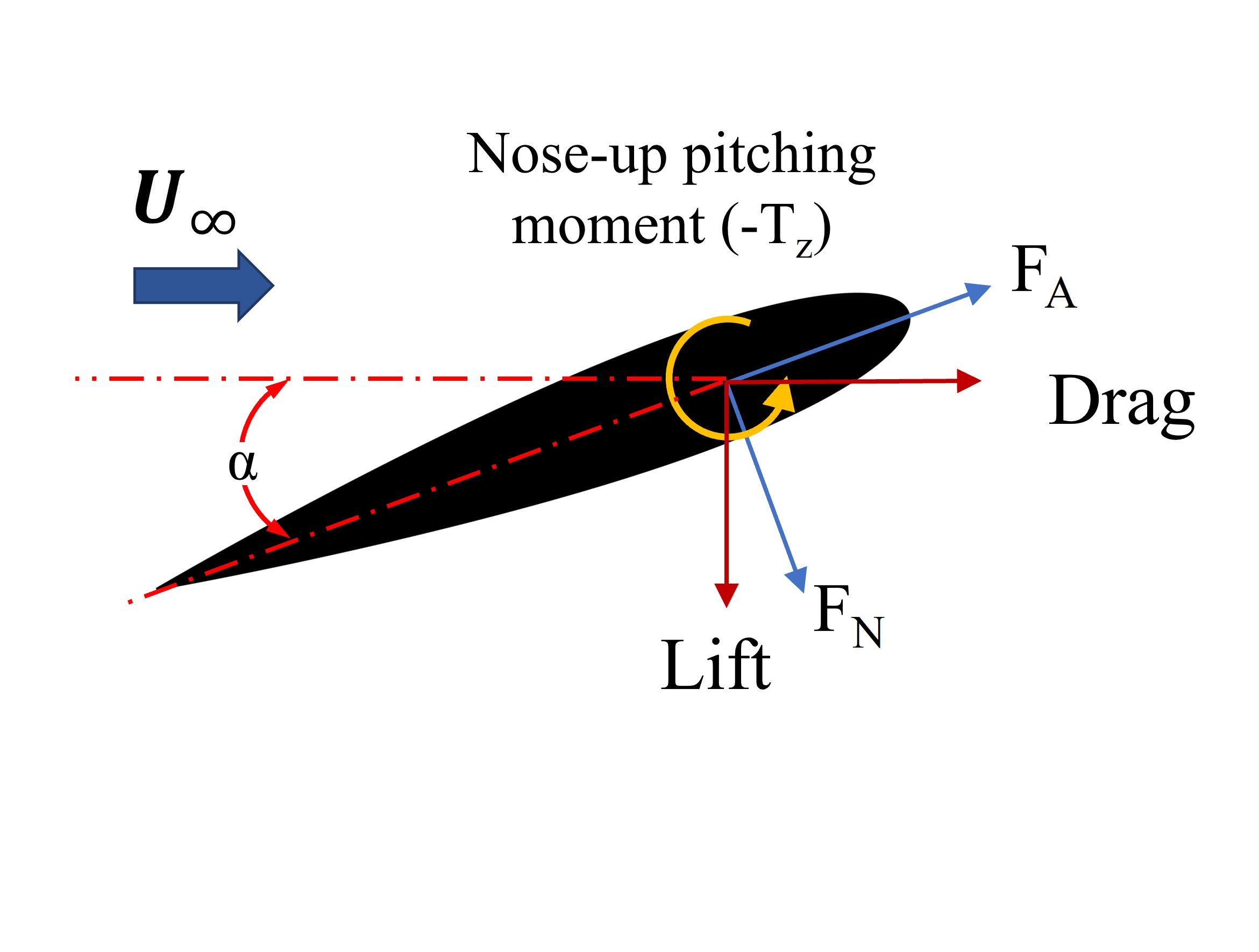}
        \caption{}
        \label{load_schm_rev}
    \end{subfigure}
    
    \vspace{0.5em}
    
    \begin{subfigure}[t]{0.8\textwidth}
        \centering
        \includegraphics[width=\linewidth]{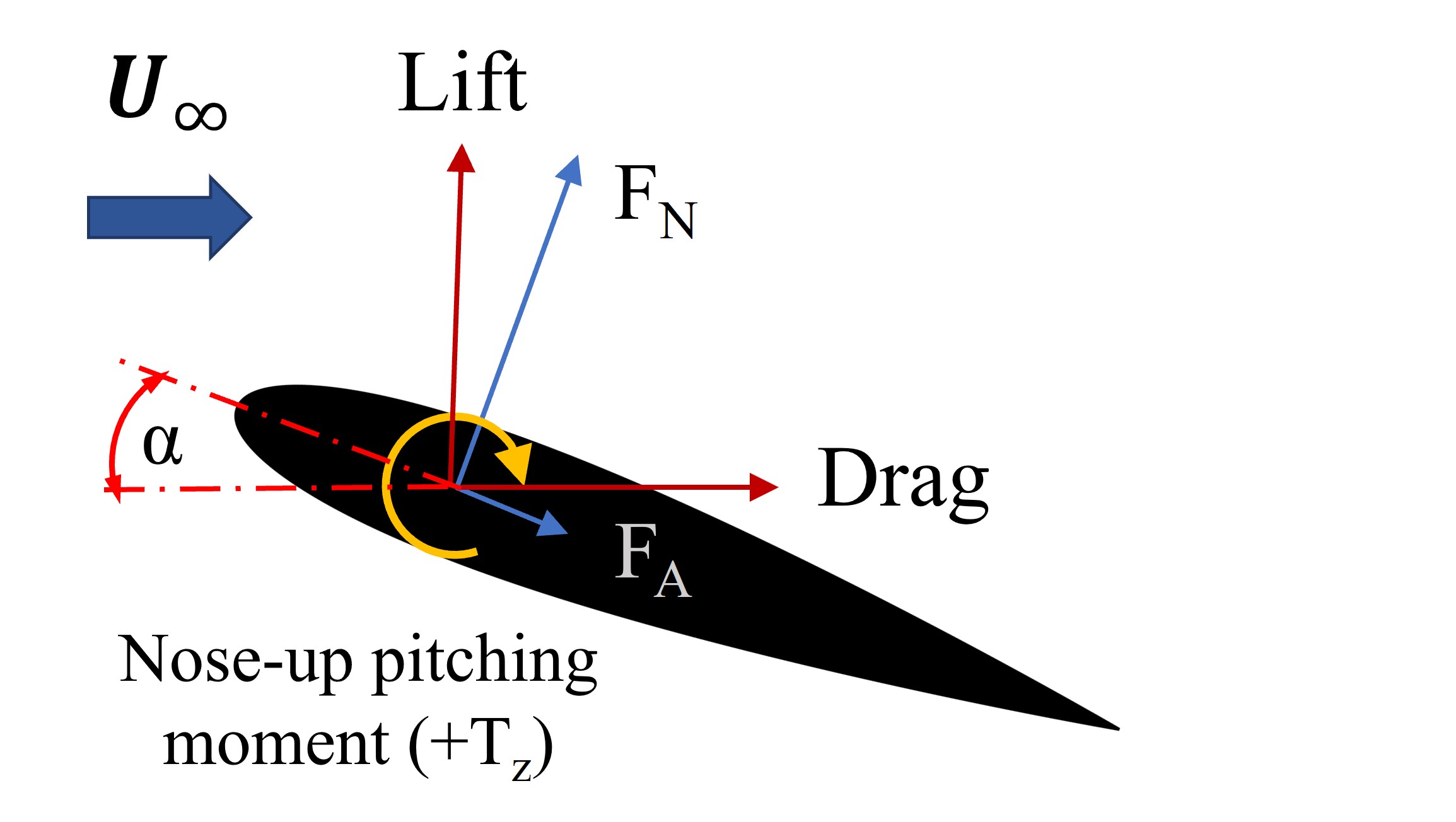}
        \caption{}
    \end{subfigure}
    
\end{minipage}
\caption{Schematics of the experimental setup showing front view of the model mounted in the test section (a), and the orientation of the load cell axes for the measurement of aerodynamic forces and moments under reverse flow (b), and forward flow conditions (c).}
\label{load_schm}
\end{figure}

Measurements consisted of the acquisition of the aerodynamic forces and moments experienced by the model using a six-component ATI\textsuperscript{\textregistered} Delta force/torque sensor. The sensor center was aligned with the model’s center of rotation, located at 0.3c from the geometric leading edge, which under reverse flow conditions corresponds to 0.7c from the sharp aerodynamic leading edge. 
A front view of the orientation of the model and the sensor, along with the measured forces and moments in both reverse and forward flow, is shown in Fig. \ref{load_schm}. Note that the nose-up pitching moment is clockwise ( positive, $+T_Z$ ) in forward flow and counter-clockwise (negative, $-T_Z$ ) in reverse flow conditions. To avoid confusion, the nose-up pitching moment has been considered to be positive while reporting the pitching moment coefficient, independent of the airfoil orientation.  

The orientation of the load cell relative to the airfoil chord remained fixed at all static pitch angles, as both were mounted on the turntable. $F_{A}$ and $F_{N}$ correspond to the aerodynamic axial and normal forces, respectively, whereas $T_Z$ corresponds to the pitching moment. 
The measurement resolution was 1/32 N for $F_{A}$ and $F_{N}$, and 1/528 N.m for $T_Z$. Data acquisition was conducted using a National Instruments\textsuperscript{\textregistered} USB 6210 multifunction DAQ, which was controlled using the LabVIEW software.  Measurements were acquired at 6.5 kHz over a period of 30 s for all static pitch angles. The statistical mean of the measured data was used to compute the time-averaged lift, drag, and pitching moment coefficients using MATLAB\textsuperscript{\textregistered}. All pitching moment coefficients reported in this study were calculated about the geometric quarter-chord location, 0.25c, though measurements were conducted at the geometric location of 0.3c.

\begin{figure}[H]
\begin{subfigure}[b]{0.45\textwidth}
        \centering 
        \includegraphics[width=\textwidth]{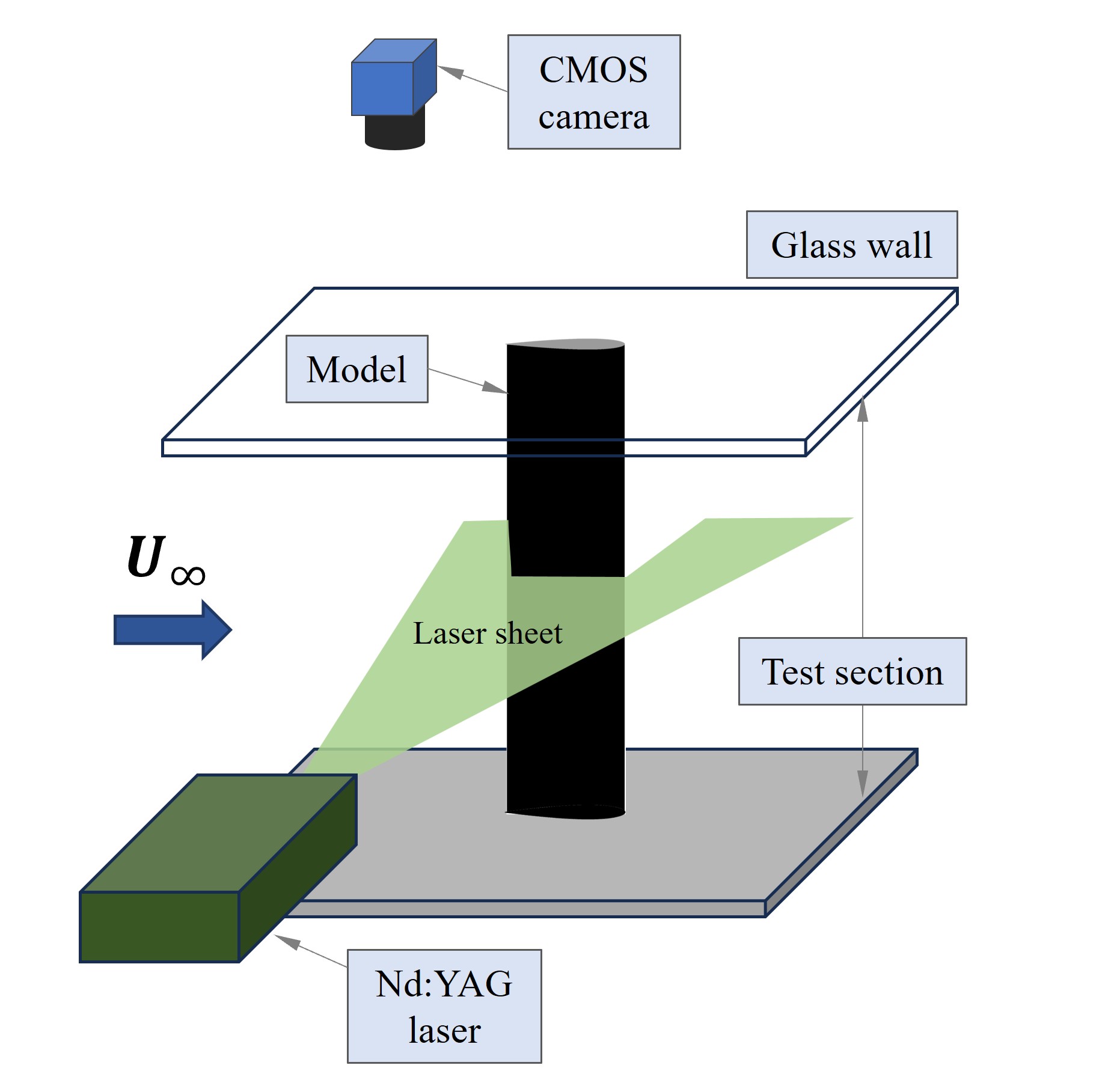}
        \caption{}
        \label{}
    \end{subfigure}
    \begin{subfigure}[b]{0.333\textwidth}
        \centering 
        \includegraphics[width=\textwidth]{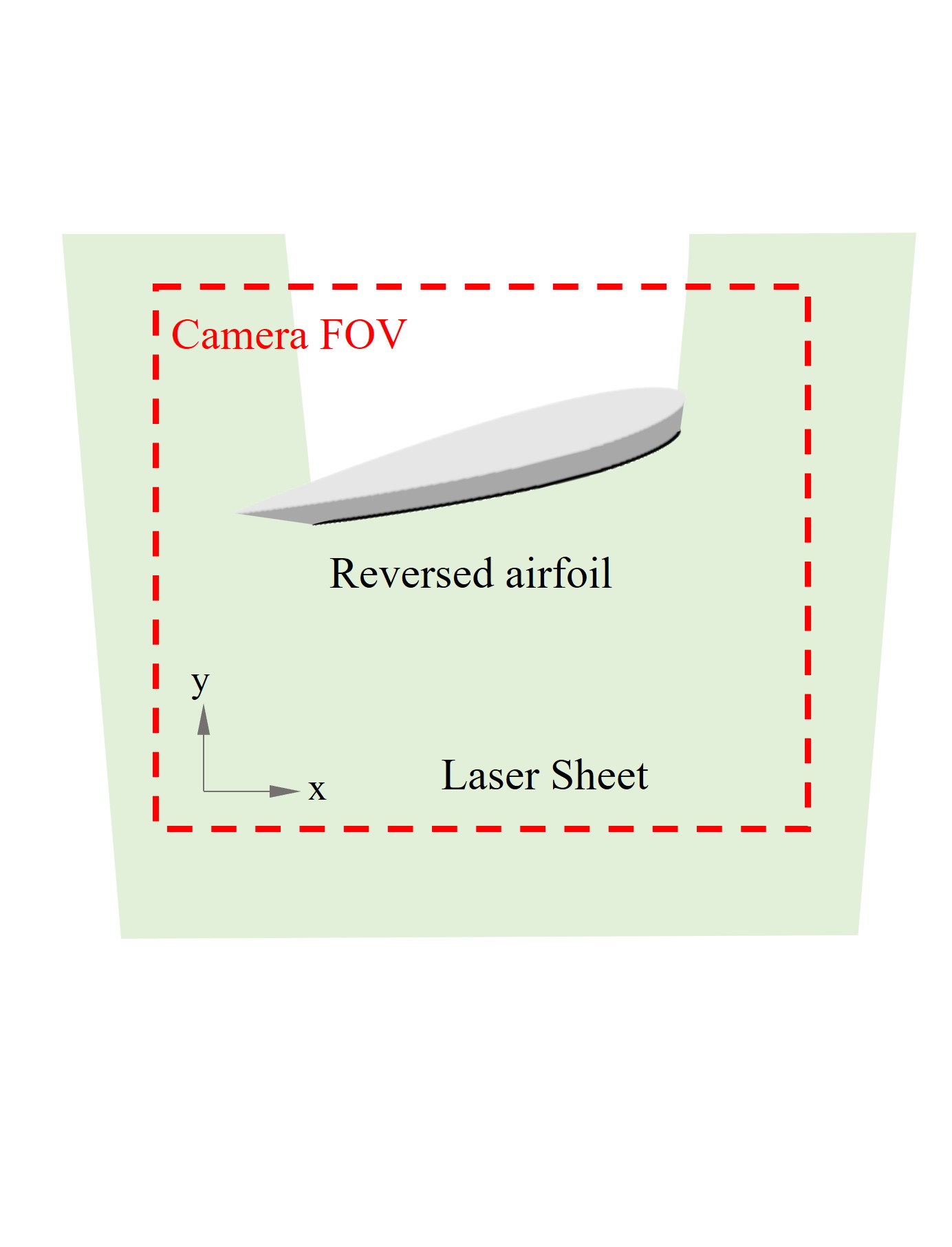}
        \caption{}
        \label{PIV_schm_fov}
    \end{subfigure}
        \begin{subfigure}[b]{0.2025\textwidth}
        \centering 
        \includegraphics[width=\textwidth]{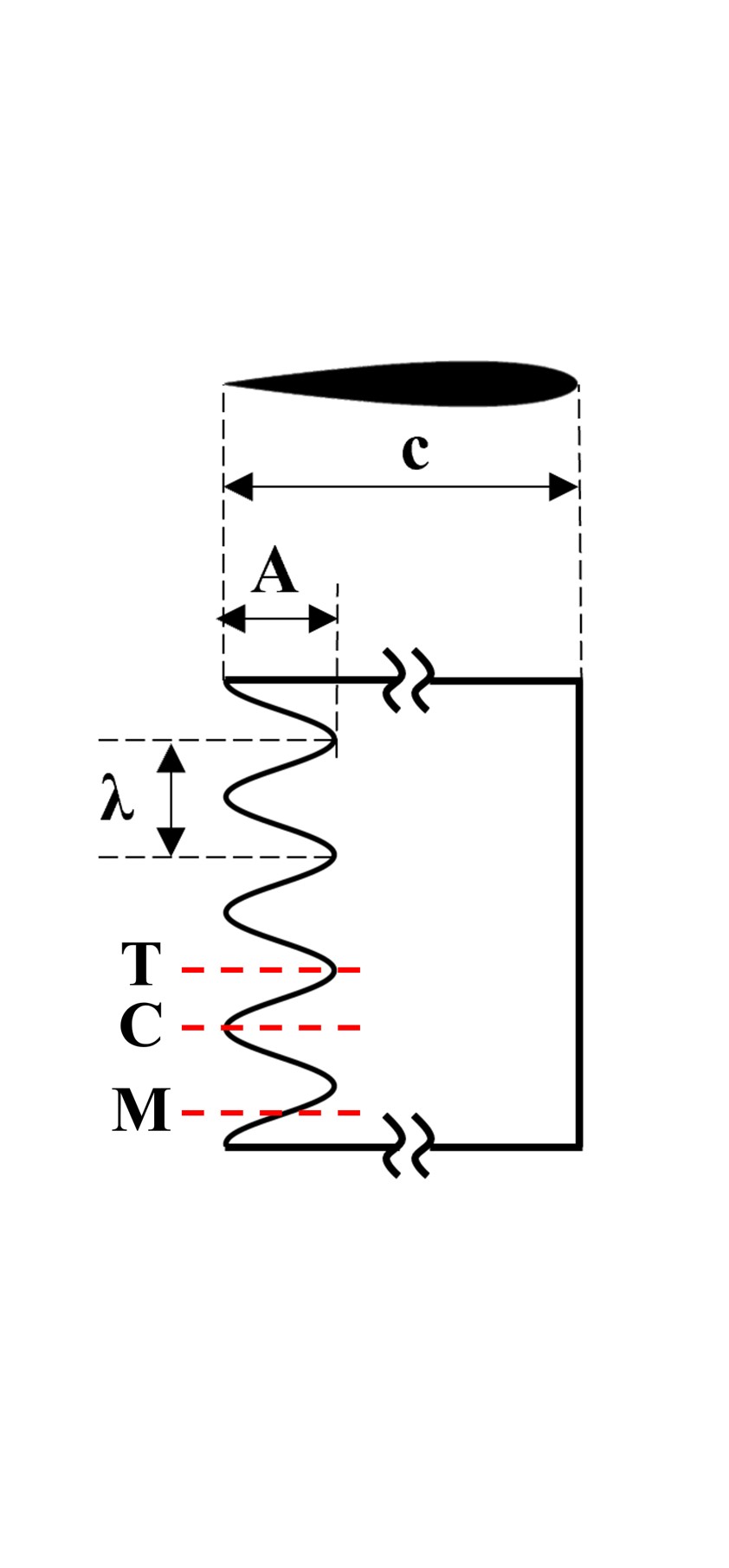}
        \caption{}
        \label{Planes_PIV}
    \end{subfigure}
    
    \caption{Schematics showing the planar PIV setup (a), camera field of view (FOV) of the interrogation window (b), and planes of interrogation for the models with a sinusoidal trailing edge (c). %laser sheet is absent on the pressure side (top in Fig 2b) due to obstruction by the model. 
    Flow is from left to right in all cases.}
    \label{PIV_schm}
\end{figure}

During reverse flow, the suction and pressure sides refer to the bottom and top sides of the airfoil, as per the orientation in Figs. \ref{load_schm_rev} (and  \ref{PIV_schm_fov}). The stagnation point is on the top side during reverse flow, resulting in higher pressure, even though the flow is primarily converging, making it the pressure side. In comparison, the pressures are lower on the bottom side, even though the flow is diverging, making it the suction side \cite{Cambermorphcomputation}. Thus, the sides are reversed relative to conventional forward flow. 
The flowfield on the suction side of the airfoil was analyzed using planar time-averaged PIV, for which the setup and camera field of view are shown in Fig. \ref{PIV_schm}. Measurements were conducted at the midspan plane for the baseline airfoil. For the sinusoidal trailing edge airfoils, measurements were conducted at three separate planes, located on a midspan sinusoid. These planes correspond to the crest, middle, and trough of a sinusoid, and are referred to as C, M, and T, respectively. Three planes were investigated, as the flowfield is expected to vary between these planes, as indicated by Fig. \ref{fig:serration vortices-schem}. A schematic of these planes of interrogation is shown in Fig. \ref{Planes_PIV}. 

During PIV, the flow was seeded using water-based droplets having a diameter on the order of 1$\mu m$.
The particles were illuminated using a 200 mJ Neodymium-doped Yttrium Aluminum Garnet (Nd: YAG) double-pulsed Quantel EverGreen\textsuperscript{\textregistered} big sky laser. The incident beam was focused to a point using an adjustable spherical lens (focal length from 300 to 4000 mm) and fanned into a sheet using a cylindrical lens (focal length of -15 mm).
Images were acquired using a five-megapixel complementary-metal oxide semiconductor (CMOS) Imager CX3 camera fitted with a 50 mm Nikon\textsuperscript{\textregistered} lens. The field of view was kept constant across all measured angles of attack. Five hundred instantaneous image pairs were acquired for all pitch angles investigated. 
Data acquisition and processing were conducted using LaVision\textsuperscript{\textregistered} DaVis 11 software to obtain the resultant vector fields. Vector processing was done using a multi-pass algorithm with decreasing pass sizes, with initial and final window sizes of 48$\times$48 and 24$\times$24, respectively, and overlaps of 50\% and 75\% for initial and final passes, respectively. Adaptive weighting was used for the final two passes to account for the strong velocity gradients associated with the sharp edge separation. The final vector resolution of the flowfield is 0.44 mm, which is 0.44\% of the aerodynamic chord. The velocity contours presented in the results were generated using MATLAB\textsuperscript{\textregistered}. 
Uncertainties in PIV measurements can be attributed to factors such as particle size, seeding inhomogeneity, laser speckle, and particle displacements due to large velocity gradients in regions of high shear. In the present study, uncertainty fields for the planar PIV measurements were obtained from DaVIS, which employs the correlation statistics methods suggested by Wieneke \cite{wieneke2015piv}. The uncertainty levels inside and outside the regions of shear are less than 3\% and 1\% of the mean freestream velocity, respectively. 
\section{Results}
The results section is divided into two subsections. The first subsection presents the time-averaged PIV flowfields, which are used to explain the time-averaged static forces and moments in the second. 
\subsection{Static Pitch: PIV}
\label{PIV}
\subsubsection{Baseline airfoil}
The time-averaged streamwise velocity contours on the suction side of the baseline airfoil at selected static pitch angles ($\alpha = 180^\circ \, \text{to}\, 194^\circ\, \text{in steps of } 2^\circ$) are shown in Fig.\ref{Baseline_PIV}. These angles correspond to $\alpha = 0^\circ\, \text{to } 14^\circ$ in reverse flow. The contours are normalized with the freestream velocity and overlaid with in-plane streamlines. The airfoil geometry is also overlaid within the masked region, primarily for visual reference of the sharp leading edge. A white dashed contour line is used to represent the location where the mean streamwise velocity component is zero. Within this region, the mean streamwise velocity is negative. Data on the pressure side of the airfoil (top of the contour plots in Fig.\ref{Baseline_PIV}) are unavailable due to the obstruction of the laser sheet by the model. A schematic of the airfoil leading edge and a blue line representing the plane of investigation are also included for reference.
\begin{figure}[H]
\centering
\begin{subfigure}[b]{0.44\textwidth}
    \includegraphics[width=\textwidth]{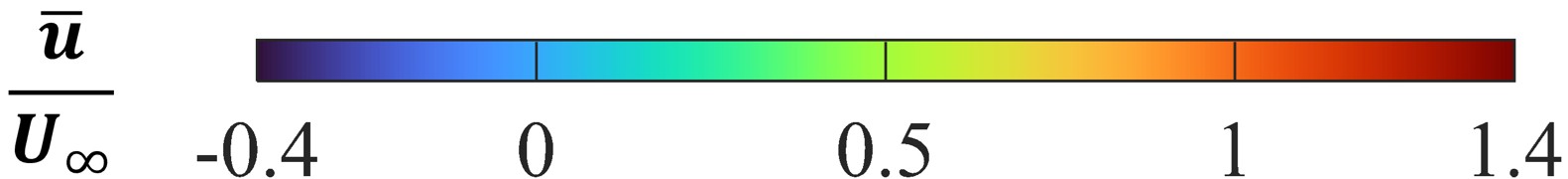}
\end{subfigure}
\hspace{6em}
\begin{subfigure}[b]{0.09\textwidth}
    \centering 
    \includegraphics[trim=0 12 0 0,clip,width=\textwidth]{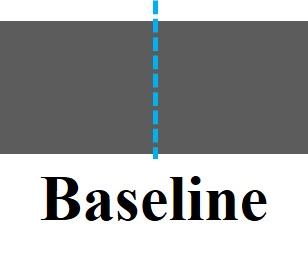}
\end{subfigure}
\vskip\baselineskip
\begin{subfigure}[b]{0.351\textwidth}
    \centering 
    \includegraphics[trim=0 34 0 0,clip,width=\textwidth]{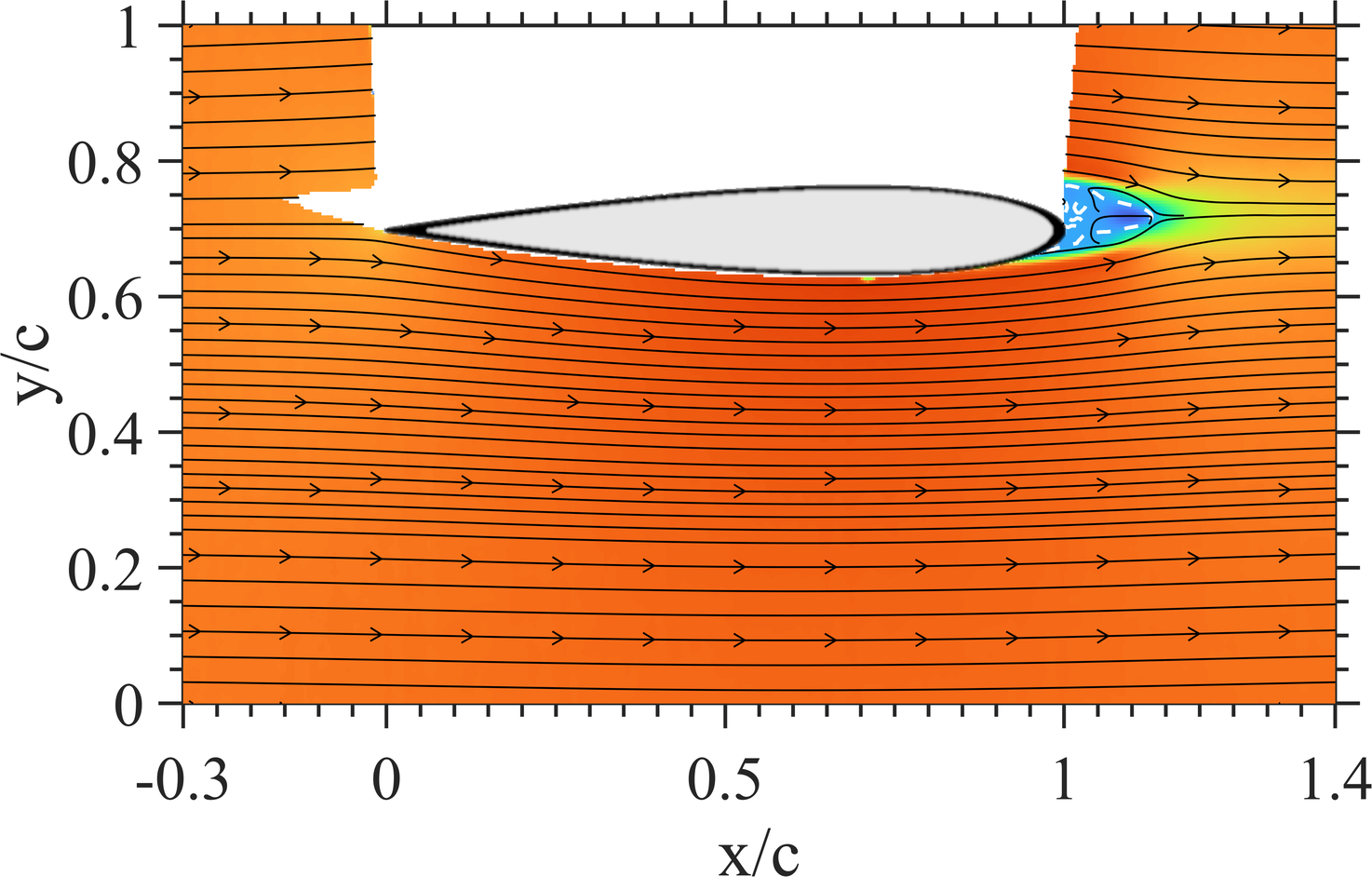}
    \caption{$\alpha$ = 180$^\circ$}
    \label{}
\end{subfigure}
% \hspace{\linewidth}
\begin{subfigure}[b]{0.315\textwidth}
    \centering 
    \includegraphics[trim=33 34 0 0,clip,width=\textwidth]{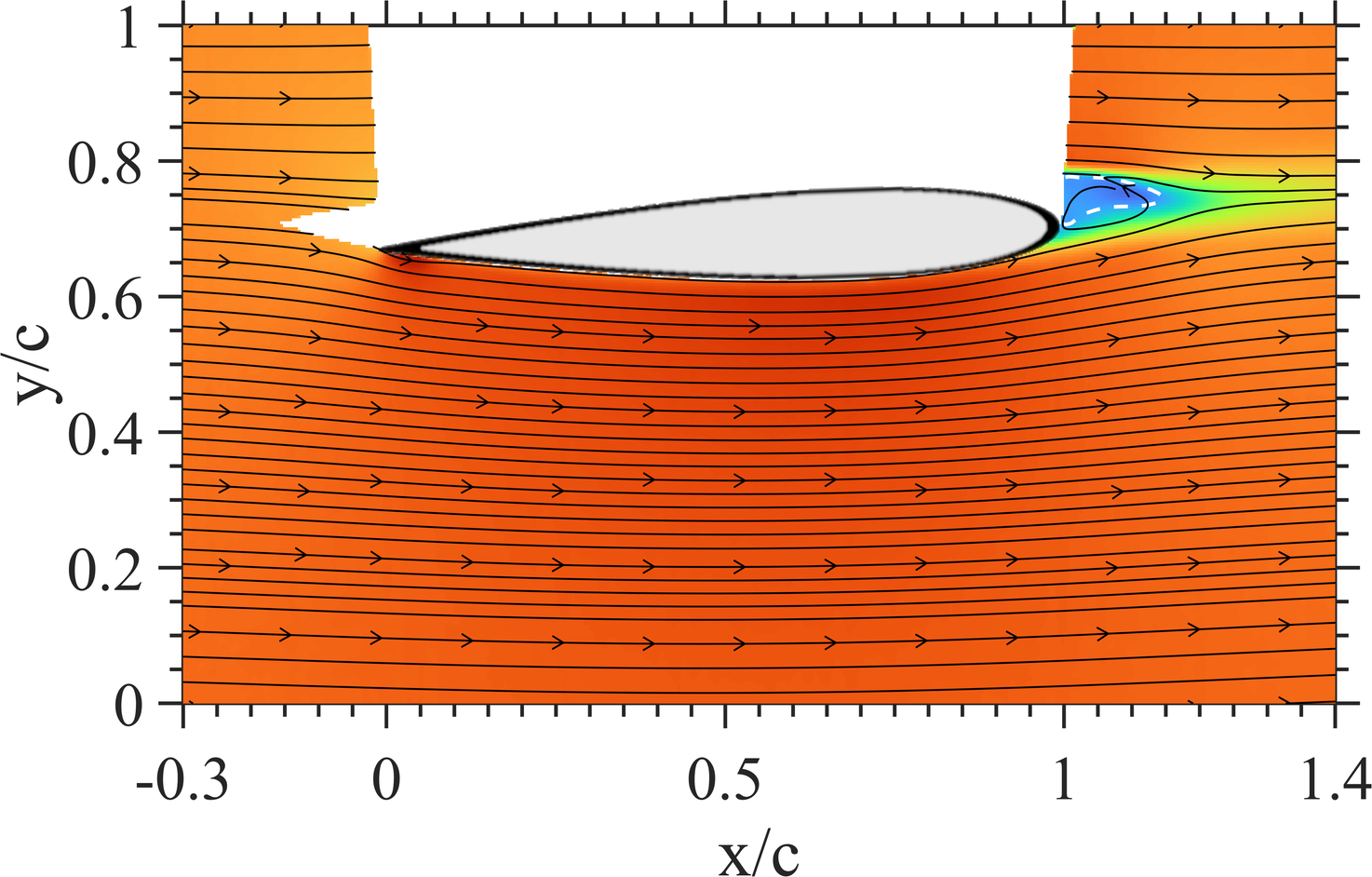}
    \caption{$\alpha$ = 182$^\circ$}
    \label{}
\end{subfigure}
% \hspace{0.5\linewidth}
\begin{subfigure}[b]{0.315\textwidth}
    \centering 
    \includegraphics[trim=33 34 0 0,clip,width=\textwidth]{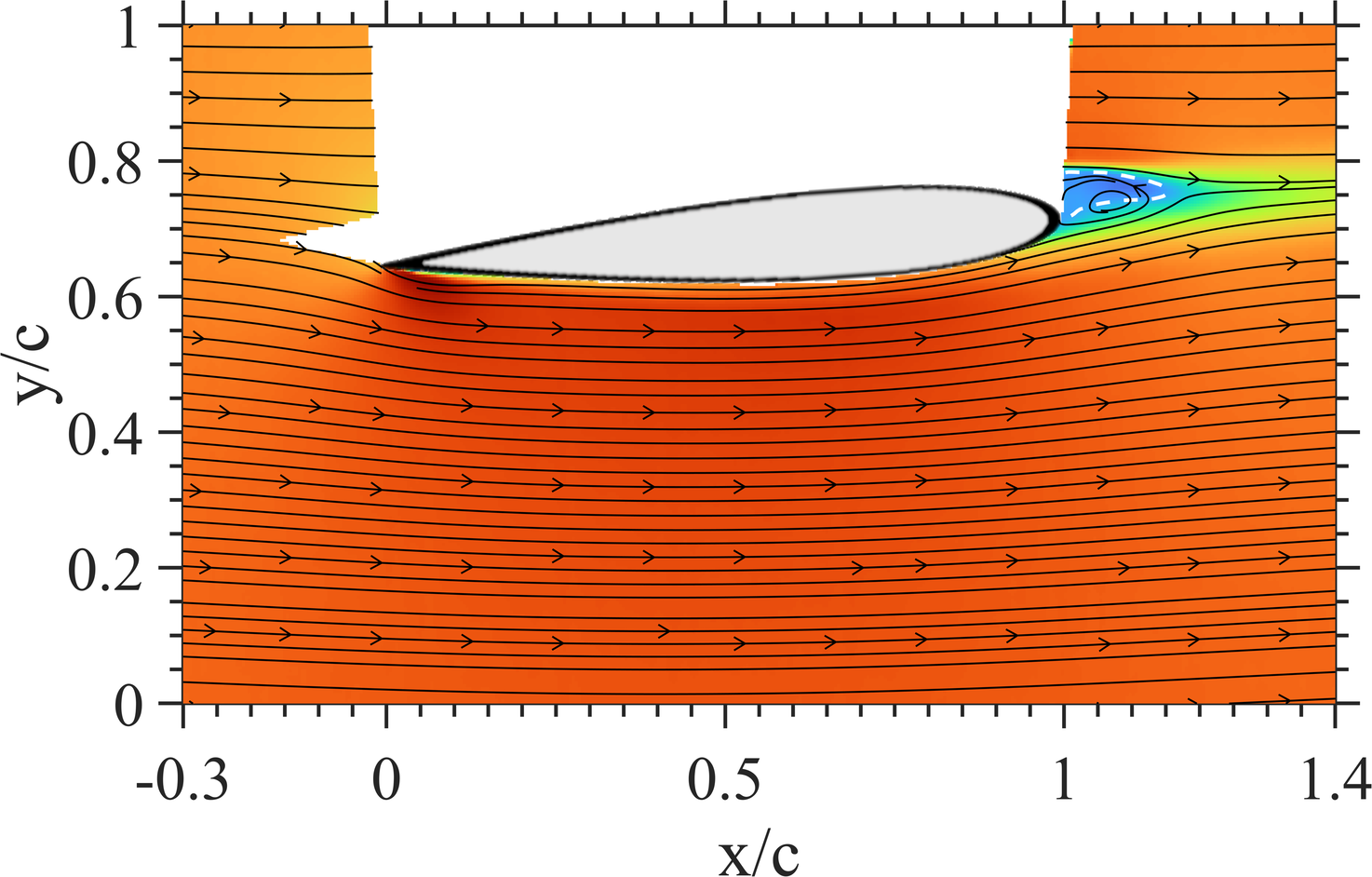}
    \caption{$\alpha$ = 184$^\circ$}
    \label{}
\end{subfigure}
\vskip\baselineskip
\begin{subfigure}[b]{0.351\textwidth}
    \centering 
    \includegraphics[trim=0 0 0 0,clip,width=\textwidth]{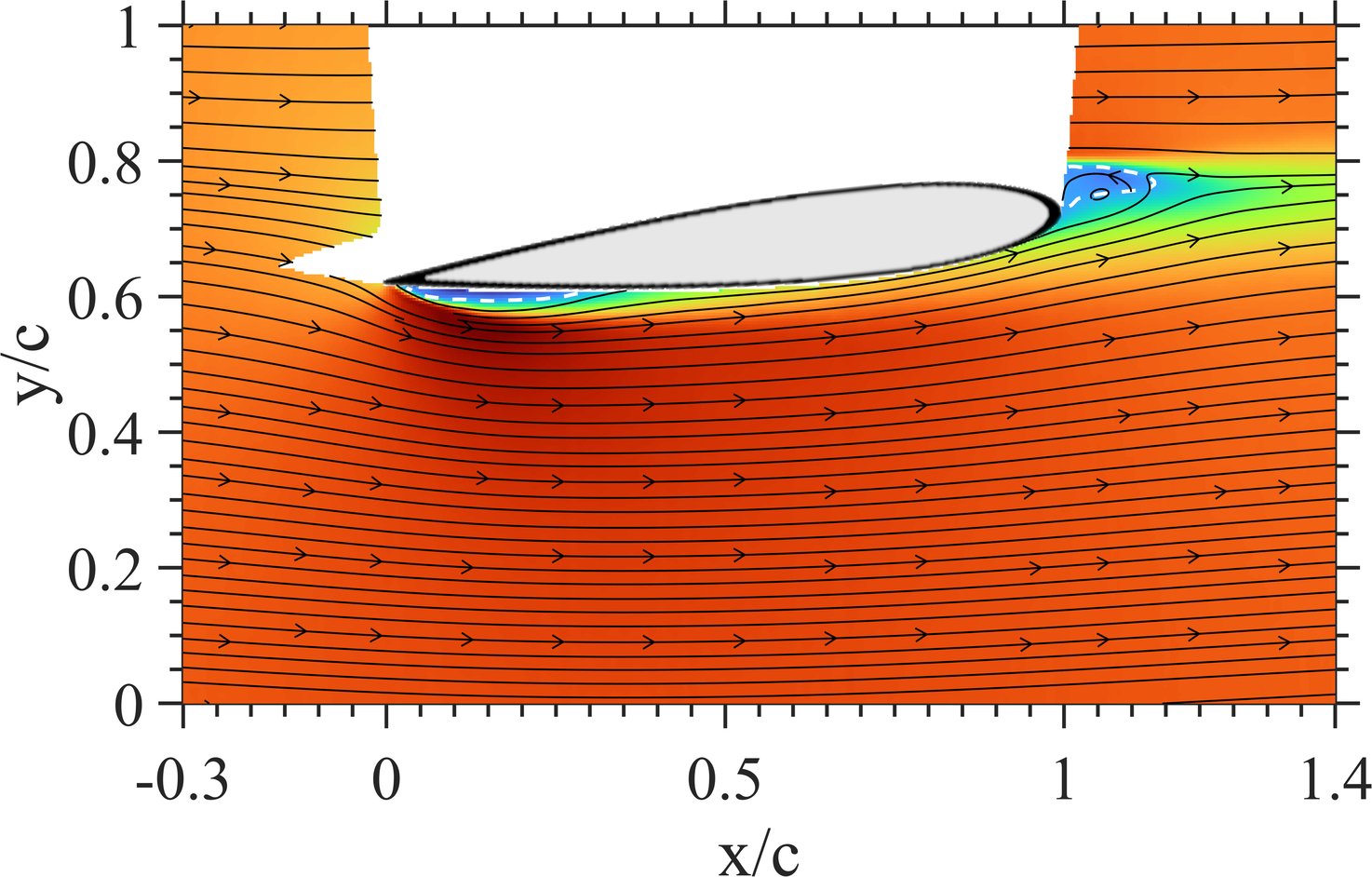}
    \caption{$\alpha$ = 186$^\circ$}
    \label{}
\end{subfigure}
\begin{subfigure}[b]{0.315\textwidth}
    \centering 
    \includegraphics[trim=33 0 0 0,clip,width=\textwidth]{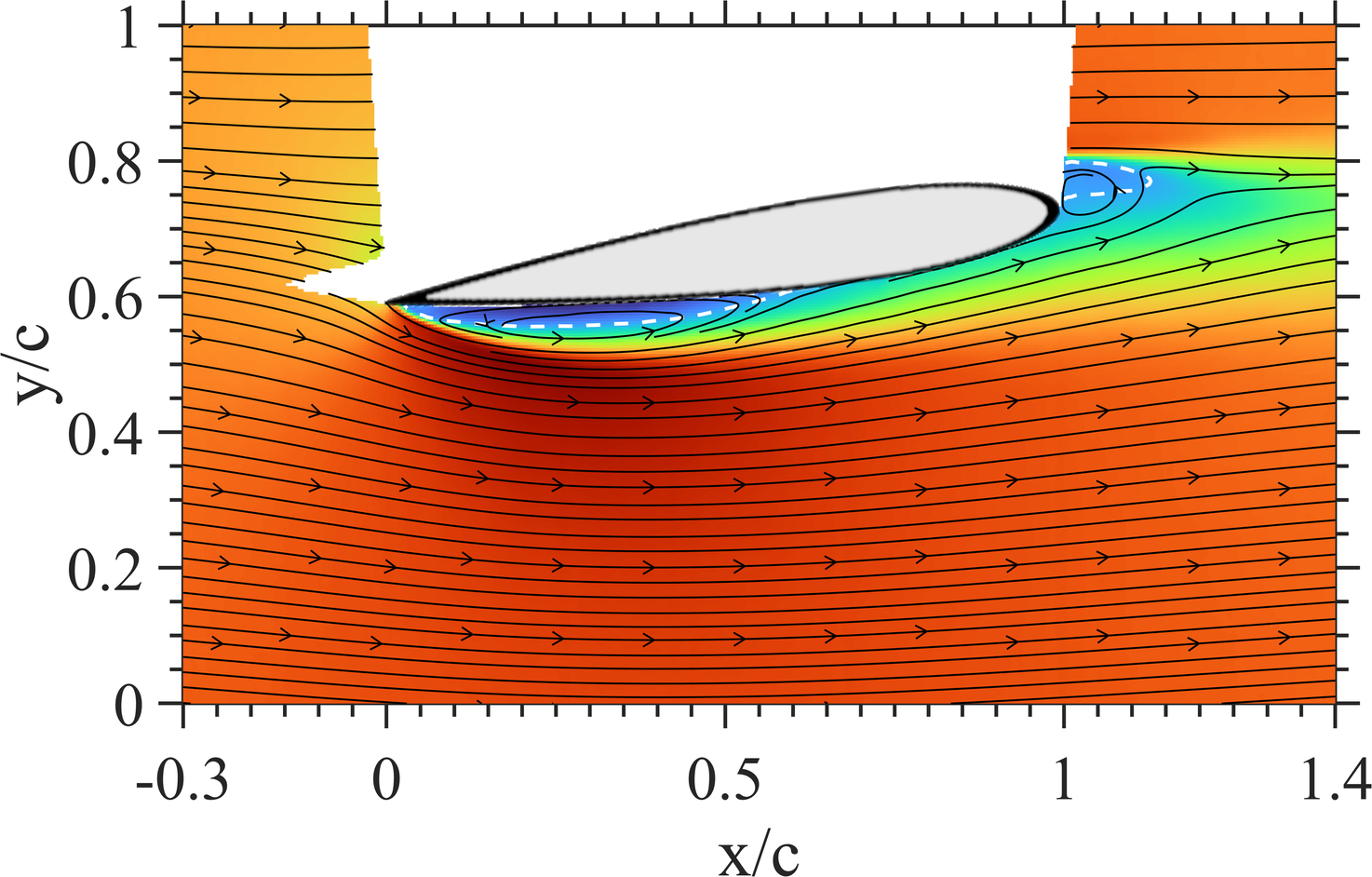}
    \caption{$\alpha$ = 188$^\circ$}
    \label{}
\end{subfigure}
\begin{subfigure}[b]{0.315\textwidth}
    \centering 
    \includegraphics[trim=33 0 0 0,clip,width=\textwidth]{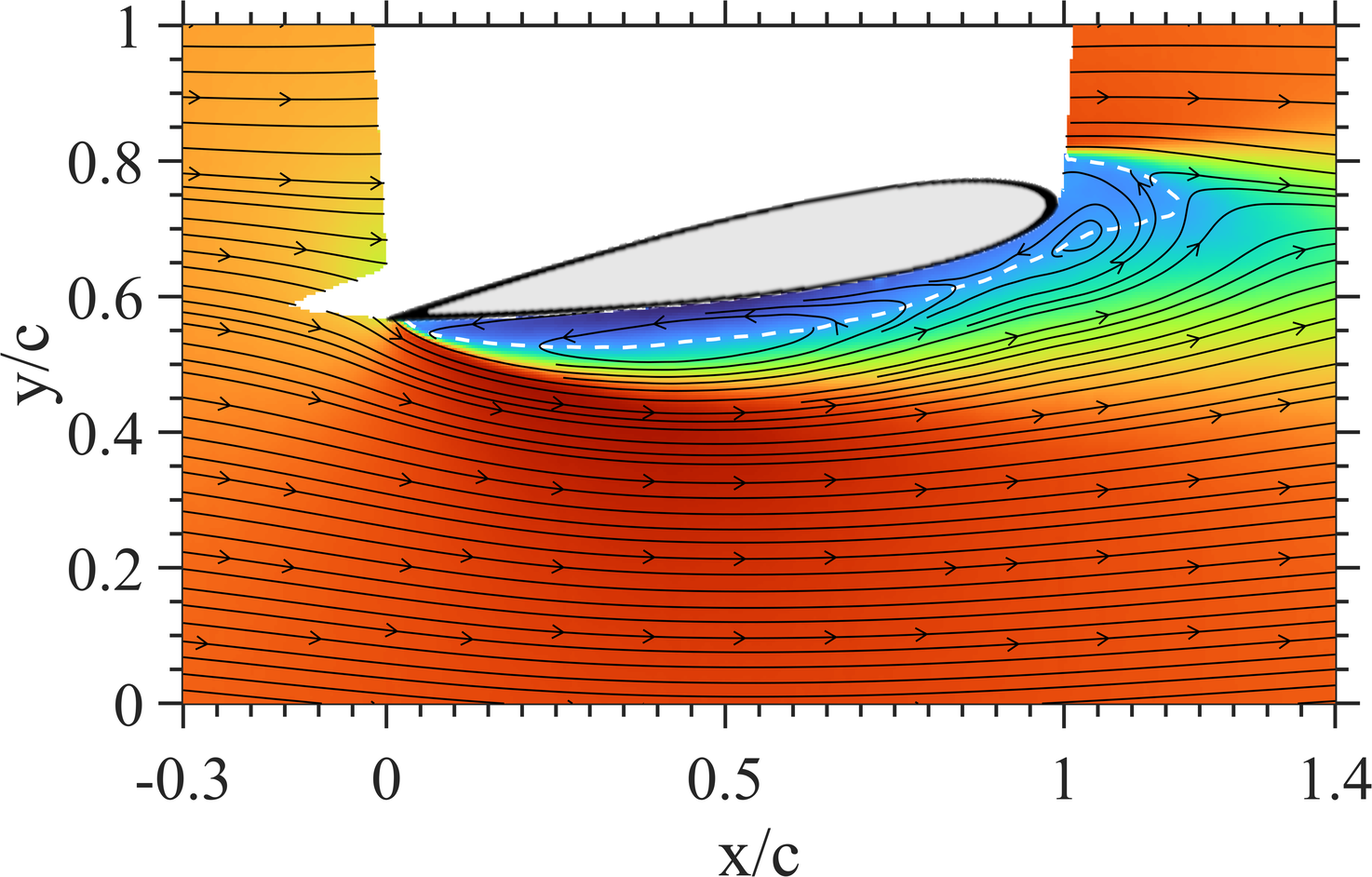}
    \caption{$\alpha$ = 190$^\circ$}
    \label{}
\end{subfigure}
% \vskip\baselineskip
\end{figure}
\begin{figure}[H]\ContinuedFloat
\hspace{7em}
\begin{subfigure}[b]{0.351\textwidth}
    \centering 
    \includegraphics[trim=0 0 0 0,clip,width=\textwidth]{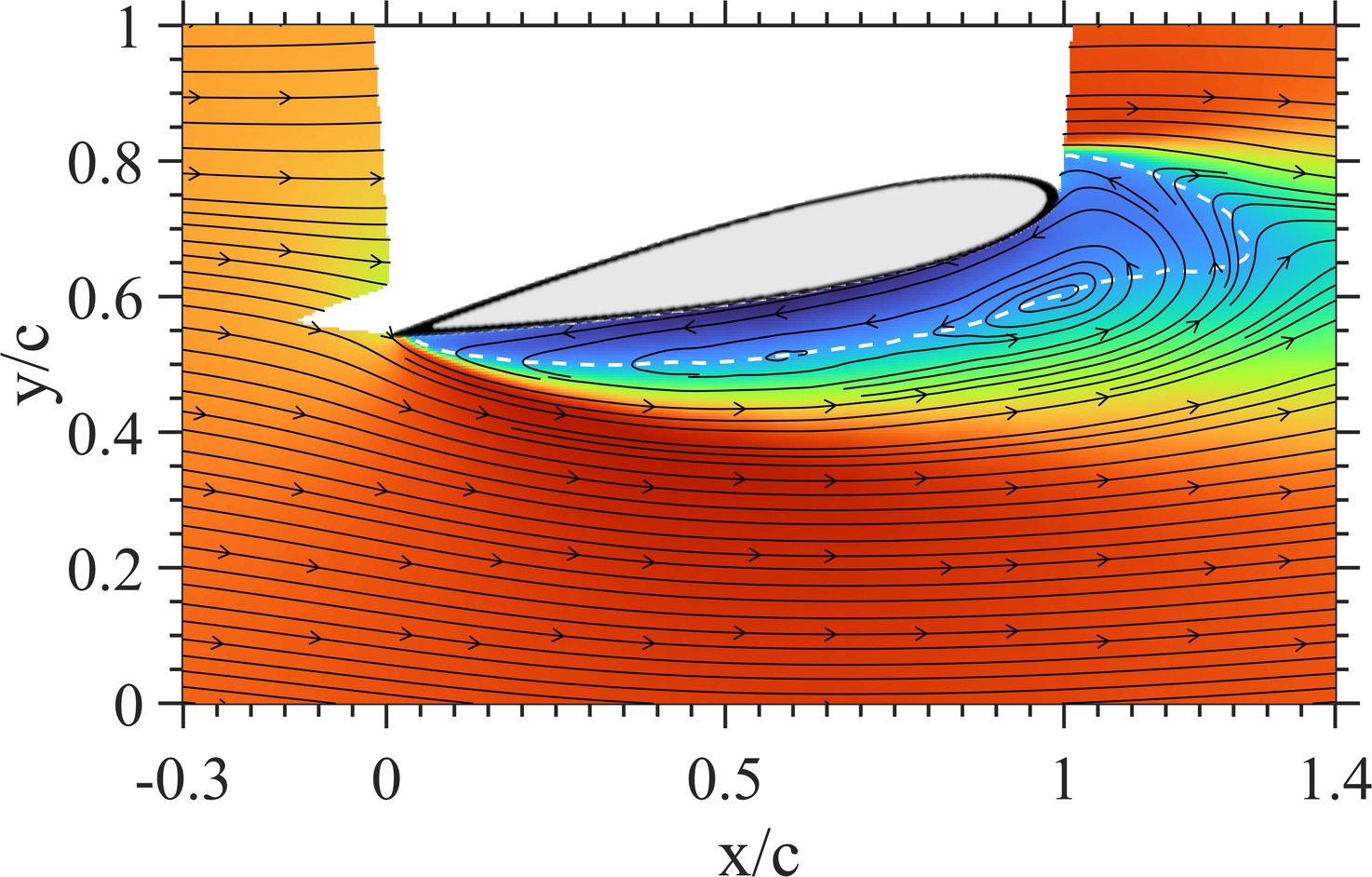}
    \caption{$\alpha$ = 192$^\circ$}
    \label{}
\end{subfigure}
\begin{subfigure}[b]{0.315\textwidth}
    \centering 
    \includegraphics[trim=33 0 0 0,clip,width=\textwidth]{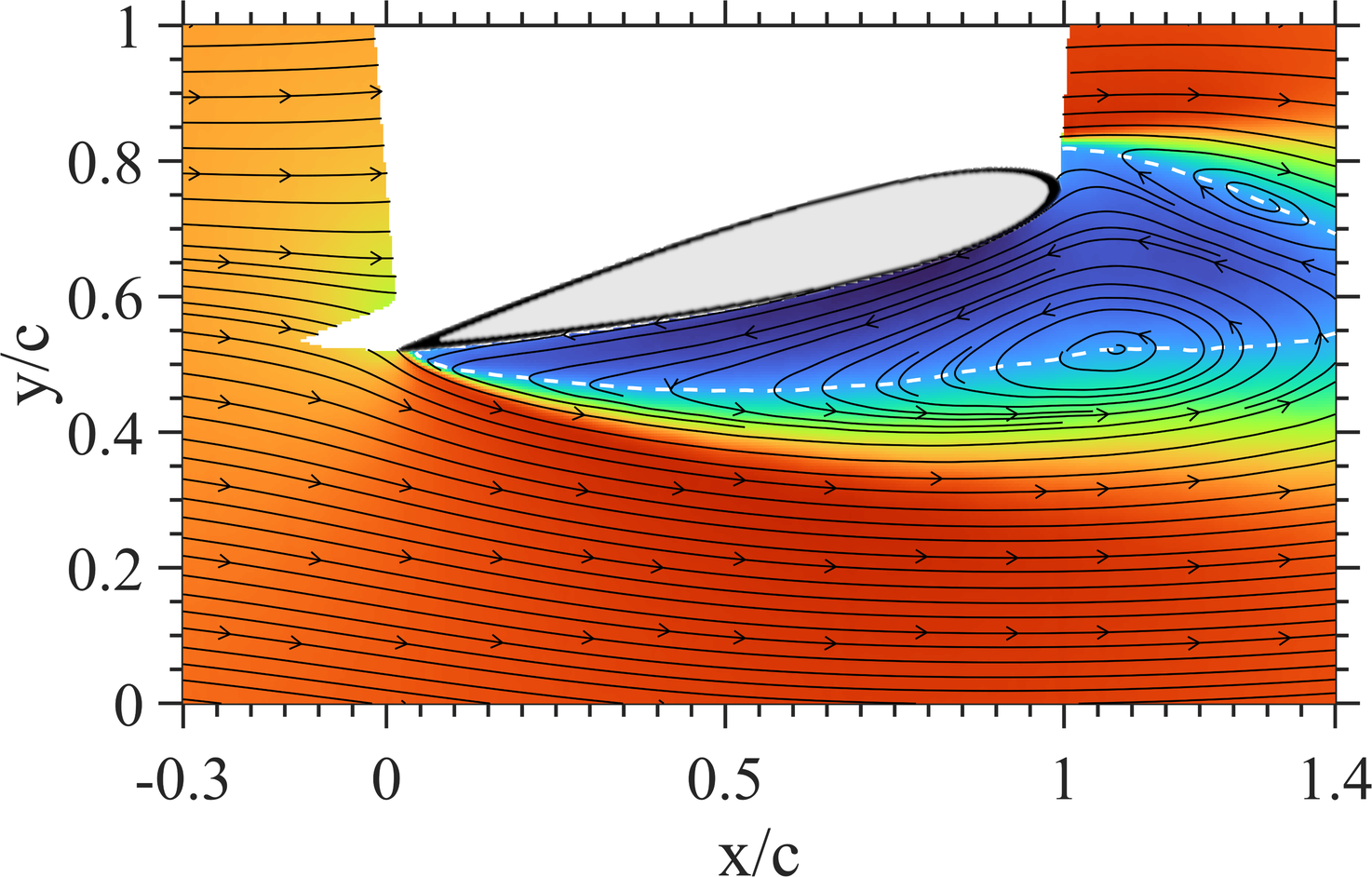}
    \caption{$\alpha$ = 194$^\circ$}
    \label{}
\end{subfigure}
\caption{Contour plots of the time-averaged streamwise velocity component, normalized with freestream velocity for the baseline airfoil at the midspan plane, at static pitch angles of 180$^\circ$ (a), 182$^\circ$ (b), 184$^\circ$ (c), 186$^\circ$ (d), 188$^\circ$ (e), 190$^\circ$ (f), 192$^\circ$ (g), and 194$^\circ$ (h), overlaid with in-plane streamlines. The white dashed contour line represents the location where mean streamwise velocity is zero. Flow direction is from left to right and $\mathbf{Re_c = 1.4\times10^5}$.}
\label{Baseline_PIV}
\end{figure}
For the baseline airfoil at $\alpha = 180^\circ$, the flow remains fully attached until near the blunt trailing edge, where it separates to form a bluff-body type wake. At $\alpha=182^\circ$, a region of accelerated flow is seen over the sharp leading edge, which may suggest a small separation bubble, an observation that is later corroborated by the load measurements. A small separation bubble is identifiable at $\alpha=184^\circ$, and a clear bubble is observed at $\alpha=186^\circ$. 
The separation bubble grows in chordwise extent with increasing incidence. The zero streamwise-velocity (white) contour suggests that the bubble extends up to $\sim$0.25c at $\alpha = 186^\circ$ and to $\sim$0.6c at $\alpha = 188^\circ$. The cross-stream (y/c) extent of the separation bubble also grows with increasing incidence, extending to $\sim$0.06c in width at $\alpha = 188^\circ$. This flow-separation bubble has a low-pressure vortex core which creates a suction that accentuates the problems of negative lift and nose-up pitching moment, as discussed previously. The sudden increase in the moment also causes a pitching moment impulse.
Small variations are also observed in the size of the wake recirculation region between $\alpha=180^\circ$ and $188^\circ$. In general, a marginal increase in length and a decrease in width are observed for $\alpha>180^\circ$.
The flow separation on the suction side increases significantly as the angle of attack is increased from $\alpha = 188^\circ$ to $190^\circ$. At $\alpha = 190^\circ$, the leading-edge separation bubble merges with the wake recirculation region to form an open recirculation region with two distinct vortex cores. At $\alpha = 192^\circ$, this region grows in size, and the two cores move farther away from the airfoil surface. At $\alpha = 194^\circ$, the two vortex cores merge to form a single core, with its centre at a distance of $\sim$0.1c from the airfoil surface.  
In this deep stall regime, as the angle of attack continues to increase, the mean recirculation region grows larger, and the corresponding vortex core moves farther away from the airfoil surface. At $\alpha = 194^\circ$, a second recirculation region, counter-rotating with respect to the primary separation vortex, is observed in the wake. This is the time-averaged recirculation region that develops from the shear layer separating from the pressure side of the blunt trailing edge, as is expected in a bluff-body type wake. Similar flow development with angle of attack has also been observed in previous studies \cite{DeannaPassive, ThomasRF}.
%%%%%%%%%%%%%%%%%%%%%%%%%%%%%%%%%%%%%%%
\vspace{3em}
\subsubsection{Sinusoidal trailing edge airfoil: sinusoidal crest plane}
Figures \ref{A10P05_crest} and \ref{A10P10_crest} show time-averaged normalized streamwise velocity contours for the $A10\lambda05$ and $A10\lambda10$ airfoils, respectively, on a midspan plane passing through the crest of a sinusoid. The cross-section at the crest, which is an airfoil with a sharp geometric trailing edge, is overlaid on the flowfields. The static angles of attack in these figures are the same as those reported for the baseline airfoil in Fig. \ref{Baseline_PIV}.

The results for the baseline airfoil suggested that flow separation started as early as $\alpha = 182^\circ$, and progressed into fully separated flow by $\alpha = 190^\circ$. In contrast, both the modified airfoils show that the flow remains almost completely attached over the airfoil surface till $\alpha = 190^\circ$. Flow separates only near the trailing edge to form a bluff-body type wake. Therefore, the flow is able to turn at the leading edge in tandem with the angle of attack, keeping it attached to the surface. Near the leading edge, a region of accelerated flow is observed for $\alpha \geq 184^\circ$. At $\alpha =190^\circ$, the chordwise extent of this region is higher for the $A10\lambda10$ airfoil, extending to $\sim$0.2c as compared to $\sim$0.1c for the $A10\lambda05$ airfoil. A similar behavior is observed at higher pitch angles, suggesting that a greater energization of the flow occurs over the leading edge of the $A10\lambda10$ airfoil.
This region, for both airfoils, is followed by a region of decelerated but attached flow that separates at the blunt trailing edge to form the wake recirculation region. At $\alpha=192^\circ$, the flow separation point shifts slightly upstream to $\sim$0.9c for both airfoils.
A further increase in pitch angle to $\alpha = 194^\circ$ shows a drastic upstream shift of the separation point, shifting to $\sim$0.6c for both cases. The wake recirculation region, inside the white dotted contour line, is fairly the same till $\alpha = 186^\circ$ for both airfoils. Beyond this angle, the region extends further downstream for the $A10\lambda05$ airfoil in comparison to the $A10\lambda10$ airfoil. Thus, the wake characteristics vary with trailing-edge (aerodynamic leading-edge) sinusoidal geometry. The shear layer separating from the pressure side of the airfoil at the blunt trailing edge forms a second recirculation zone in the wake, similar to the baseline. A portion of this recirculation region is visible at $\alpha = 194^\circ$ for both airfoils. 

\begin{figure}[H]
\centering
\begin{subfigure}[b]{0.44\textwidth}
    \includegraphics[width=\textwidth]{Images/PIV_plots/Avg_Vx_png/legend_horz.jpg}
\end{subfigure}
\hspace{6em}
\begin{subfigure}[b]{0.09\textwidth}
    \centering 
    \includegraphics[trim=0 12 0 0,clip,width=\textwidth]{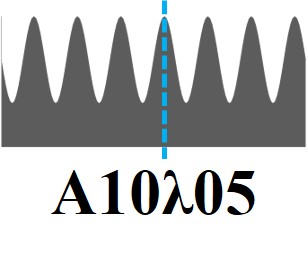}
\end{subfigure}
\vskip\baselineskip
\centering
\begin{subfigure}[b]{0.351\textwidth}
    \centering 
    \includegraphics[trim=0 34 0 0,clip,width=\textwidth]{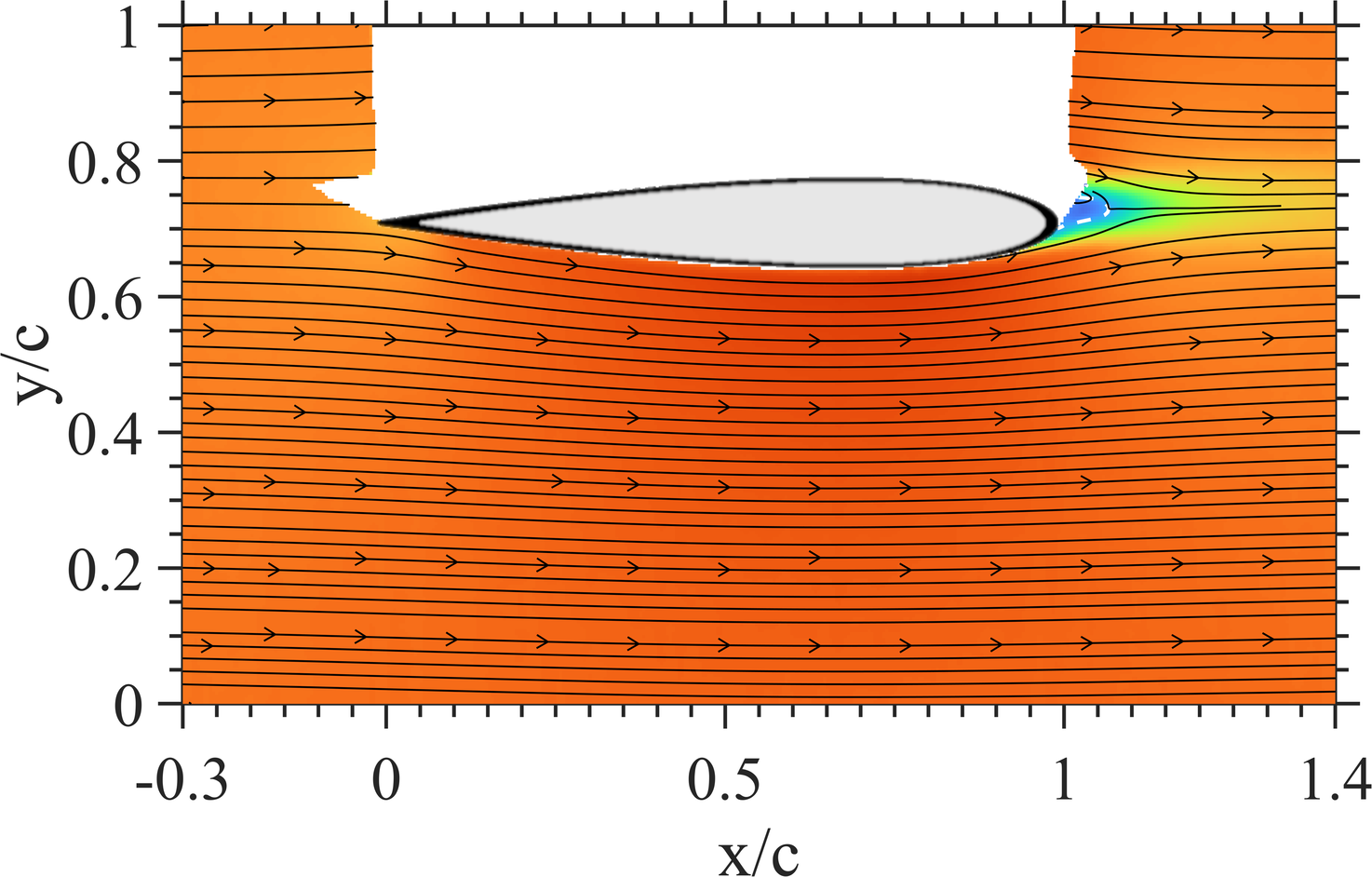}
    \caption{$\alpha$ = 180$^\circ$}
    \label{}
\end{subfigure}
\begin{subfigure}[b]{0.315\textwidth}
    \centering 
    \includegraphics[trim=33 34 0 0,clip,width=\textwidth]{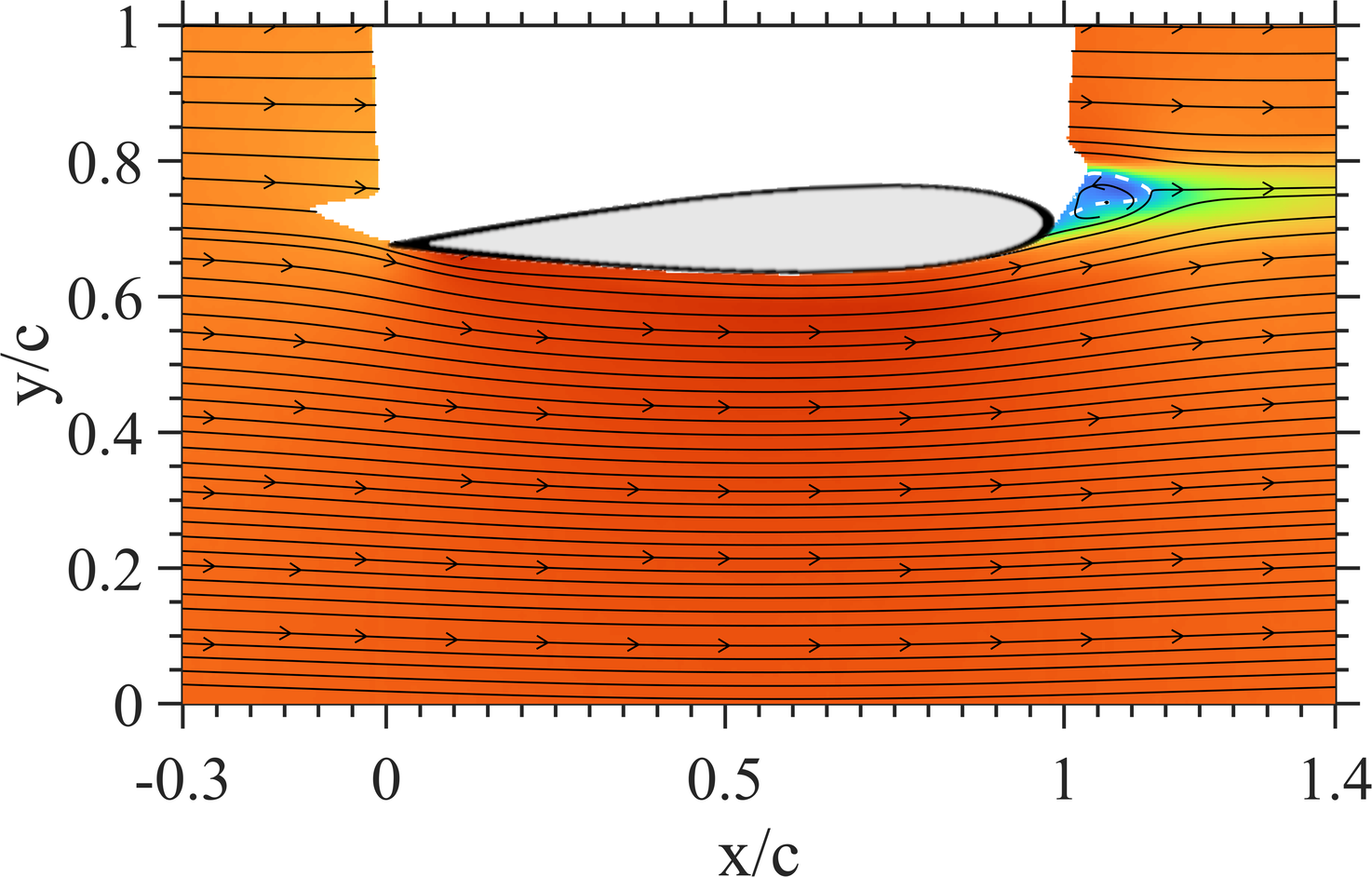}
    \caption{$\alpha$ = 182$^\circ$}
    \label{}
\end{subfigure}
\begin{subfigure}[b]{0.315\textwidth}
    \centering 
    \includegraphics[trim=33 34 0 0,clip,width=\textwidth]{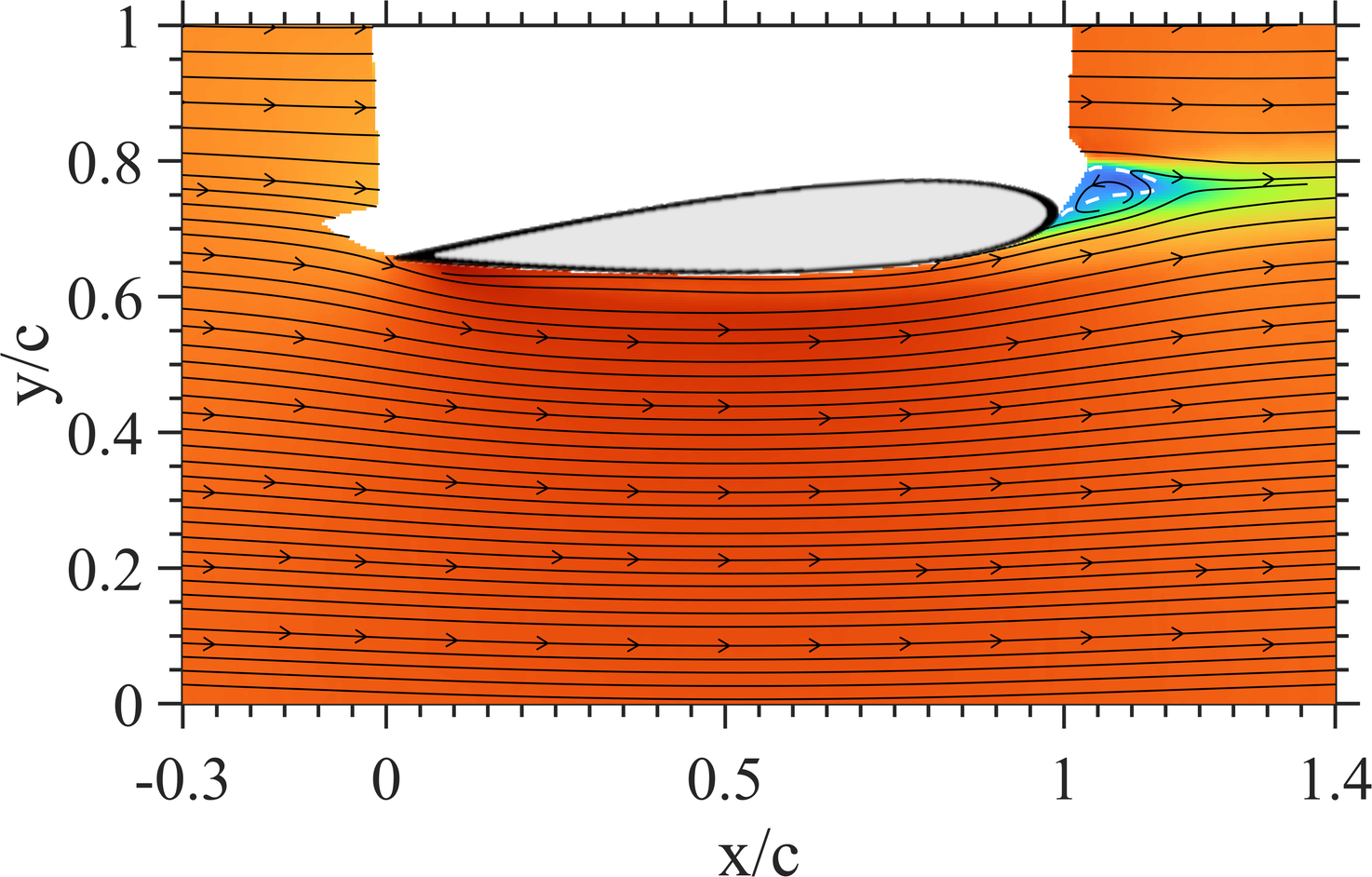}
    \caption{$\alpha$ = 184$^\circ$}
    \label{}
\end{subfigure}
% \vskip\baselineskip
\end{figure}
\begin{figure}[H]\ContinuedFloat
\begin{subfigure}[b]{0.351\textwidth}
    \centering 
    \includegraphics[trim=0 0 0 0,clip,width=\textwidth]{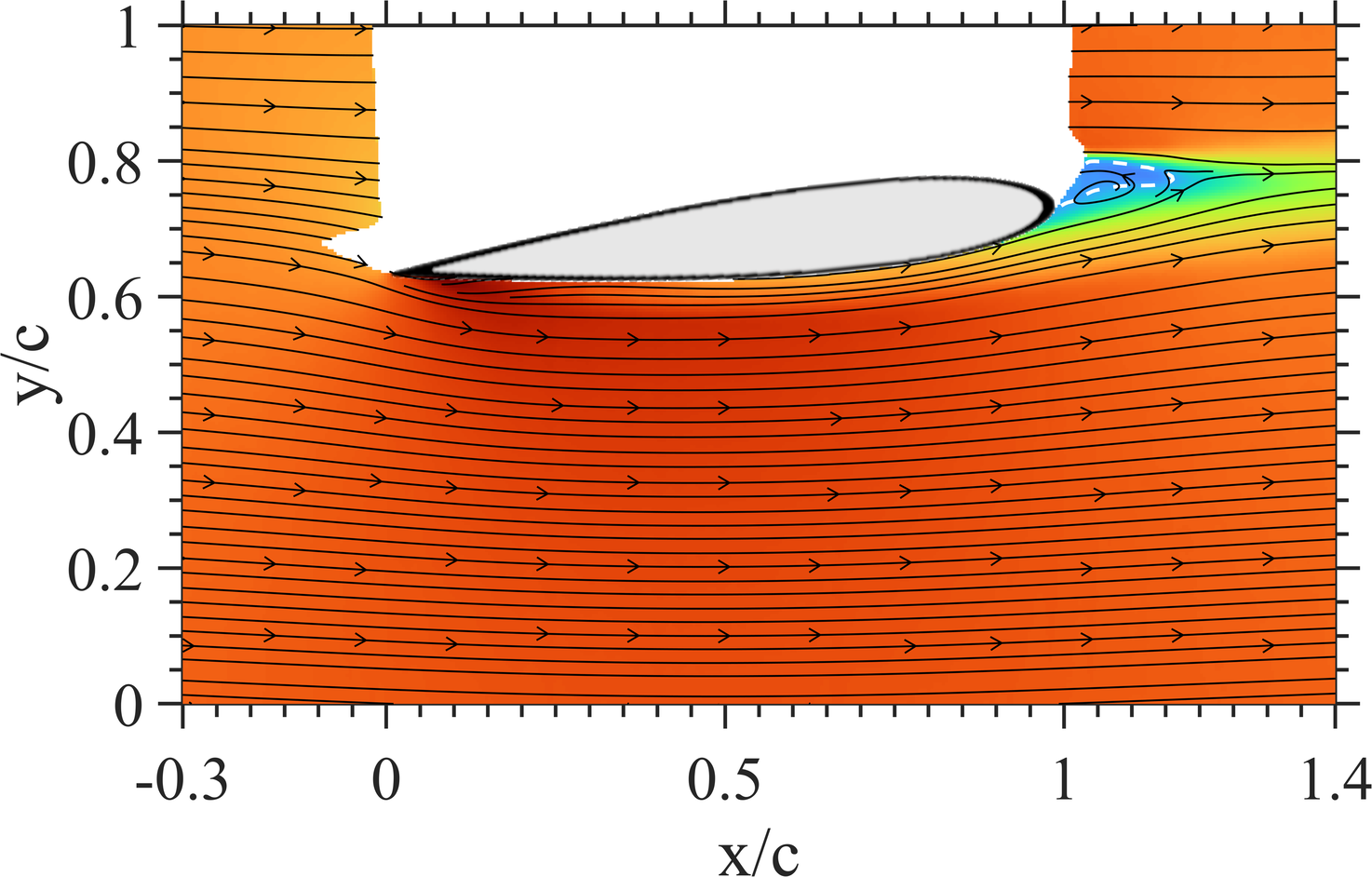}
    \caption{$\alpha$ = 186$^\circ$}
    \label{}
\end{subfigure}
\begin{subfigure}[b]{0.315\textwidth}
    \centering 
    \includegraphics[trim=33 0 0 0,clip,width=\textwidth]{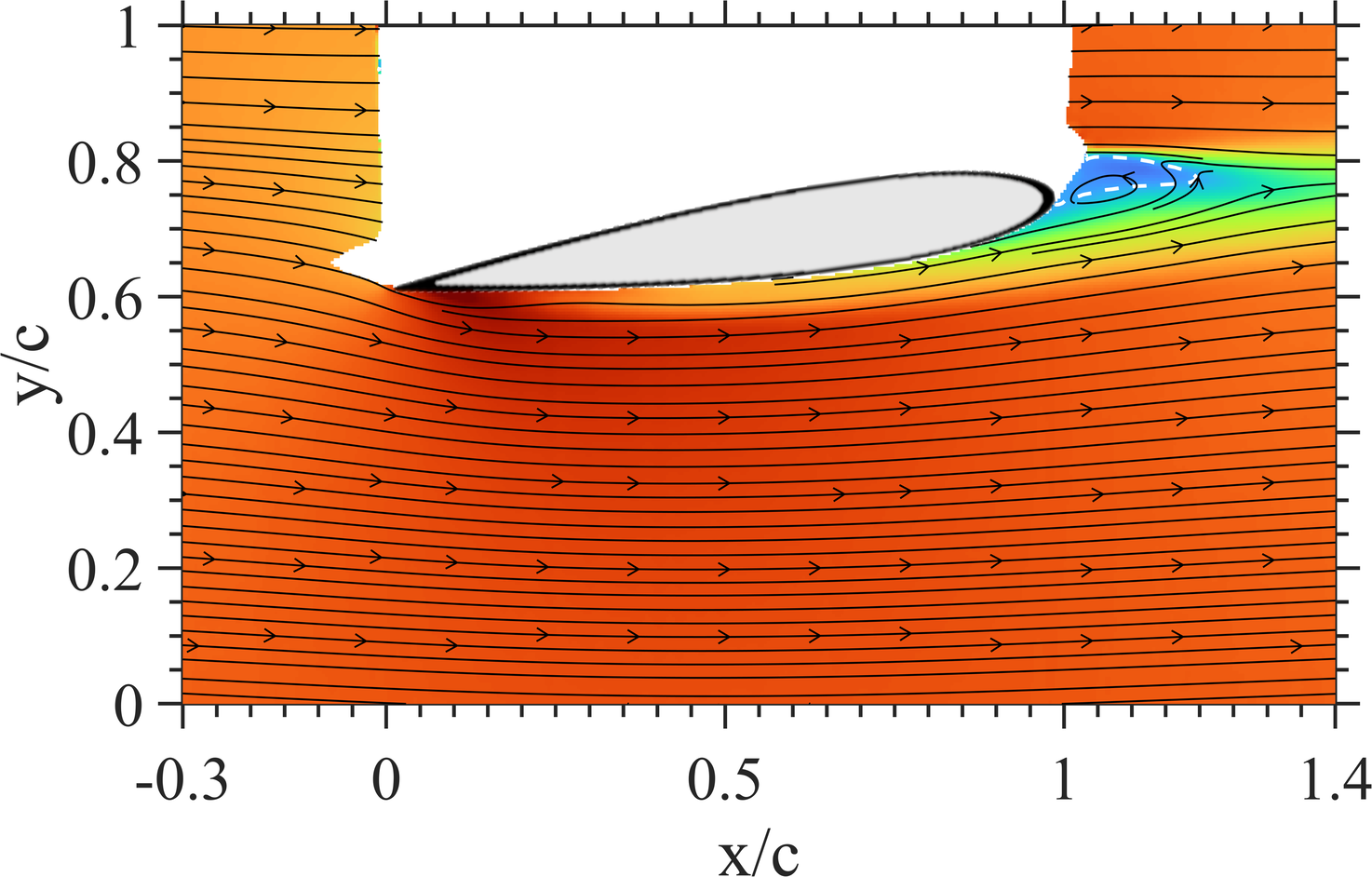}
    \caption{$\alpha$ = 188$^\circ$}
    \label{}
\end{subfigure}
\begin{subfigure}[b]{0.315\textwidth}
    \centering 
    \includegraphics[trim=33 0 0 0,clip,width=\textwidth]{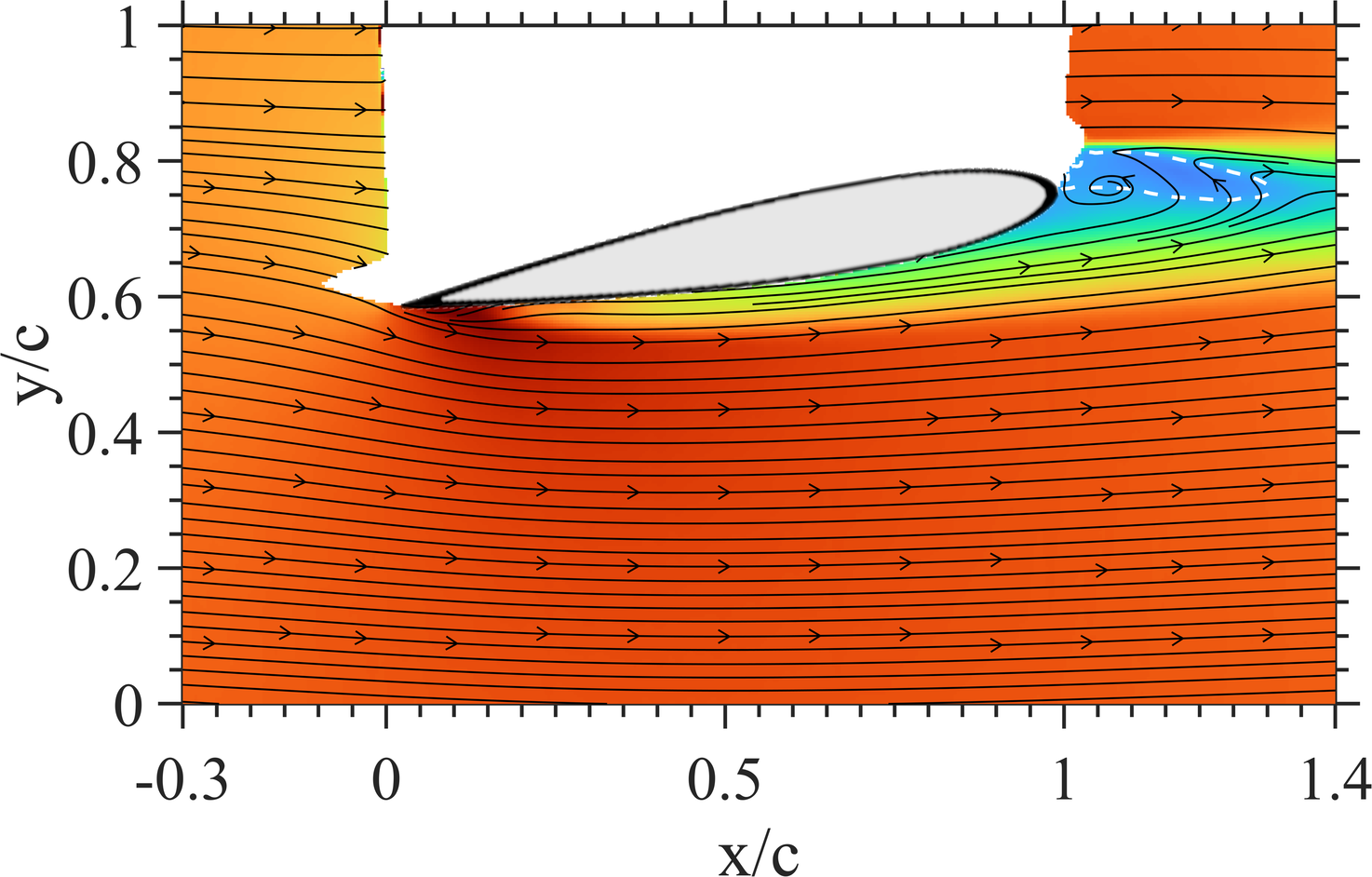}
    \caption{$\alpha$ = 190$^\circ$}
    \label{}
\end{subfigure}
\vskip\baselineskip
\hspace{7em}
\begin{subfigure}[b]{0.351\textwidth}
    \centering 
    \includegraphics[trim=0 0 0 0,clip,width=\textwidth]{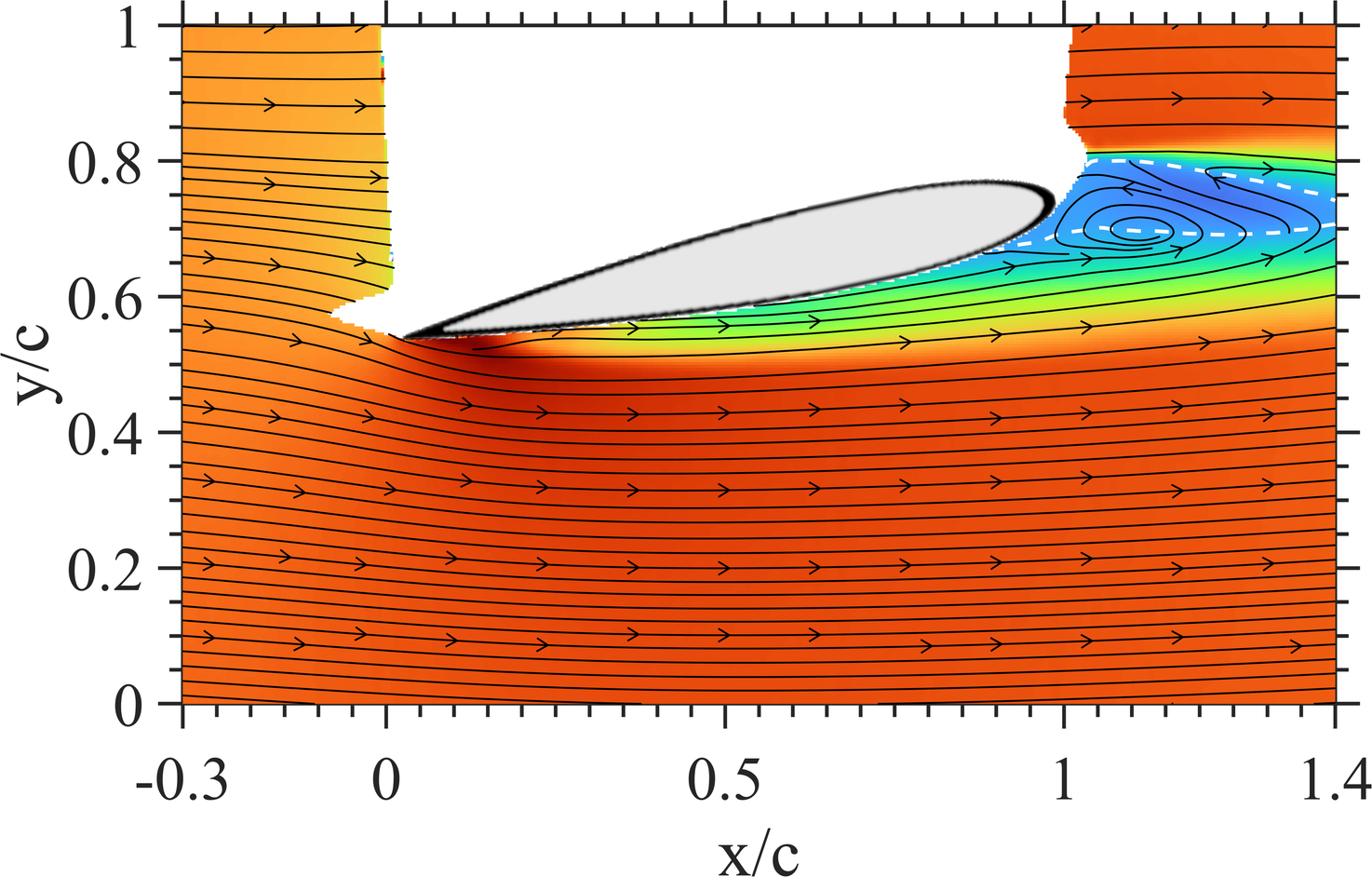}
    \caption{$\alpha$ = 192$^\circ$}
    \label{}
\end{subfigure}
\begin{subfigure}[b]{0.315\textwidth}
    \centering 
    \includegraphics[trim=33 0 0 0,clip,width=\textwidth]{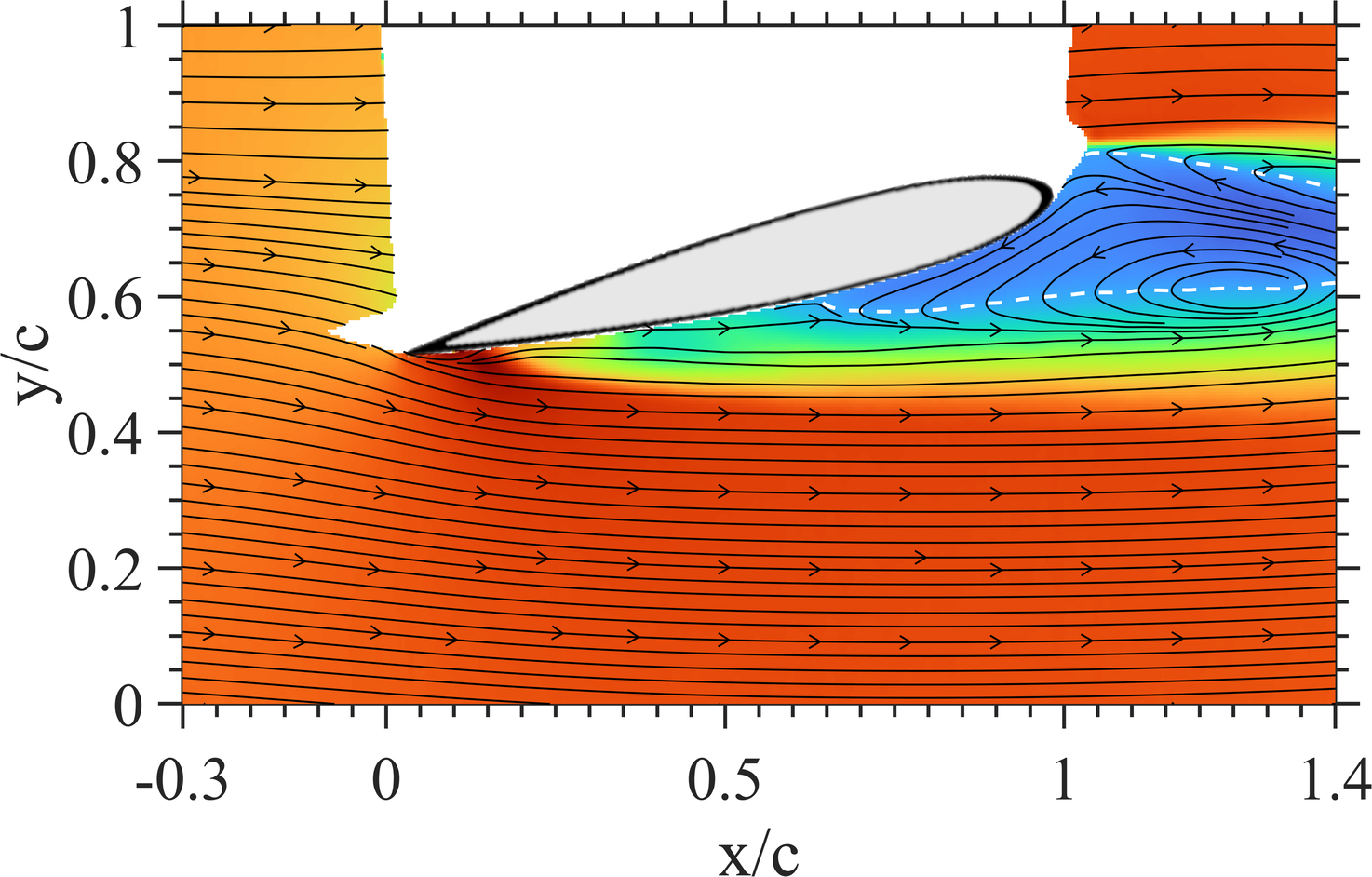}
    \caption{$\alpha$ = 194$^\circ$}
    \label{}
\end{subfigure}
\caption{Contour plots of the time-averaged streamwise velocity component, normalized with freestream velocity for the $\mathbf{A10\lambda05}$ airfoil along the crest of a midspan sinusoid, at static pitch angles of 180$^\circ$ (a), 182$^\circ$ (b), 184$^\circ$ (c), 186$^\circ$ (d), 188$^\circ$ (e), 190$^\circ$ (f), 192$^\circ$ (g), and 194$^\circ$ (h), overlaid with in-plane streamlines. The white dashed contour line represents the location where mean streamwise velocity is zero. Flow direction is from left to right and $\mathbf{Re_c = 1.4\times10^5}$.}
\label{A10P05_crest}
\end{figure}

The most important conclusion from these results is that the separation bubble developing from the sharp leading edge, which accentuates the problems associated with reverse flow, is completely mitigated for both modified cases. The delayed flow separation suggests that the geometric trailing edge sinusoids act as vortex generators in reverse flow, as was proposed previously in Fig. \ref{fig:serration vortices-schem}.  
The downwash between the counter-rotating vortex pairs, which becomes upwash in the current configuration, redirects the incoming flow and aligns it with the surface, thereby eliminating flow separation. 
The region of attached flow downstream of the leading edge suggests that the vortices are effective in energizing the boundary layer, which in turn is able to withstand the adverse pressure gradient. The exact characteristics of these vortices for various sinusoidal geometries, freestream velocities, and angles of attack require investigation using Stereoscopic PIV (SPIV). These characteristics are expected to vary with sinusoidal wavelength, thus leading to the observed variations in the flowfield between $A10\lambda05$ and $A10\lambda10$ airfoils. 
It is important to note that while the baseline airfoil has flow separation starting at the sharp leading edge, the modified airfoils show separation starting at the blunt trailing edge. The point of separation progresses upstream with increasing angle of attack. The initial flow separation in such a scenario will become Reynolds number dependent and therefore different from the classical reverse flow problem over a thin airfoil, which has been shown to be Reynolds number independent at high Reynolds numbers \cite{Reeffects}.
\begin{figure}[H]
\centering
\begin{subfigure}[b]{0.44\textwidth}
    \includegraphics[width=\textwidth]{Images/PIV_plots/Avg_Vx_png/legend_horz.jpg}
\end{subfigure}
\hspace{6em}
\begin{subfigure}[b]{0.09\textwidth}
    \centering 
    \includegraphics[trim=0 12 0 0,clip,width=\textwidth]{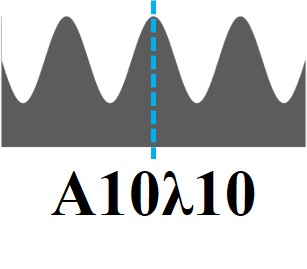}
\end{subfigure}
\vskip\baselineskip
\centering
\begin{subfigure}[b]{0.351\textwidth}
    \centering 
    \includegraphics[trim=0 34 0 0,clip,width=\textwidth]{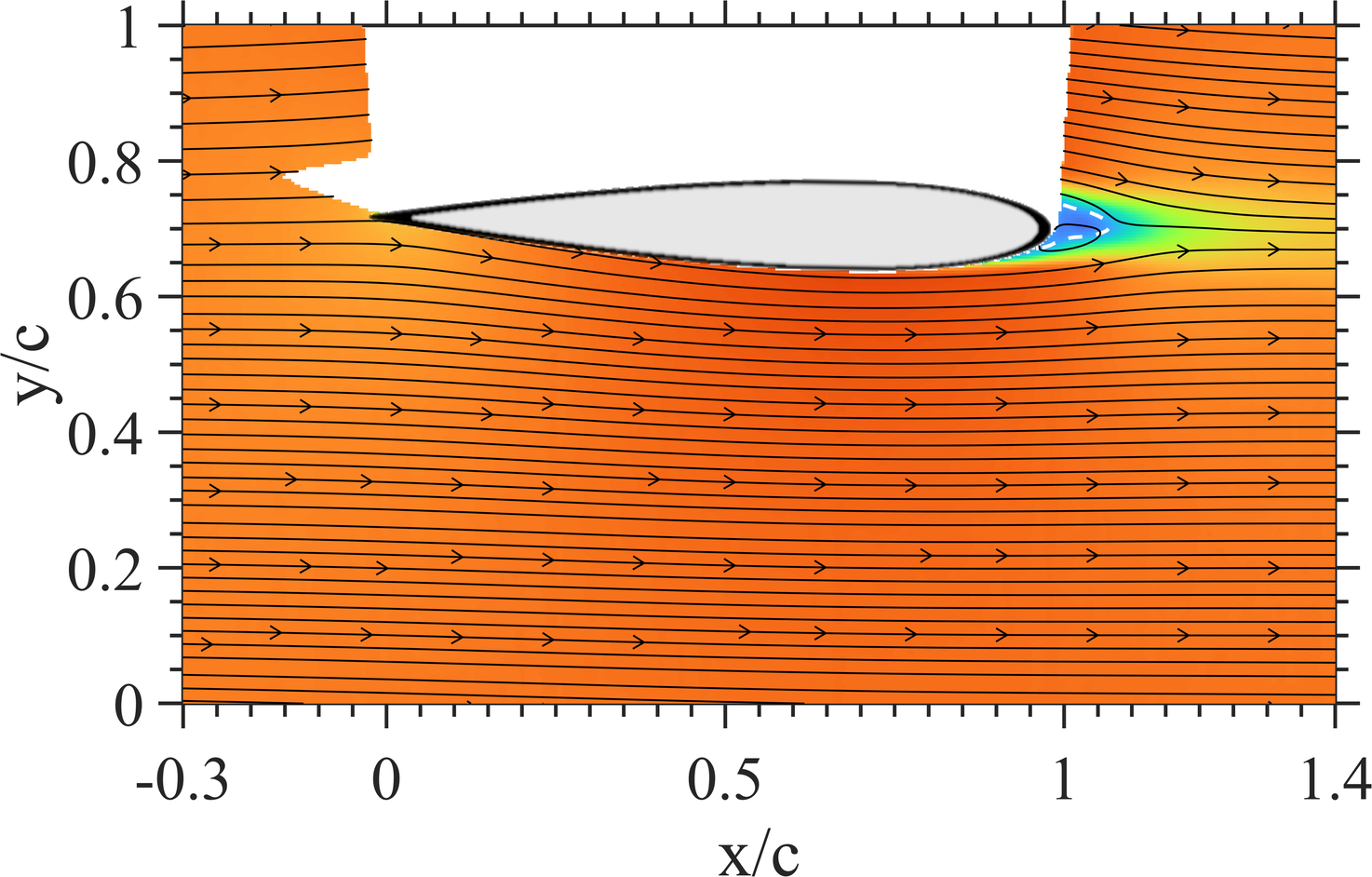}
    \caption{$\alpha$ = 180$^\circ$}
    \label{}
\end{subfigure}
\begin{subfigure}[b]{0.315\textwidth}
    \centering 
    \includegraphics[trim=33 34 0 0,clip,width=\textwidth]{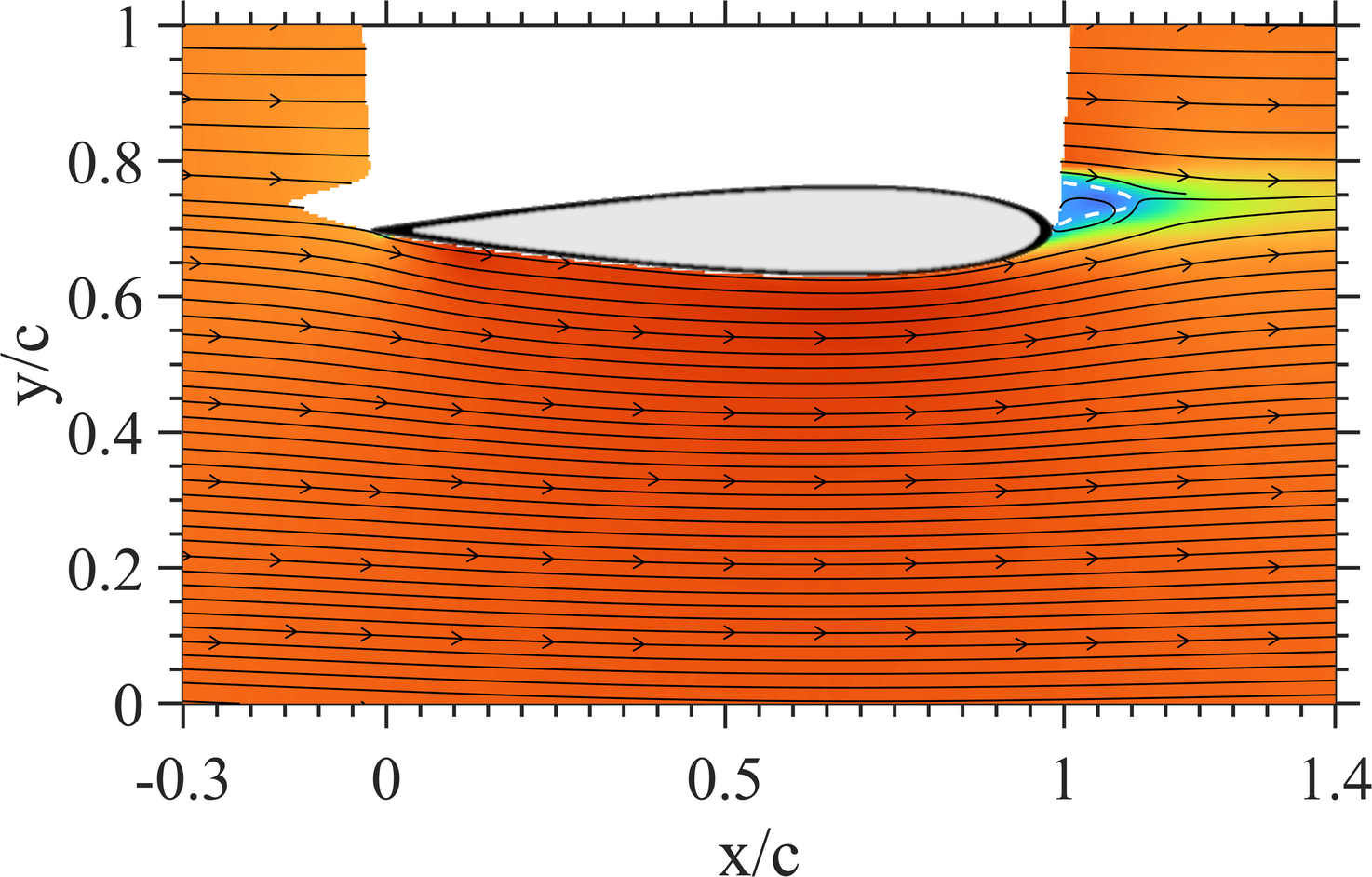}
    \caption{$\alpha$ = 182$^\circ$}
    \label{}
\end{subfigure}
\begin{subfigure}[b]{0.315\textwidth}
    \centering 
    \includegraphics[trim=33 34 0 0,clip,width=\textwidth]{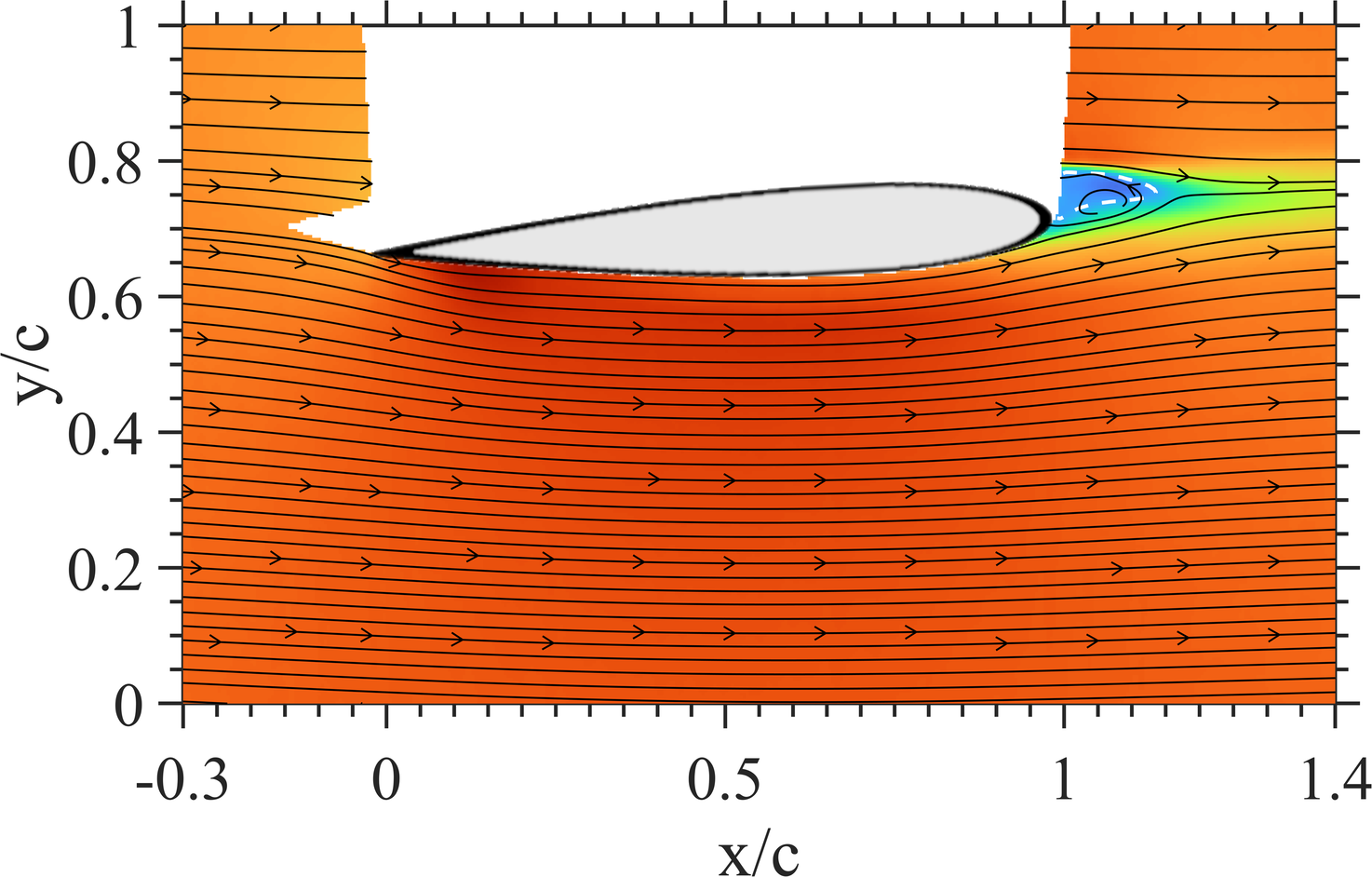}
    \caption{$\alpha$ = 184$^\circ$}
    \label{}
\end{subfigure}
\vskip\baselineskip
\end{figure}
\begin{figure}[H]\ContinuedFloat
\begin{subfigure}[b]{0.351\textwidth}
    \centering 
    \includegraphics[trim=0 0 0 0,clip,width=\textwidth]{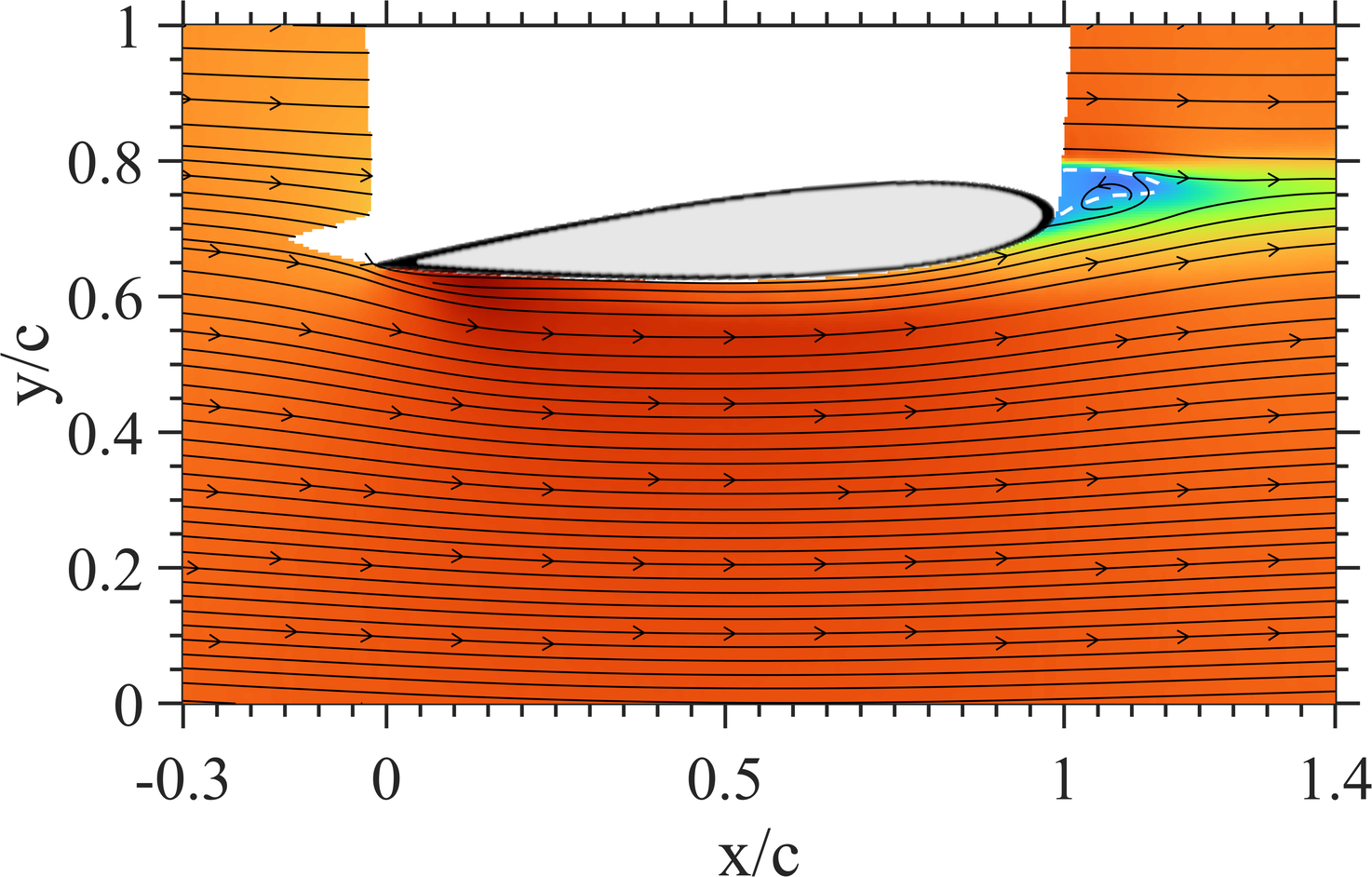}
    \caption{$\alpha$ = 186$^\circ$}
    \label{}
\end{subfigure}
\begin{subfigure}[b]{0.315\textwidth}
    \centering 
    \includegraphics[trim=33 0 0 0,clip,width=\textwidth]{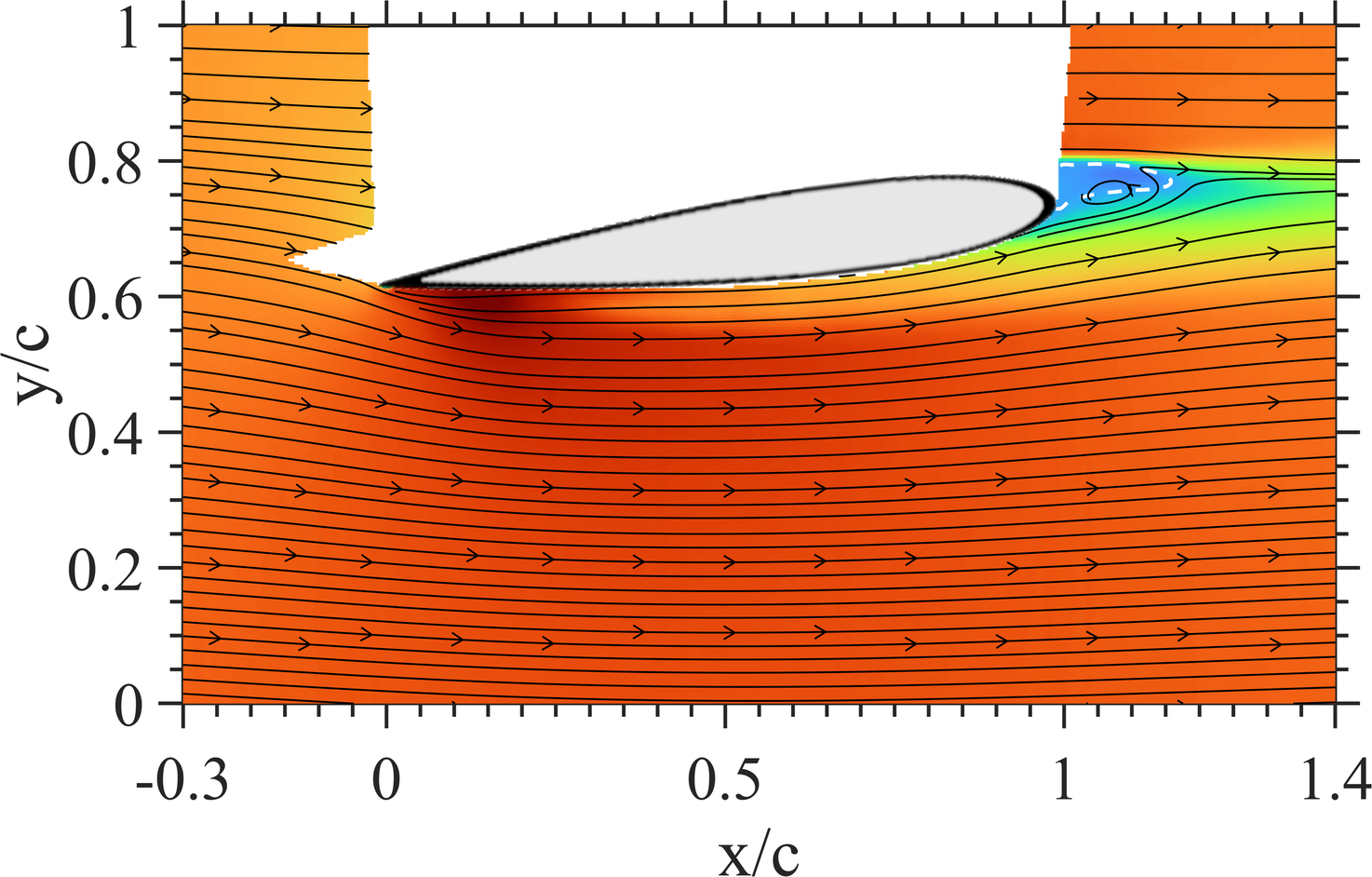}
    \caption{$\alpha$ = 188$^\circ$}
    \label{}
\end{subfigure}
\begin{subfigure}[b]{0.315\textwidth}
    \centering 
    \includegraphics[trim=33 0 0 0,clip,width=\textwidth]{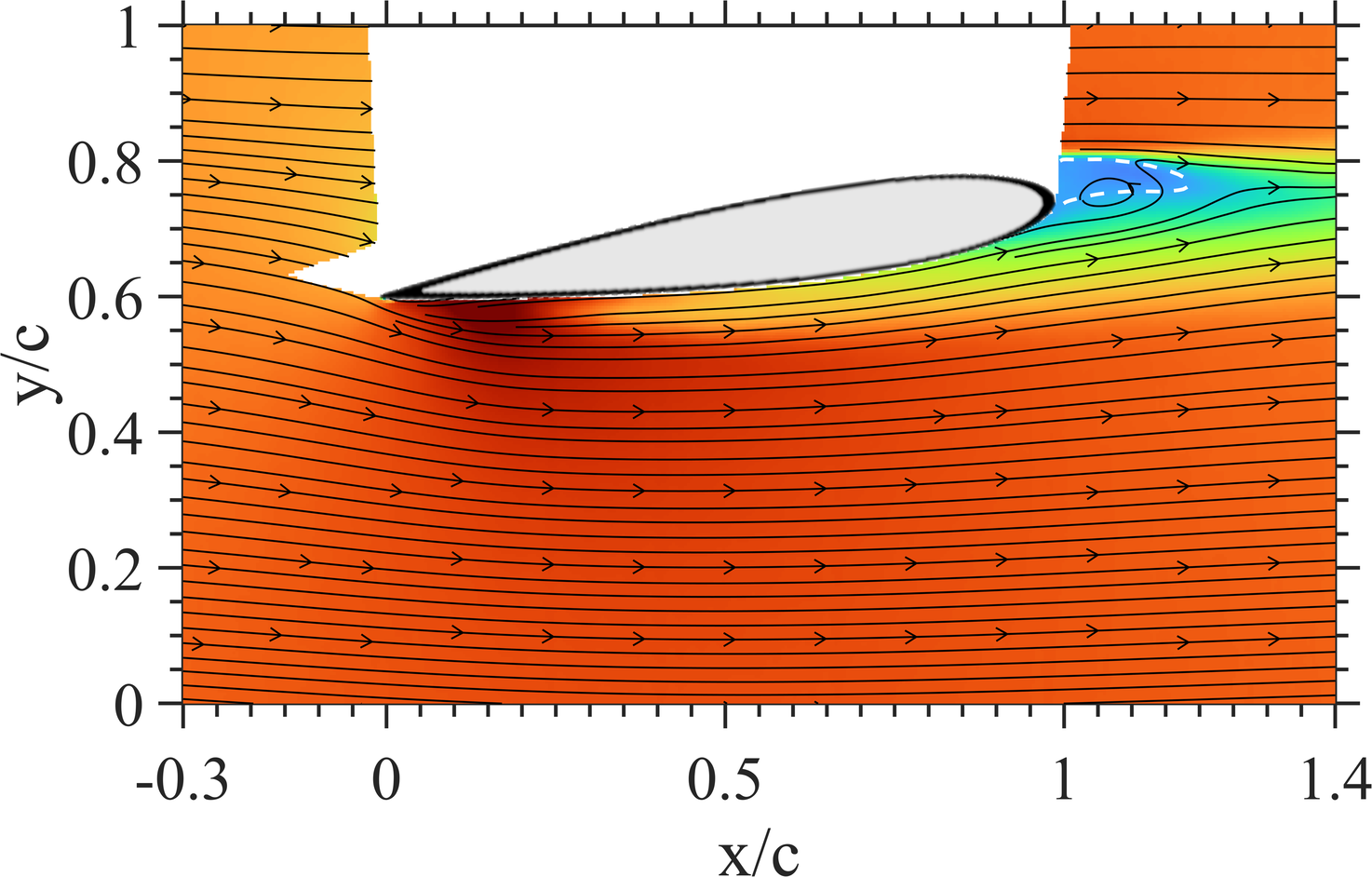}
    \caption{$\alpha$ = 190$^\circ$}
    \label{}
\end{subfigure}
\vskip\baselineskip
\hspace{7em}
\begin{subfigure}[b]{0.351\textwidth}
    \centering 
    \includegraphics[trim=0 0 0 0,clip,width=\textwidth]{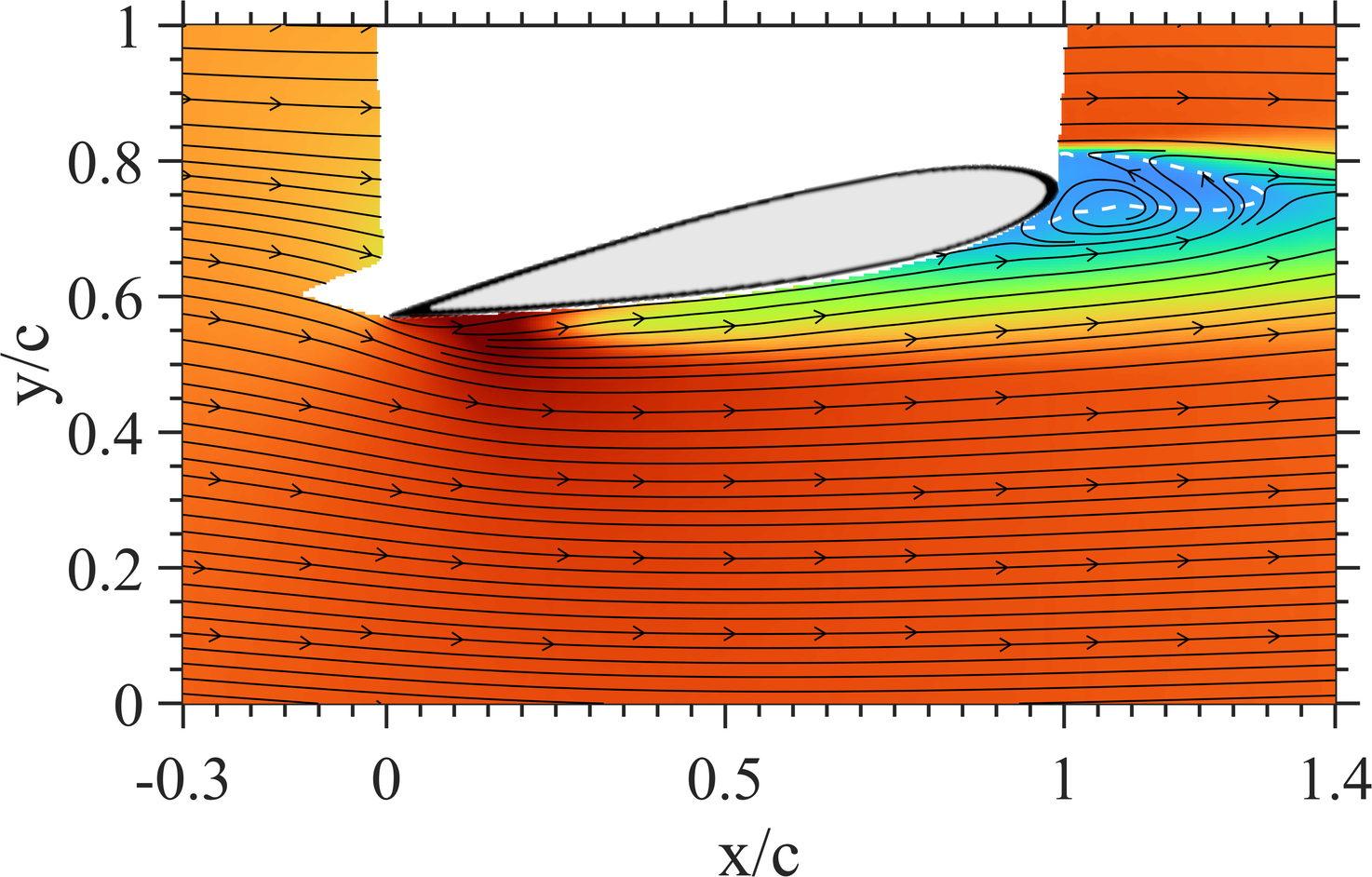}
    \caption{$\alpha$ = 192$^\circ$}
    \label{}
\end{subfigure}
\begin{subfigure}[b]{0.315\textwidth}
    \centering 
    \includegraphics[trim=33 0 0 0,clip,width=\textwidth]{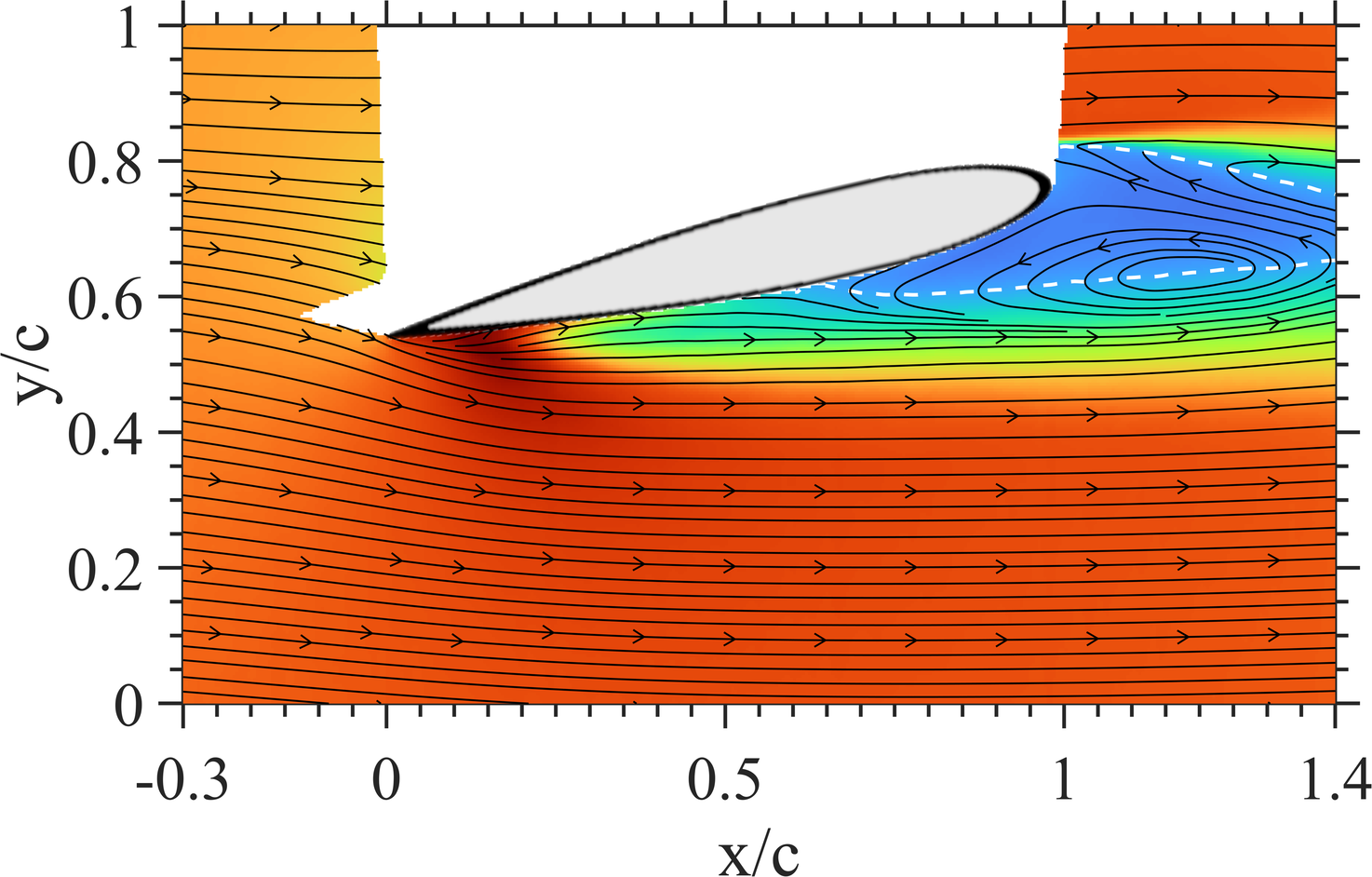}
    \caption{$\alpha$ = 194$^\circ$}
    \label{}
\end{subfigure}
\caption{Contour plots of the time-averaged streamwise velocity component, normalized with freestream velocity for the $\mathbf{A10\lambda10}$ airfoil along the crest of a midspan sinusoid, at static pitch angles of 180$^\circ$ (a), 182$^\circ$ (b), 184$^\circ$ (c), 186$^\circ$ (d), 188$^\circ$ (e), 190$^\circ$ (f), 192$^\circ$ (g), and 194$^\circ$ (h), overlaid with in-plane streamlines. The white dashed contour line represents the location where mean streamwise velocity is zero. Flow direction is from left to right and $\mathbf{Re_c = 1.4\times10^5}$.}
\label{A10P10_crest}
\end{figure}

%%%%%%%%%%%%%%%%%%%%%%%%%%%%%%%%%%%%%%%%%%%%%%%%%%%%%

\subsubsection{Sinusoidal trailing edge airfoil: sinusoidal middle plane}
\begin{figure}[H]
\centering
\begin{subfigure}[b]{0.44\textwidth}
    \includegraphics[width=\textwidth]{Images/PIV_plots/Avg_Vx_png/legend_horz.jpg}
\end{subfigure}
\hspace{6em}
\begin{subfigure}[b]{0.09\textwidth}
    \centering 
    \includegraphics[trim=0 12 0 0,clip,width=\textwidth]{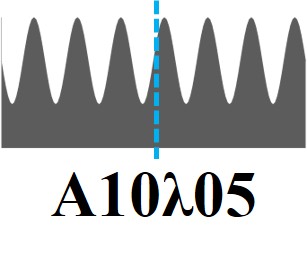}
\end{subfigure}
\vskip\baselineskip
\centering
\begin{subfigure}[b]{0.351\textwidth}
    \centering 
    \includegraphics[trim=0 34 0 0,clip,width=\textwidth]{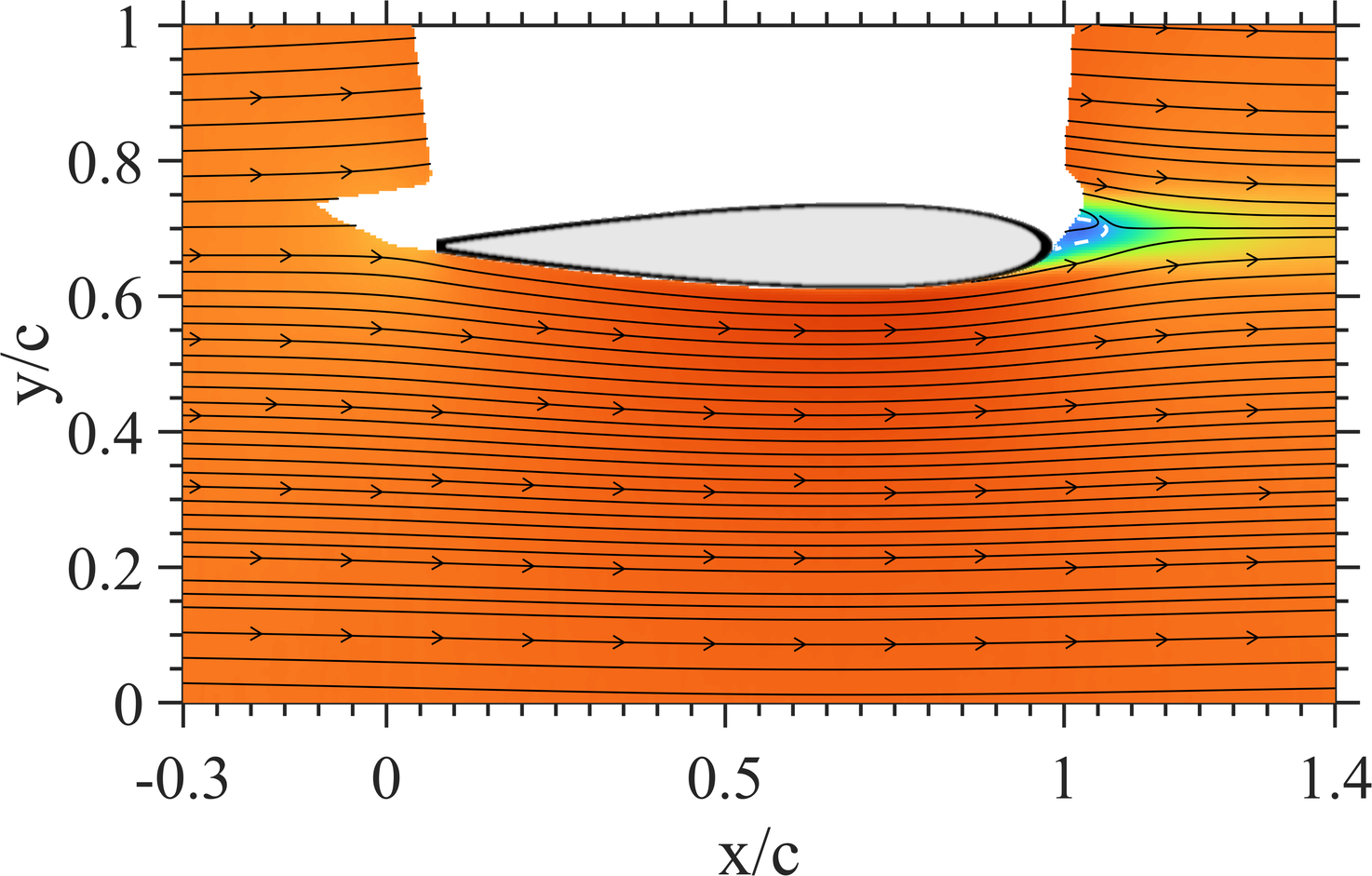}
    \caption{$\alpha$ = 180$^\circ$}
    \label{}
\end{subfigure}
\begin{subfigure}[b]{0.315\textwidth}
    \centering 
    \includegraphics[trim=33 34 0 0,clip,width=\textwidth]{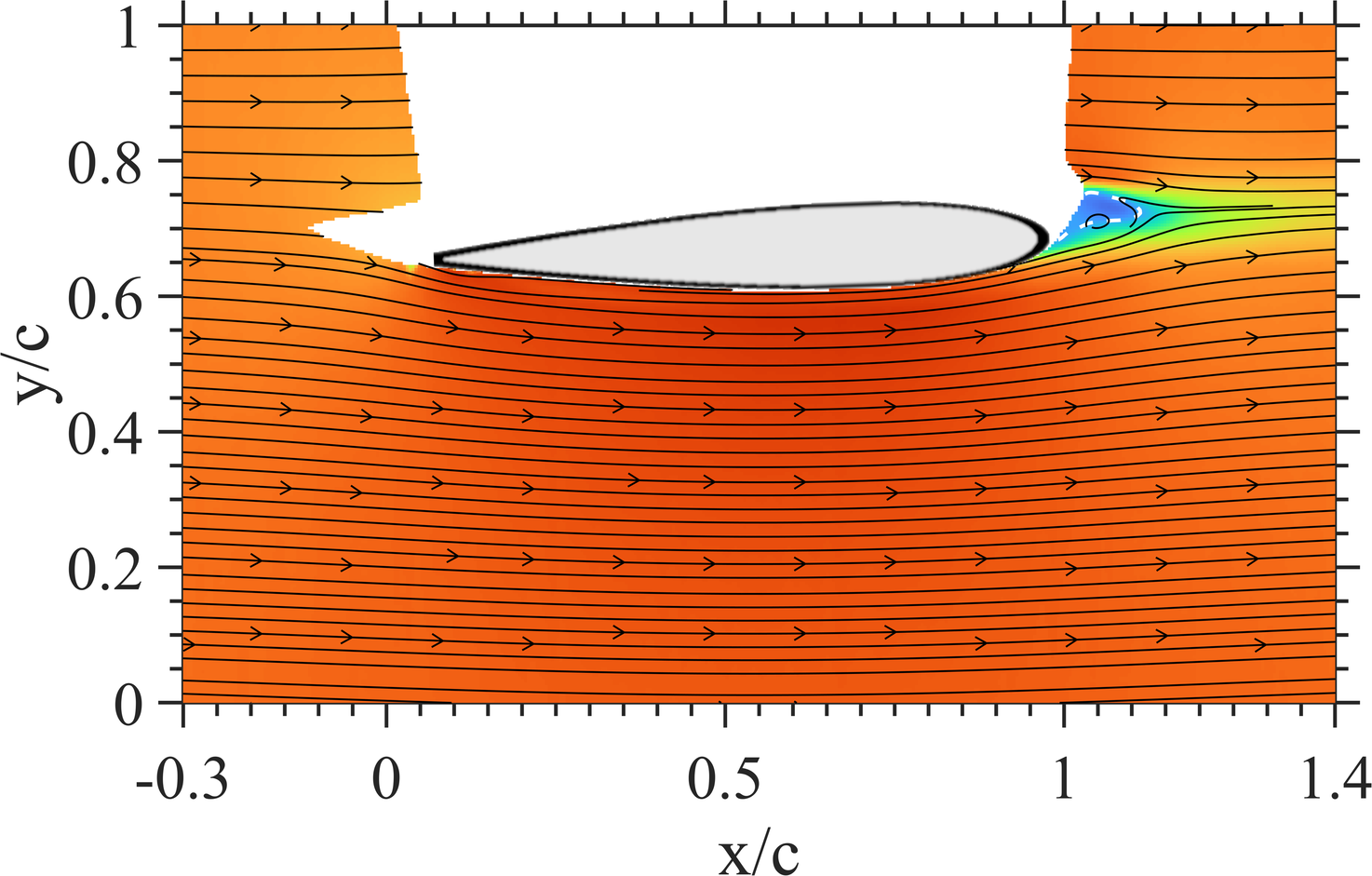}
    \caption{$\alpha$ = 182$^\circ$}
    \label{}
\end{subfigure}
\begin{subfigure}[b]{0.315\textwidth}
    \centering 
    \includegraphics[trim=33 34 0 0,clip,width=\textwidth]{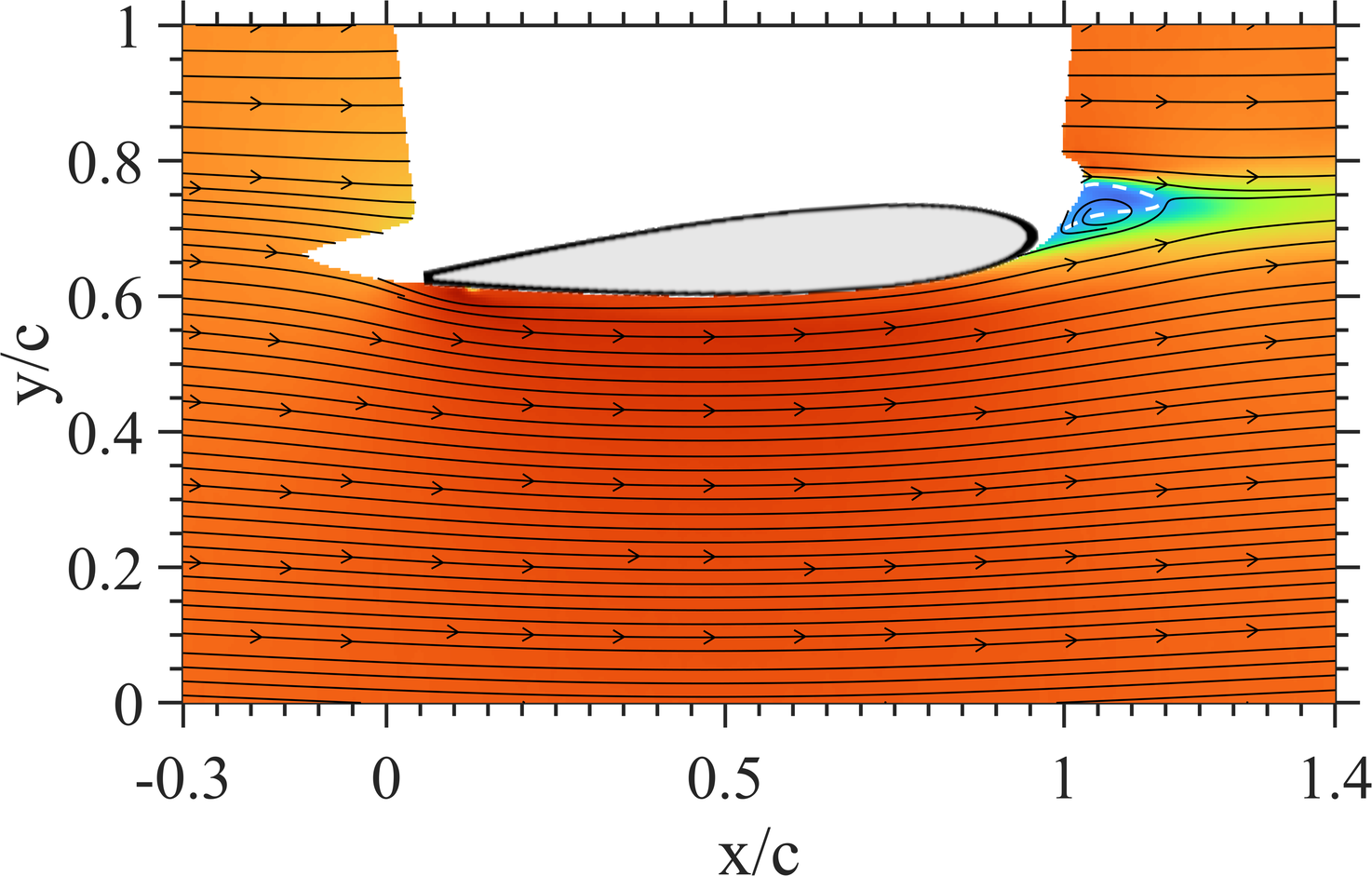}
    \caption{$\alpha$ = 184$^\circ$}
    \label{}
\end{subfigure}
\vskip\baselineskip
% \end{figure}
% \begin{figure}[H]\ContinuedFloat
\begin{subfigure}[b]{0.351\textwidth}
    \centering 
    \includegraphics[trim=0 0 0 0,clip,width=\textwidth]{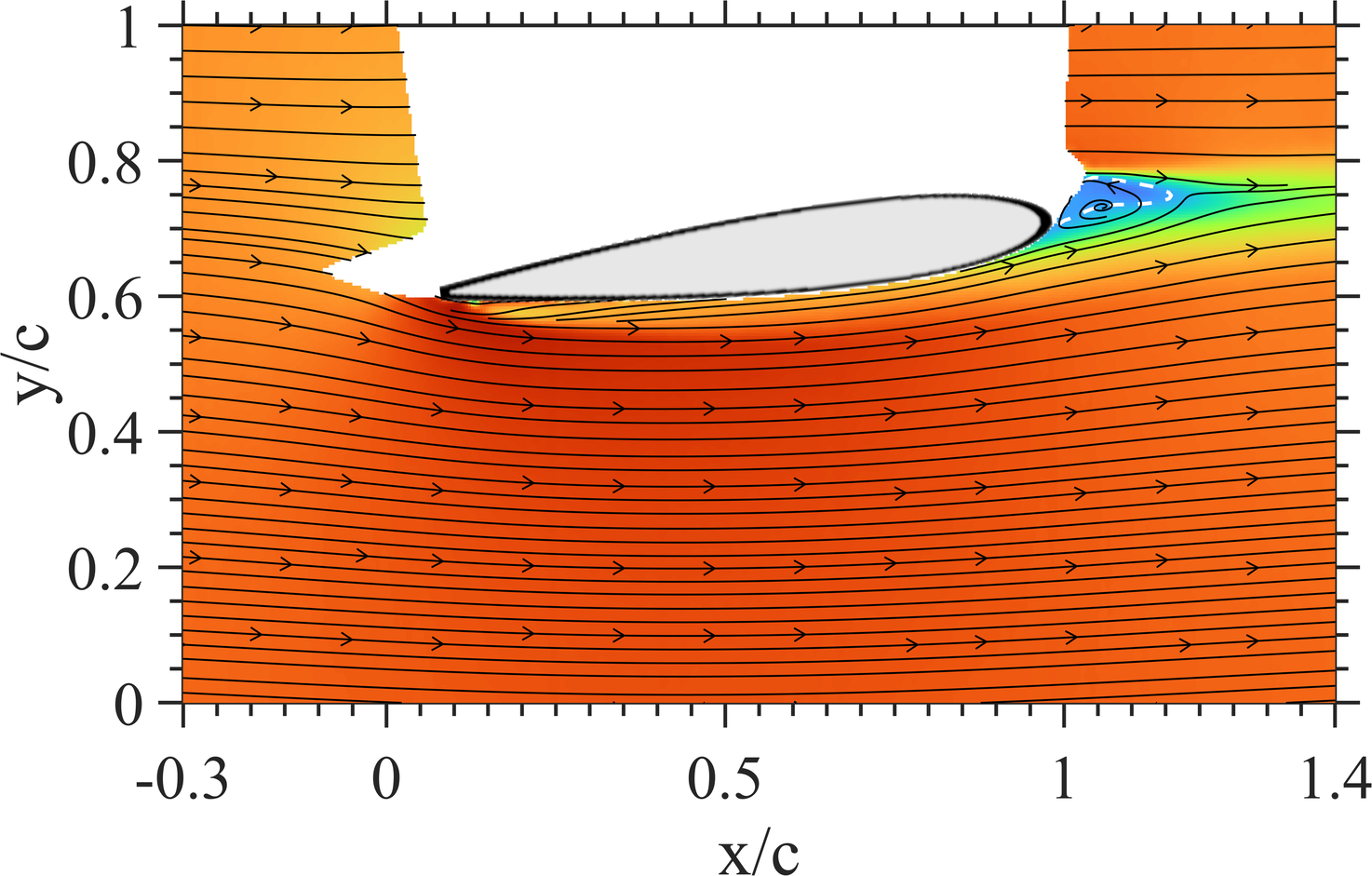}
    \caption{$\alpha$ = 186$^\circ$}
    \label{}
\end{subfigure}
\begin{subfigure}[b]{0.315\textwidth}
    \centering 
    \includegraphics[trim=33 0 0 0,clip,width=\textwidth]{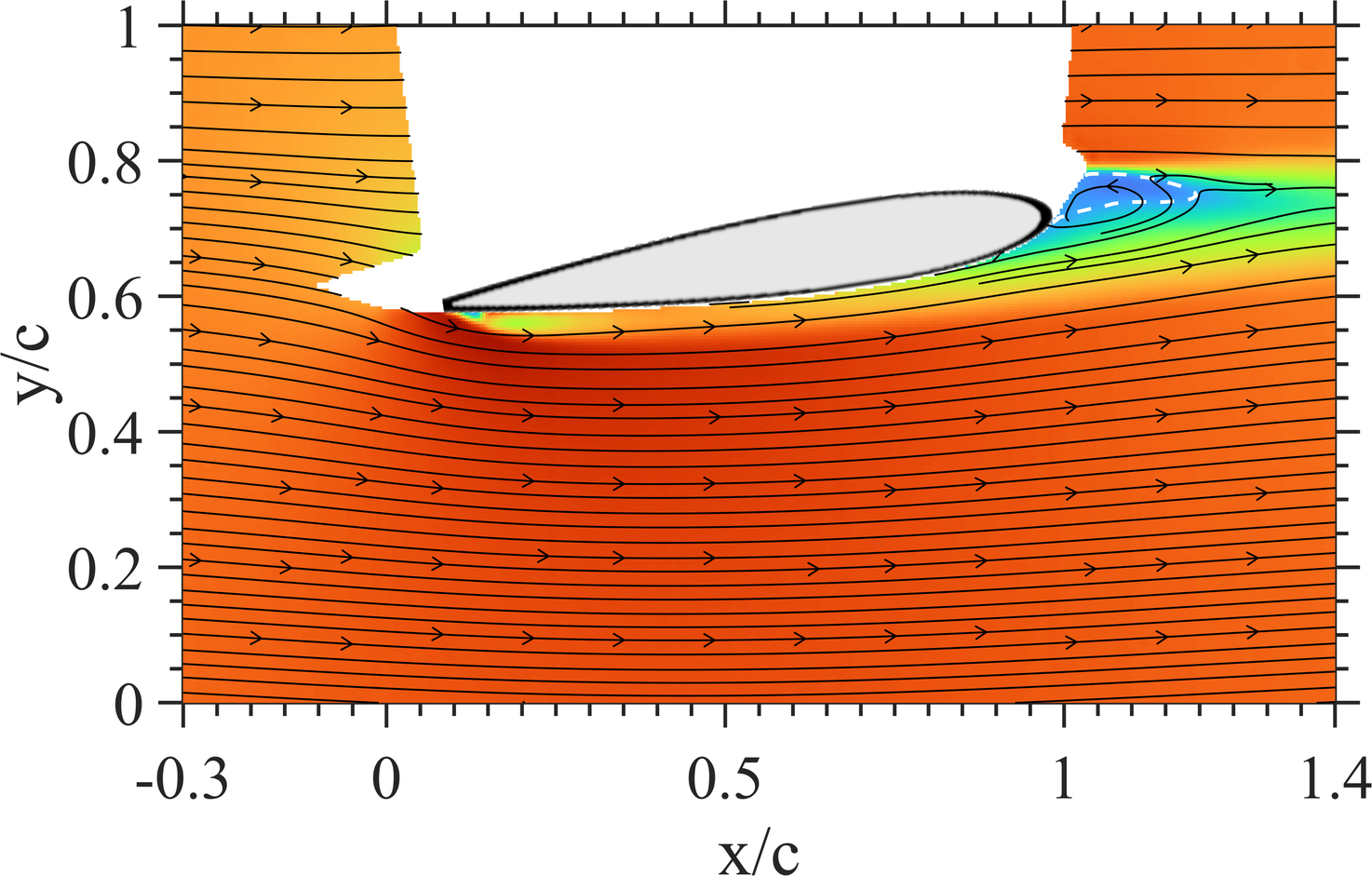}
    \caption{$\alpha$ = 188$^\circ$}
    \label{}
\end{subfigure}
\begin{subfigure}[b]{0.315\textwidth}
    \centering 
    \includegraphics[trim=33 0 0 0,clip,width=\textwidth]{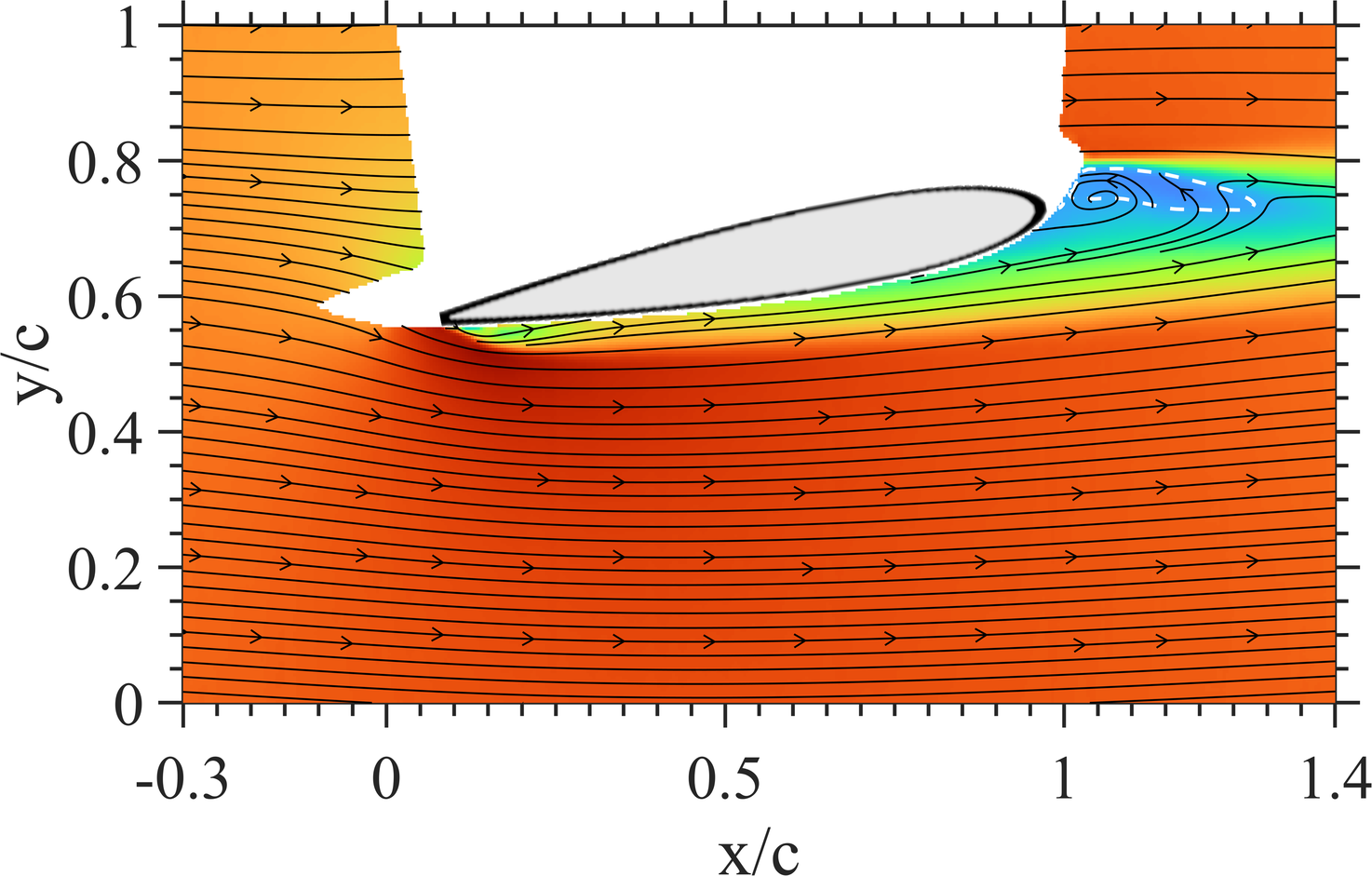}
    \caption{$\alpha$ = 190$^\circ$}
    \label{}
\end{subfigure}
\vskip\baselineskip
% \hspace{7em}
\begin{subfigure}[b]{0.351\textwidth}
    \centering 
    \includegraphics[trim=0 0 0 0,clip,width=\textwidth]{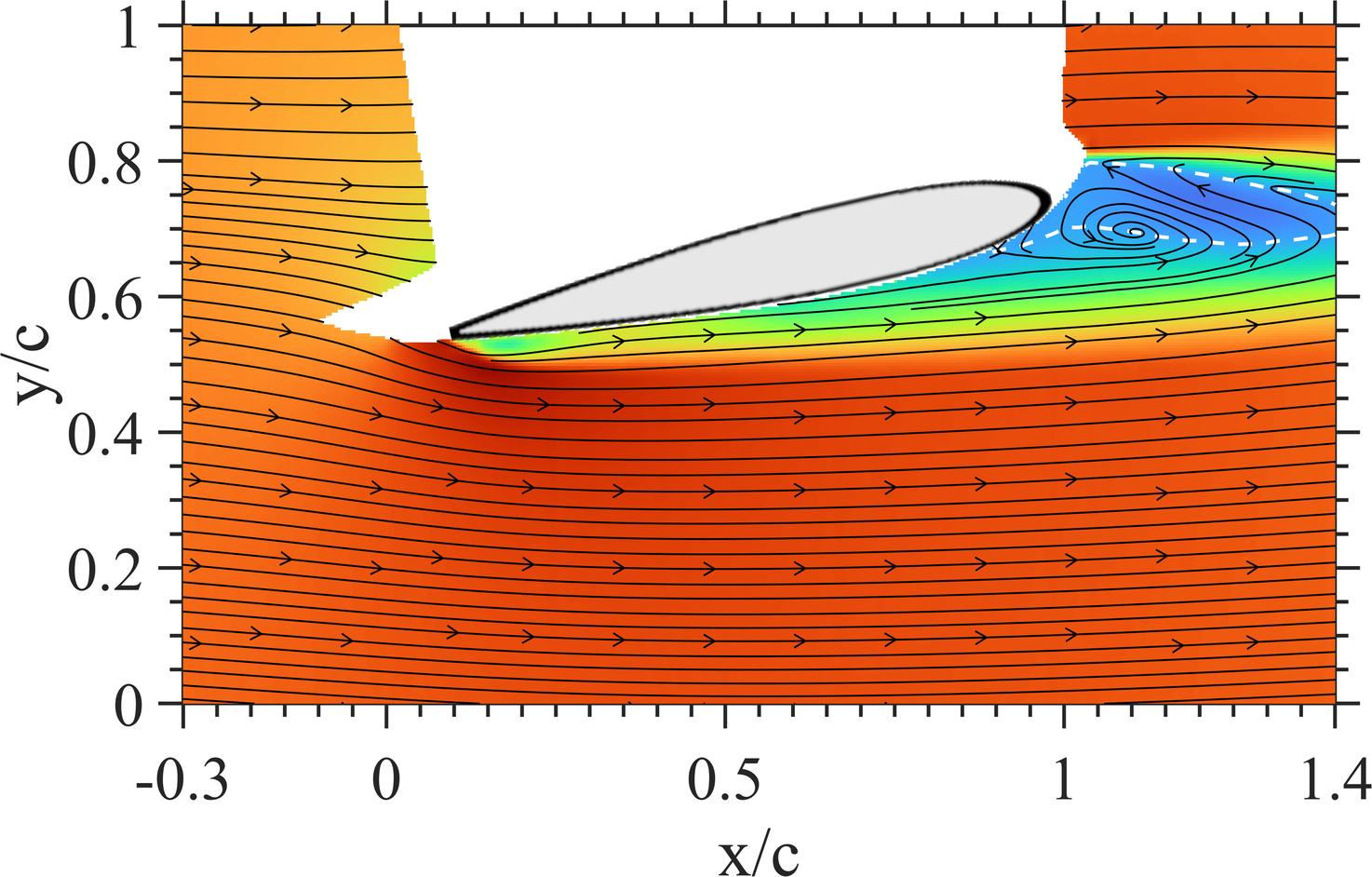}
    \caption{$\alpha$ = 192$^\circ$}
    \label{}
\end{subfigure}
\begin{subfigure}[b]{0.315\textwidth}
    \centering 
    \includegraphics[trim=33 0 0 0,clip,width=\textwidth]{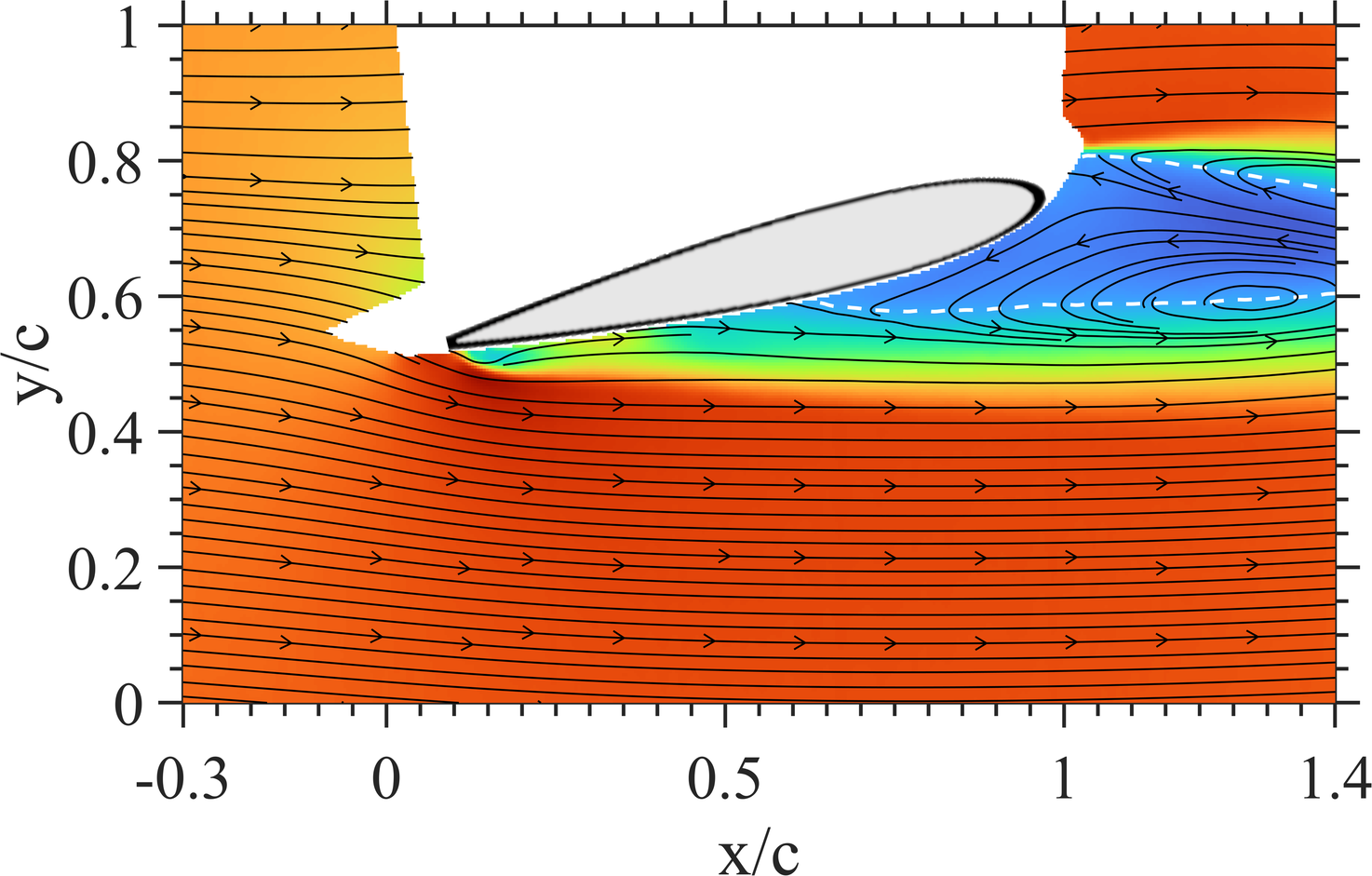}
    \caption{$\alpha$ = 194$^\circ$}
    \label{}
\end{subfigure}
\caption{Contour plots of the time-averaged streamwise velocity component, normalized with freestream velocity for the $\mathbf{A10\lambda05}$ airfoil along the middle of a midspan sinusoid, at static pitch angles of 180$^\circ$ (a), 182$^\circ$ (b), 184$^\circ$ (c), 186$^\circ$ (d), 188$^\circ$ (e), 190$^\circ$ (f), 192$^\circ$ (g), and 194$^\circ$ (h), overlaid with in-plane streamlines. The white dashed contour line represents the location where mean streamwise velocity is zero. Flow direction is from left to right and $\mathbf{Re_c = 1.4\times10^5}$.}
\label{A10P05_M}
\end{figure}

Figures \ref{A10P05_M} and \ref{A10P10_M} show time-averaged normalized streamwise velocity contours for the $A10\lambda05$ and $A10\lambda10$ airfoils, respectively, on a midspan plane passing through the middle of a sinusoid. The static angles of attack are the same as in previous cases. The cross-section of the airfoil at this location, which has a blunt leading edge with a thickness equal to that of a NACA 0015 airfoil at 95\% of chord length, is overlaid on the flowfields.
The measurements on this plane show general characteristics similar to the sinusoidal crest plane. For both the airfoils, the flow is completely attached over the surface till $\alpha$ = 190$^\circ$, separating only near the blunt trailing edge to form a bluff-body type wake. 

The point of separation moves upstream to $\sim$0.9c at $\alpha = 192^\circ$, and further upstream to $\sim$0.6c at $\alpha = 194^\circ$. However, several features are different from the sinusoidal crest plane. At the crest plane, close to the surface for both airfoils, a region of flow acceleration was observed near the leading edge, followed by a region of flow deceleration. At the middle plane, the region of flow acceleration is absent. 
More interestingly, for $\alpha \geq 188^\circ$, a region of low velocity is observed near the leading edge, followed by a region of high velocity and gradual deceleration. 
For the $A10\lambda10$ airfoil, at $\alpha = 192^\circ$ and $194^\circ$, closed zero-velocity (white-dashed) contours are observable near the leading edge, suggesting small recirculation regions. A possible explanation for these observations can be obtained from Fig. \ref{fig:serration vortices-schem}. The initial segment of the sinusoidal middle plane coincides with the region of vortex formation over the sinusoid. In this region, the presence of low or negative velocities may be either due to the nature of the axial velocity characteristics within the vortex cores, or due to the two-dimensional projection of a highly three-dimensional flowfield.
\begin{figure}[H]
\centering
\begin{subfigure}[b]{0.44\textwidth}
    \includegraphics[width=\textwidth]{Images/PIV_plots/Avg_Vx_png/legend_horz.jpg}
\end{subfigure}
\hspace{6em}
\begin{subfigure}[b]{0.09\textwidth}
    \centering 
    \includegraphics[trim=0 12 0 0,clip,width=\textwidth]{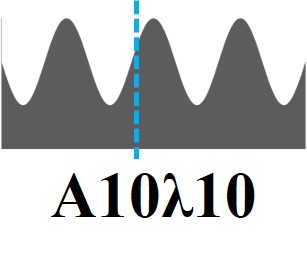}
\end{subfigure}
\vskip\baselineskip
\centering
\begin{subfigure}[b]{0.351\textwidth}
    \centering 
    \includegraphics[trim=0 34 0 0,clip,width=\textwidth]{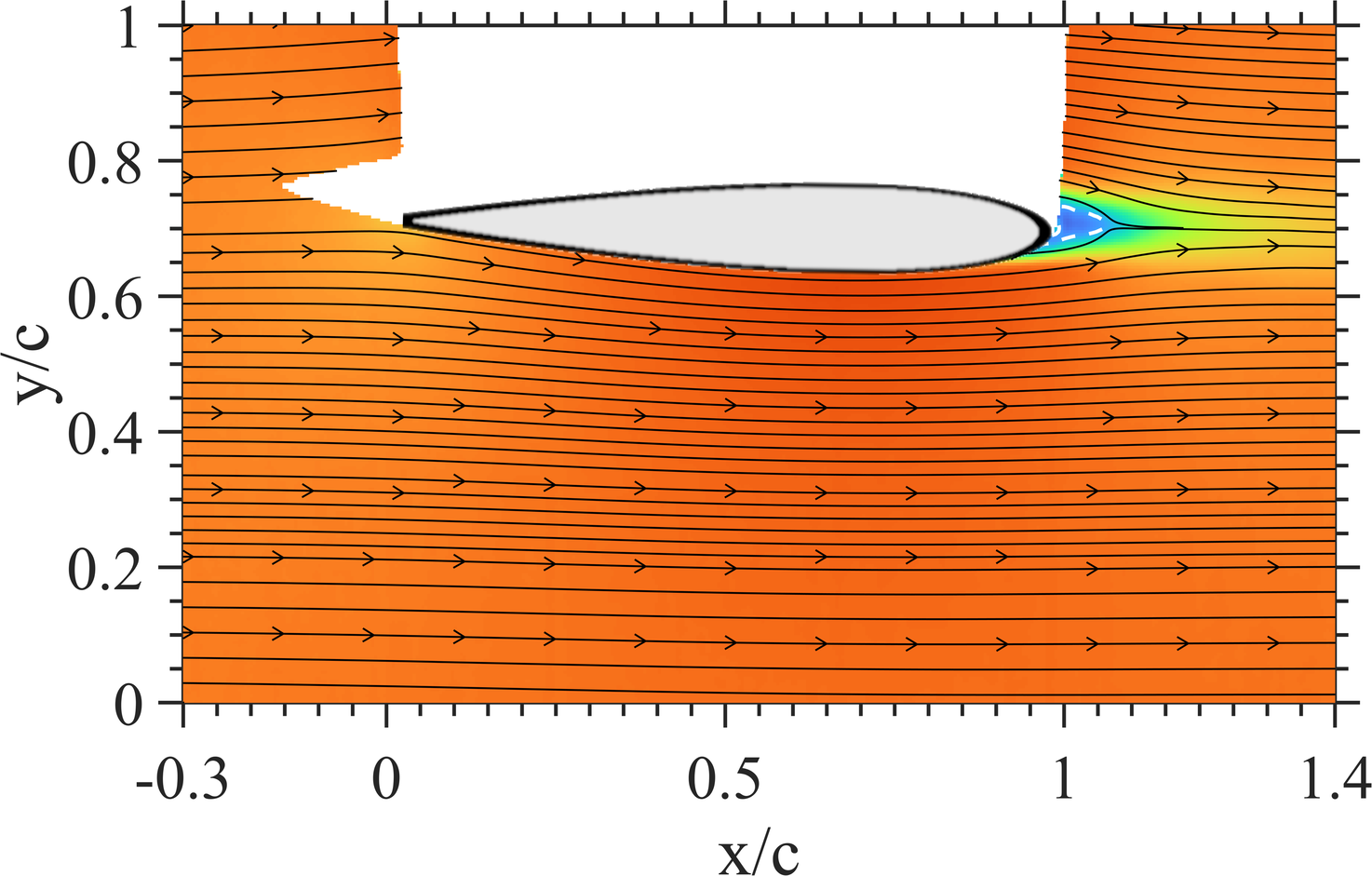}
    \caption{$\alpha$ = 180$^\circ$}
    \label{}
\end{subfigure}
\begin{subfigure}[b]{0.315\textwidth}
    \centering 
    \includegraphics[trim=33 34 0 0,clip,width=\textwidth]{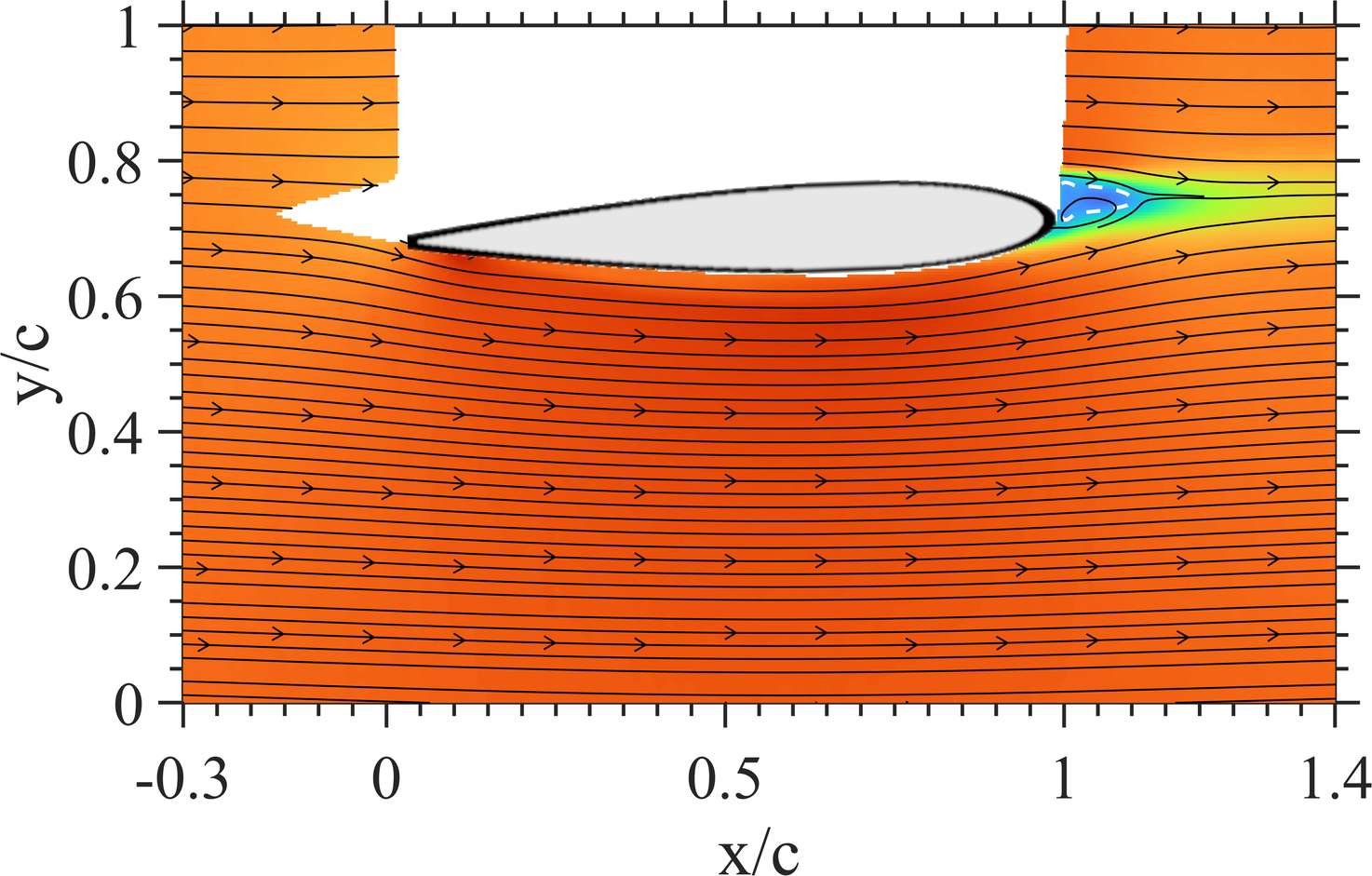}
    \caption{$\alpha$ = 182$^\circ$}
    \label{}
\end{subfigure}
\begin{subfigure}[b]{0.315\textwidth}
    \centering 
    \includegraphics[trim=33 34 0 0,clip,width=\textwidth]{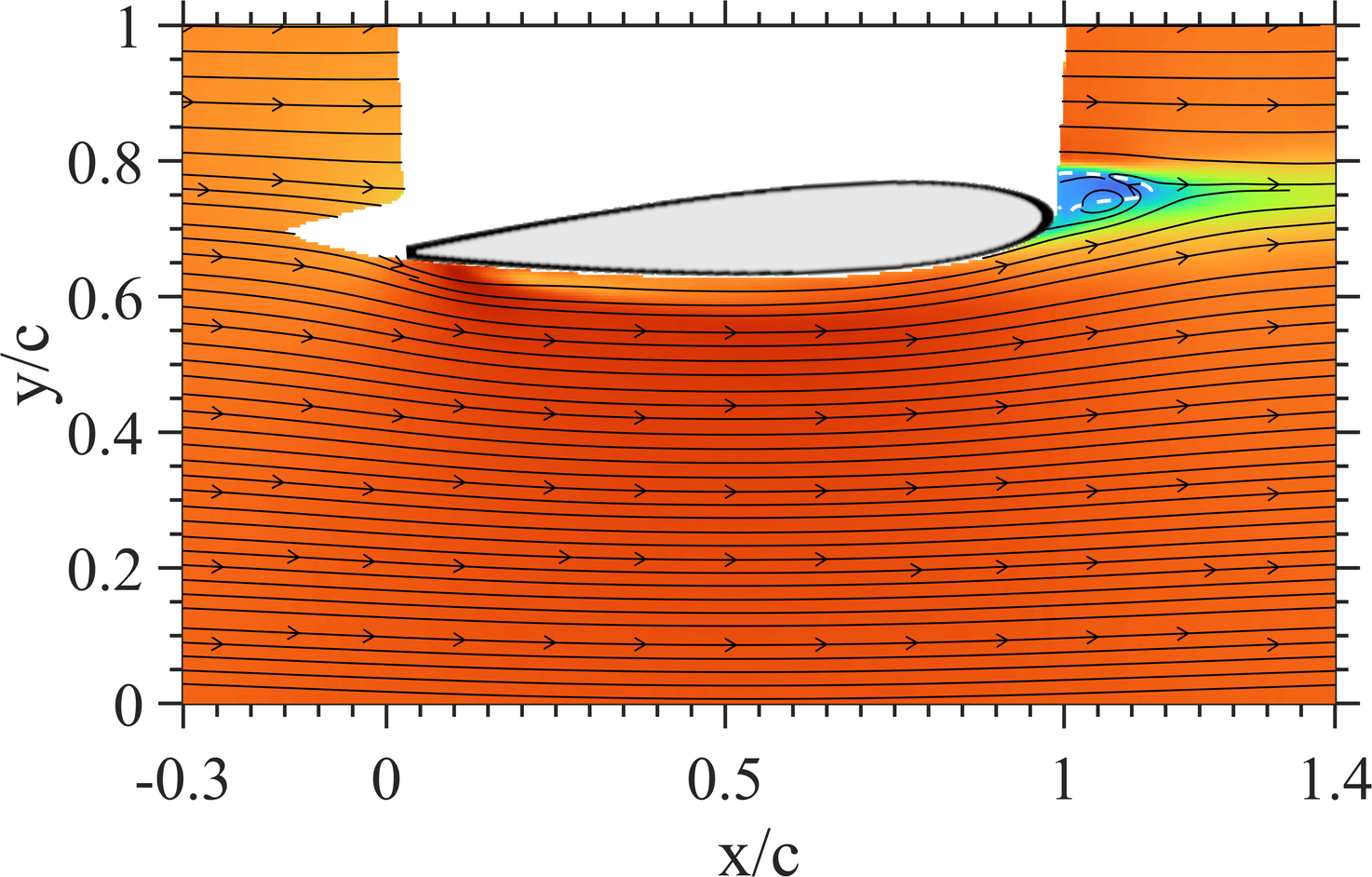}
    \caption{$\alpha$ = 184$^\circ$}
    \label{}
\end{subfigure}
\vskip\baselineskip
\begin{subfigure}[b]{0.351\textwidth}
    \centering 
    \includegraphics[trim=0 0 0 0,clip,width=\textwidth]{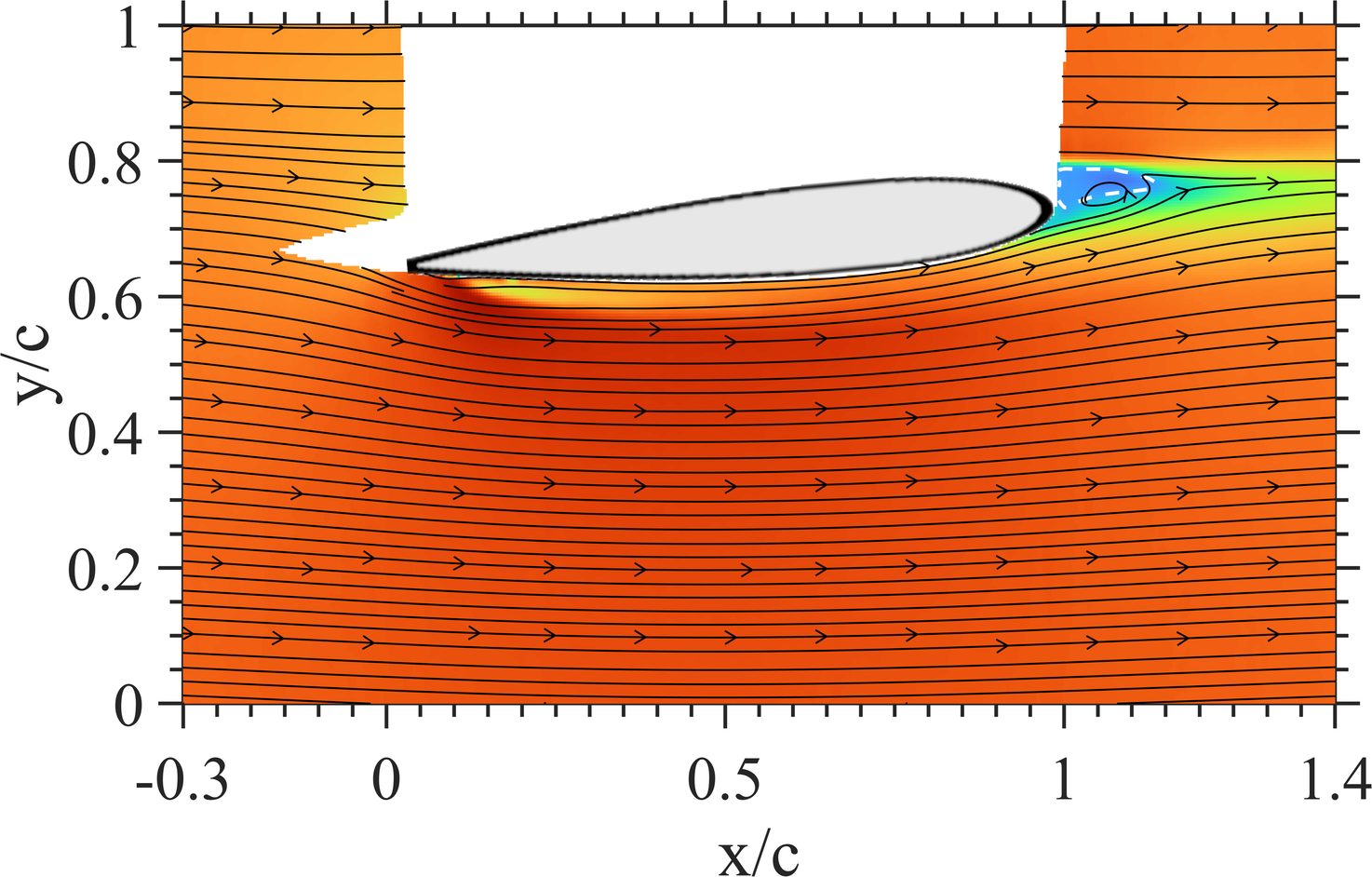}
    \caption{$\alpha$ = 186$^\circ$}
    \label{}
\end{subfigure}
\begin{subfigure}[b]{0.315\textwidth}
    \centering 
    \includegraphics[trim=33 0 0 0,clip,width=\textwidth]{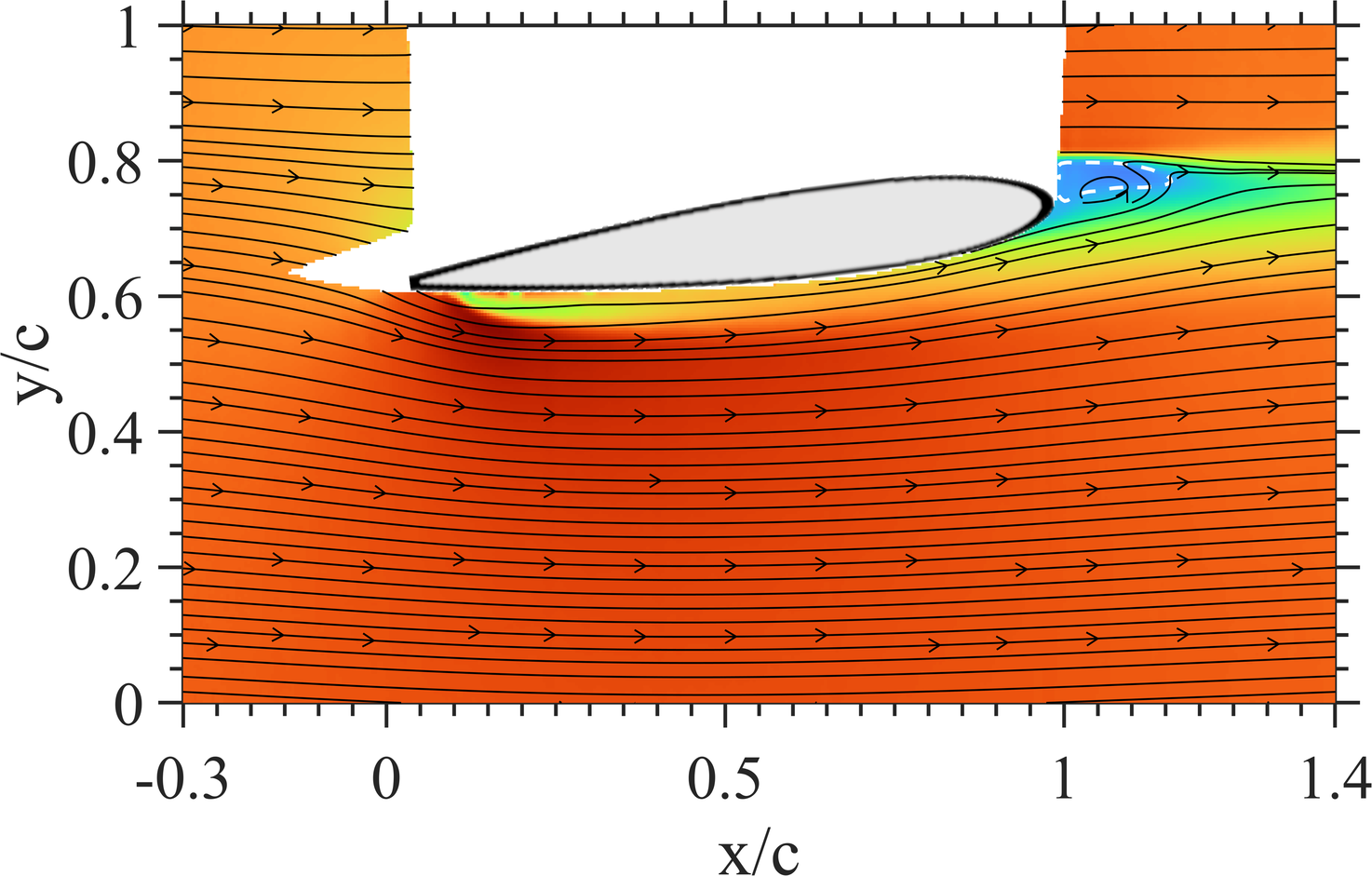}
    \caption{$\alpha$ = 188$^\circ$}
    \label{}
\end{subfigure}
\begin{subfigure}[b]{0.315\textwidth}
    \centering 
    \includegraphics[trim=33 0 0 0,clip,width=\textwidth]{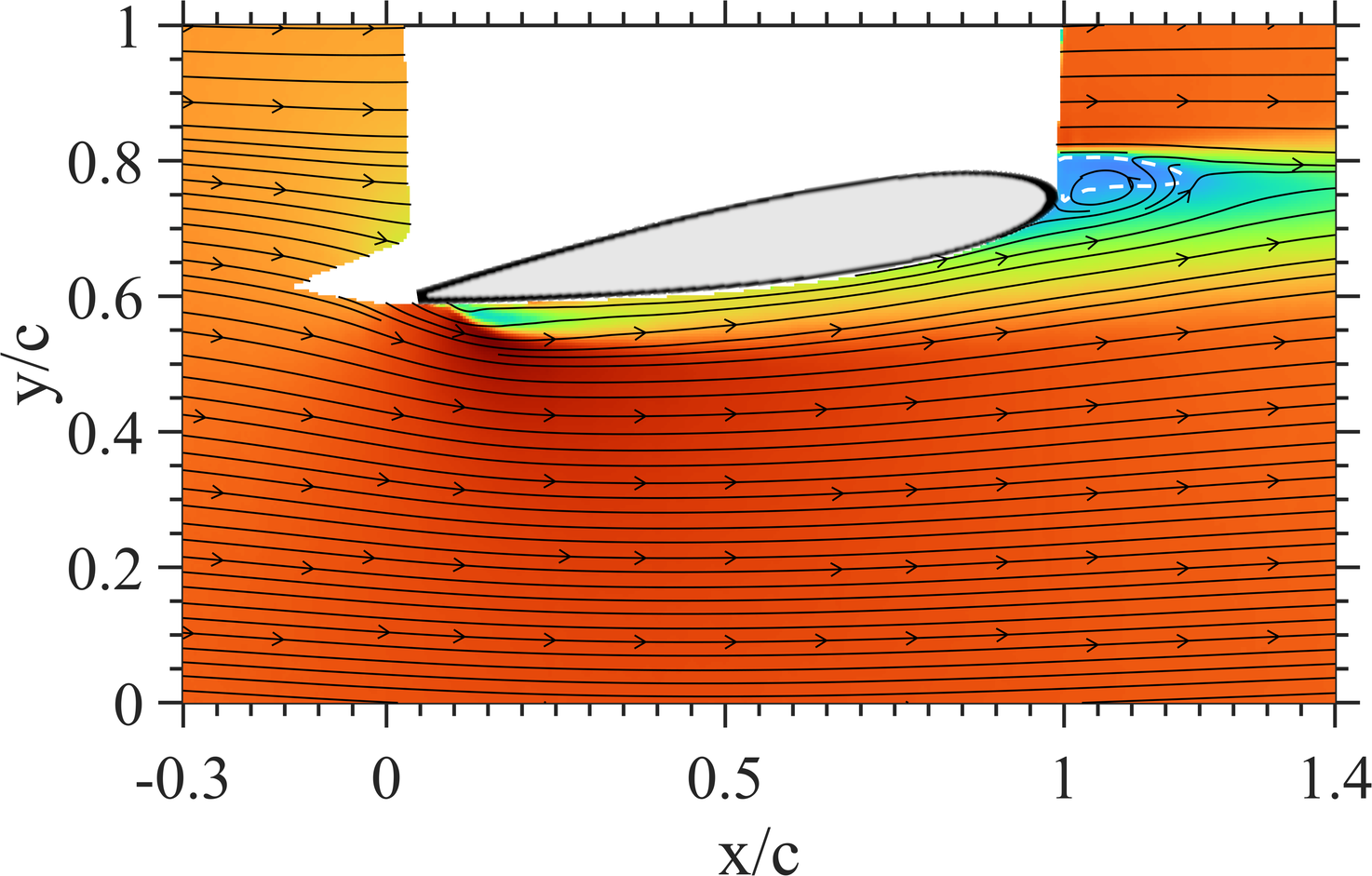}
    \caption{$\alpha$ = 190$^\circ$}
    \label{}
\end{subfigure}
% \vskip\baselineskip
\end{figure}
\begin{figure}[H]\ContinuedFloat
\hspace{7em}
\begin{subfigure}[b]{0.351\textwidth}
    \centering 
    \includegraphics[trim=0 0 0 0,clip,width=\textwidth]{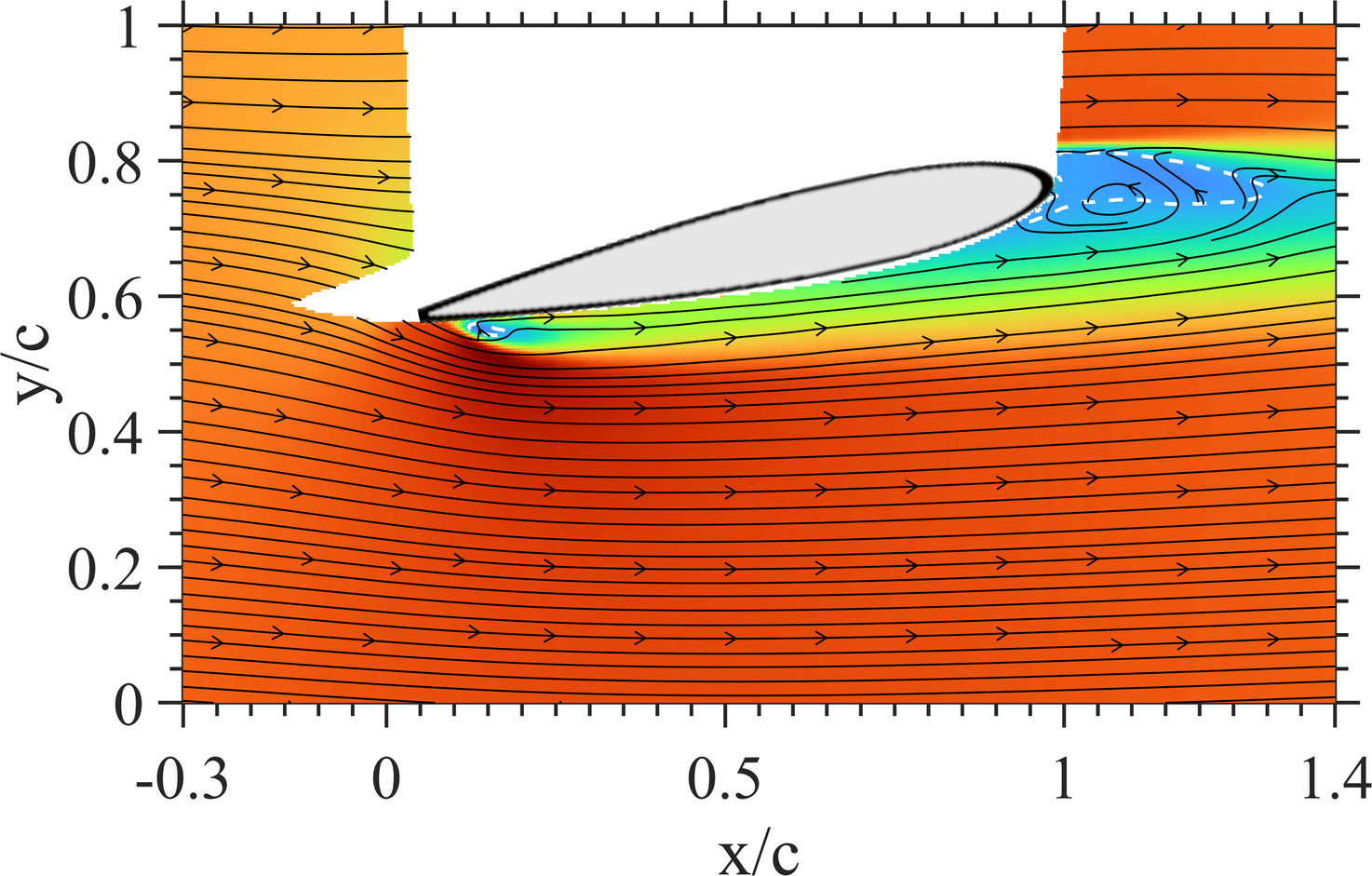}
    \caption{$\alpha$ = 192$^\circ$}
    \label{}
\end{subfigure}
\begin{subfigure}[b]{0.315\textwidth}
    \centering 
    \includegraphics[trim=33 0 0 0,clip,width=\textwidth]{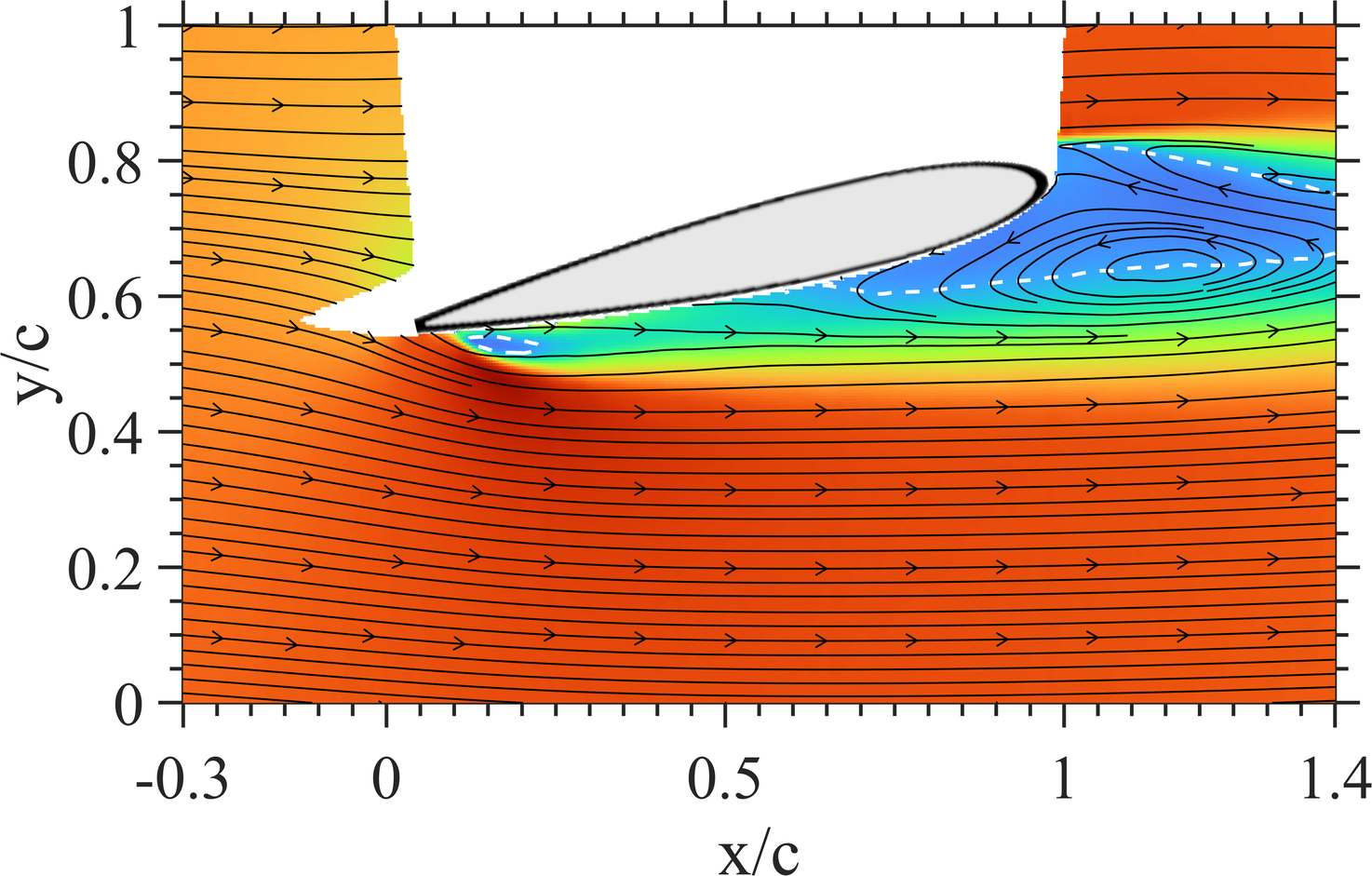}
    \caption{$\alpha$ = 194$^\circ$}
    \label{}
\end{subfigure}
\caption{Contour plots of the time-averaged streamwise velocity component, normalized with freestream velocity for the $\mathbf{A10\lambda10}$ airfoil along the middle of a midspan sinusoid, at static pitch angles of 180$^\circ$ (a), 182$^\circ$ (b), 184$^\circ$ (c), 186$^\circ$ (d), 188$^\circ$ (e), 190$^\circ$ (f), 192$^\circ$ (g), and 194$^\circ$ (h), overlaid with in-plane streamlines. White dashed contour lines represent the locations where mean streamwise velocity is zero. Flow direction is from left to right and $\mathbf{Re_c = 1.4\times10^5}$.}
\label{A10P10_M}
\end{figure}
The remaining portion of the plane corresponds to the region where the energized flow between the vortices gradually decelerates and eventually separates due to the adverse pressure gradient. The adverse pressure gradient intensifies with increasing angle of attack; correspondingly, near-surface velocities decrease with increasing incidence.

%%%%%%%%%%%%%%%%%%%%%%%%%%%%%%%%%%%%%%%%%%%%%%%%%%%%%

\subsubsection{Sinusoidal trailing edge airfoil: Sinusoidal trough plane}
Figures \ref{A10P05_T} and \ref{A10P10_T} show time-averaged normalized streamwise velocity contours for the $A10\lambda05$ and $A10\lambda10$ airfoils, respectively, on a midspan plane passing through the trough of a sinusoid. The cross-section of the airfoil at this location, which has a blunt leading edge with a thickness equal to that of a NACA 0015 airfoil at 90\% of chord length, is overlaid on the flowfields.

The flow is fully attached over the surface till $\alpha = 190^\circ$ for both airfoils, separating only near the blunt leading edge to form the bluff-body type wake. For the $A10\lambda05$ airfoil, the observations are similar to those in the previous cases, with flow separation occurring at $\sim$0.9c and $\sim$0.7c for $\alpha = 192^\circ$ and $194^\circ$, respectively. A portion of the secondary wake recirculation region, developing from the pressure side, is visible at $\alpha = 194^\circ$. For the $A10\lambda10$ airfoil, flow separates slightly earlier, at $\sim$0.8c for $\alpha = 192^\circ$. It separates at $\sim$0.7c for $\alpha = 194^\circ$, similar to $A10\lambda05$ airfoil. The secondary recirculation region in the wake is observed at $\alpha = 192^\circ$, but is beyond the field of view at $\alpha = 194^\circ$. Interesting to note that, flow separates slightly earlier along the trough plane for the $A10\lambda10$ airfoil at $\alpha = 192^\circ$, in comparison to all other modified airfoil cases. It is also the only case where a secondary recirculation is not observed in the wake at $\alpha = 194^\circ$. 
This suggests that along the span the flow is three-dimensional in nature, with possible existence of spanwise pressure gradients, both on the airfoil surface and in the wake.  The flow near the surface shows similar characteristics to those observed at the sinusoidal-middle plane. A region of low velocity near the leading edge, followed by a recovery and a gradual dissipation. In general, the cross-stream extent of this region is greater for the $A10\lambda10$ airfoil in comparison to the $A10\lambda05$ airfoil. The $A10\lambda10$ airfoil also shows a significantly more prominent low-velocity region near the leading edge, starting from a very small angle of $182^\circ$.

\begin{figure}[H]
\centering
\begin{subfigure}[b]{0.44\textwidth}
    \includegraphics[width=\textwidth]{Images/PIV_plots/Avg_Vx_png/legend_horz.jpg}
\end{subfigure}
\hspace{6em}
\begin{subfigure}[b]{0.09\textwidth}
    \centering 
    \includegraphics[trim=0 12 0 0,clip,width=\textwidth]{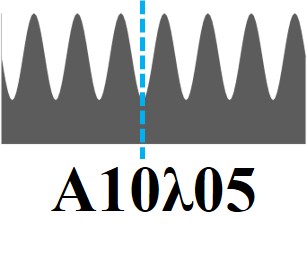}
\end{subfigure}
\vskip\baselineskip
\centering
\begin{subfigure}[b]{0.351\textwidth}
    \centering 
    \includegraphics[trim=0 34 0 0,clip,width=\textwidth]{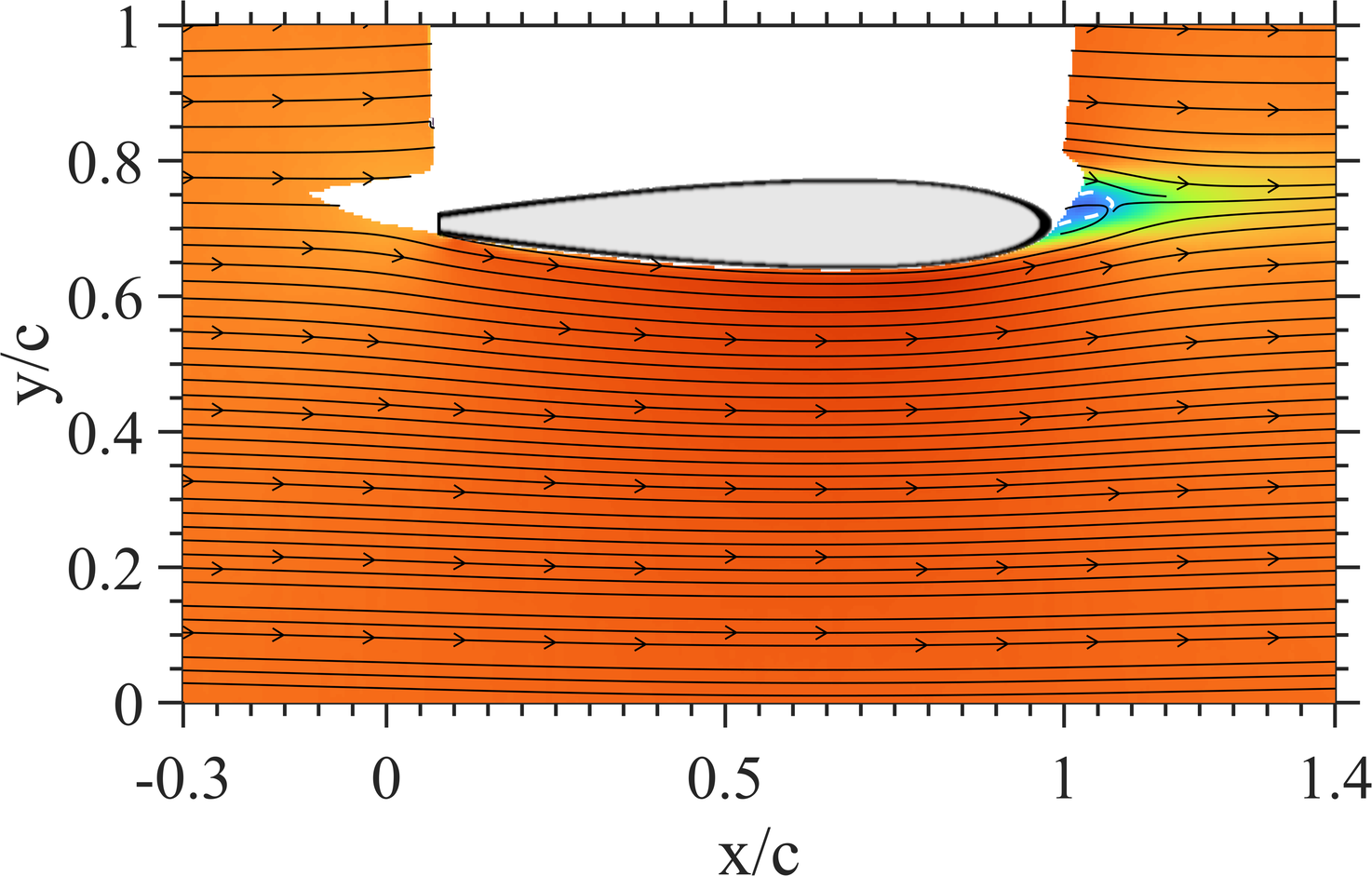}
    \caption{$\alpha$ = 180$^\circ$}
    \label{}
\end{subfigure}
\begin{subfigure}[b]{0.315\textwidth}
    \centering 
    \includegraphics[trim=33 34 0 0,clip,width=\textwidth]{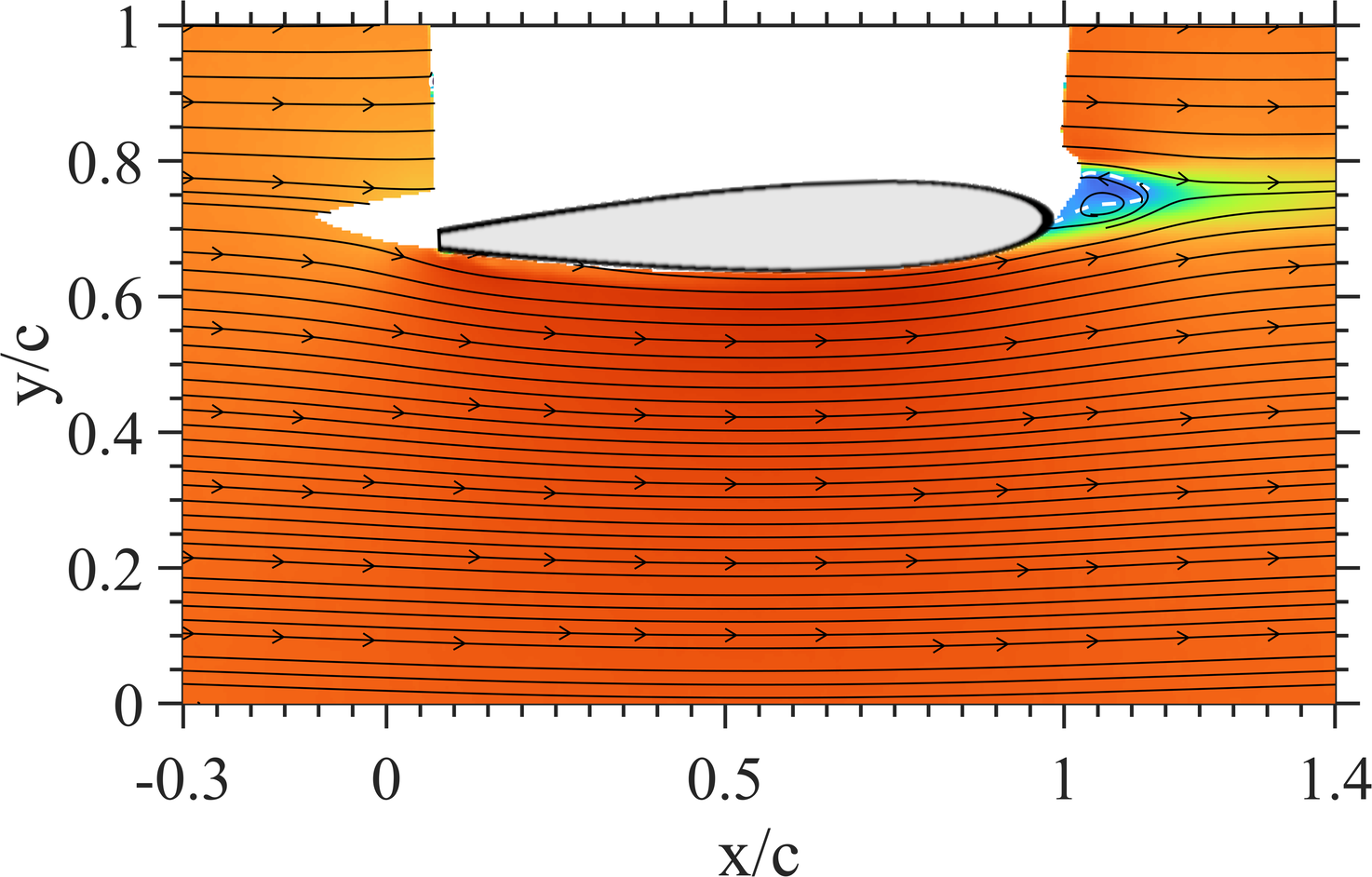}
    \caption{$\alpha$ = 182$^\circ$}
    \label{}
\end{subfigure}
\begin{subfigure}[b]{0.315\textwidth}
    \centering 
    \includegraphics[trim=33 34 0 0,clip,width=\textwidth]{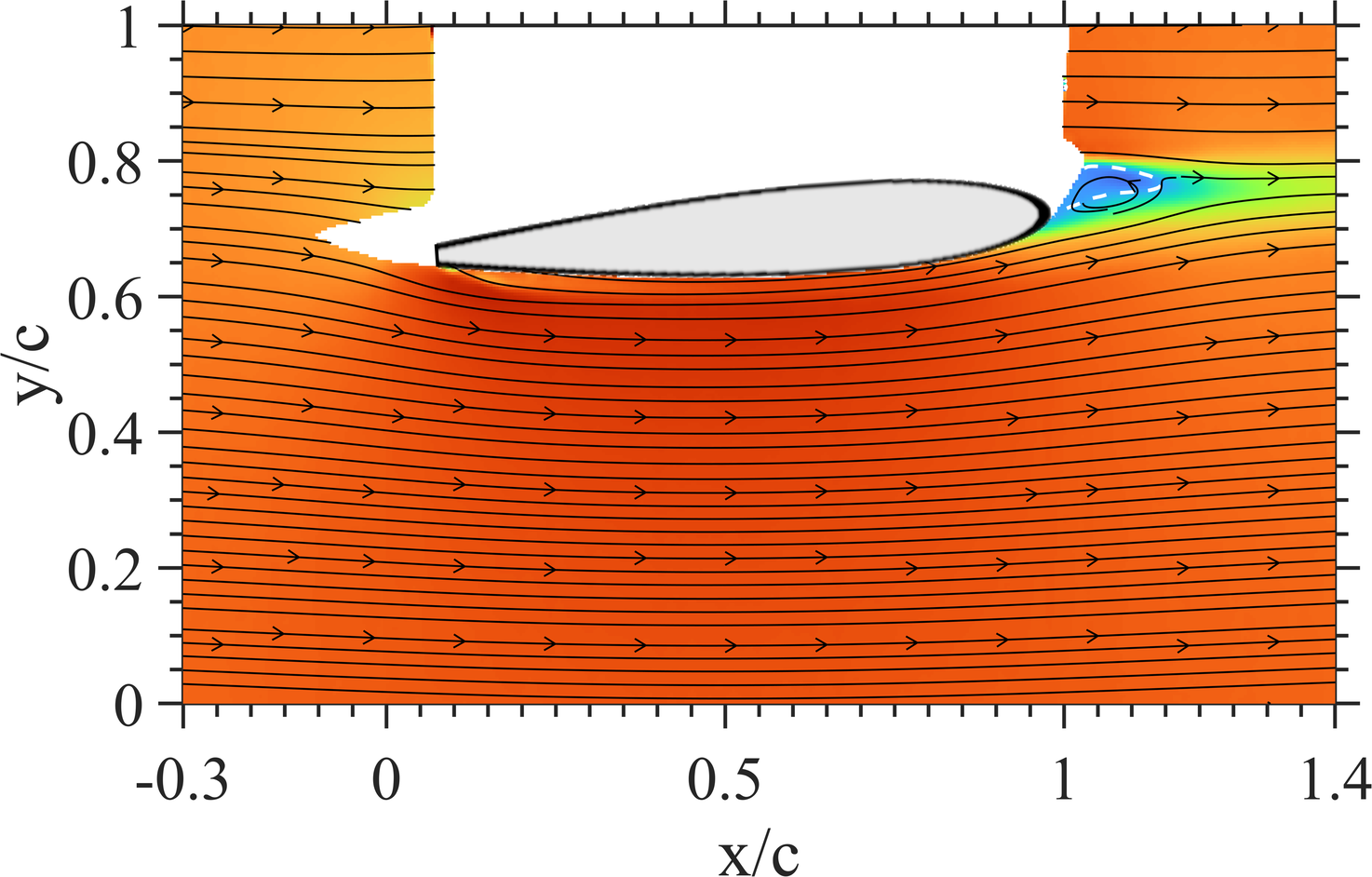}
    \caption{$\alpha$ = 184$^\circ$}
    \label{}
\end{subfigure}
% \vskip\baselineskip
\end{figure}
\begin{figure}[H]\ContinuedFloat
\begin{subfigure}[b]{0.351\textwidth}
    \centering 
    \includegraphics[trim=0 0 0 0,clip,width=\textwidth]{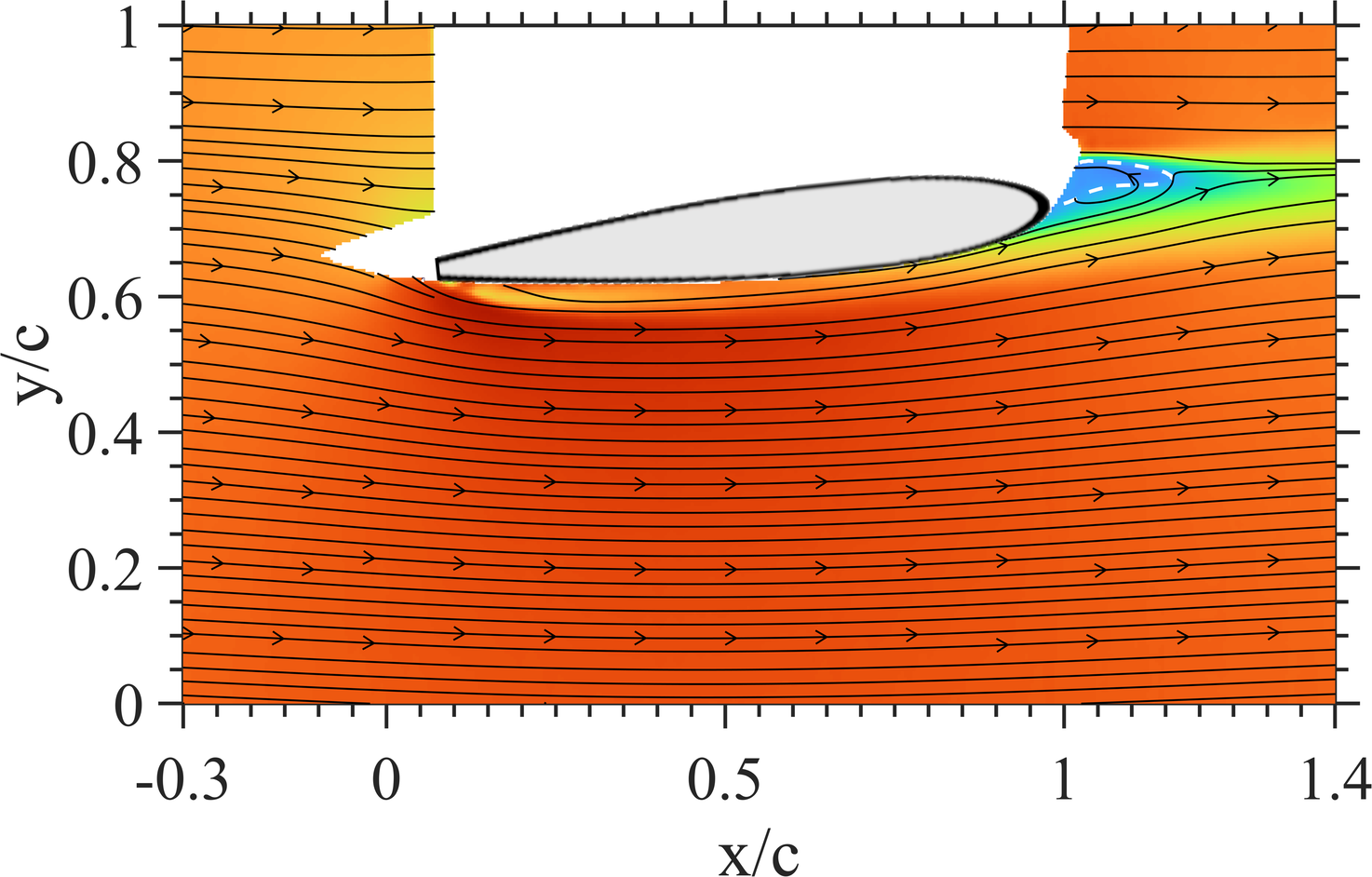}
    \caption{$\alpha$ = 186$^\circ$}
    \label{}
\end{subfigure}
\begin{subfigure}[b]{0.315\textwidth}
    \centering 
    \includegraphics[trim=33 0 0 0,clip,width=\textwidth]{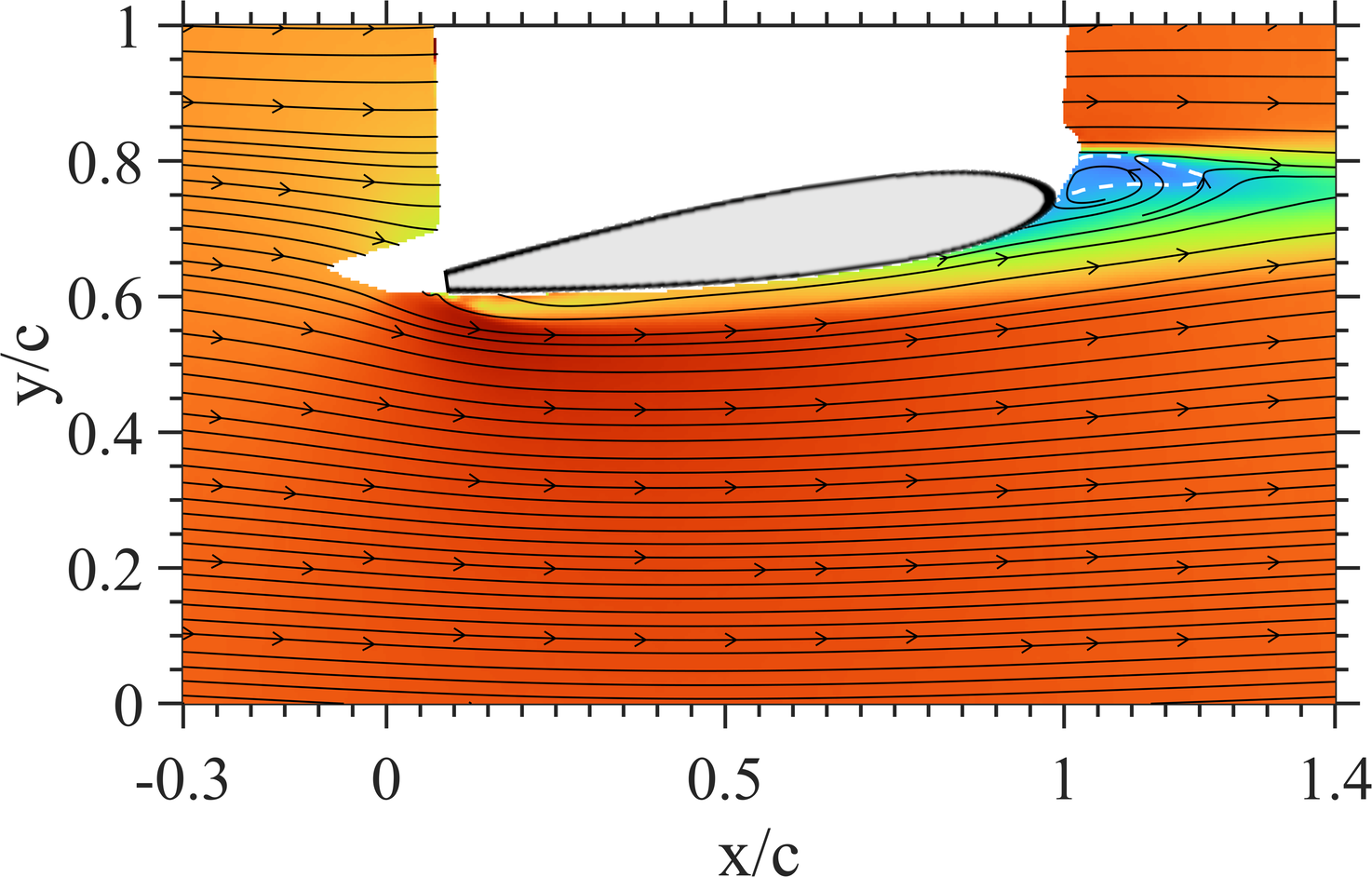}
    \caption{$\alpha$ = 188$^\circ$}
    \label{}
\end{subfigure}
\begin{subfigure}[b]{0.315\textwidth}
    \centering 
    \includegraphics[trim=33 0 0 0,clip,width=\textwidth]{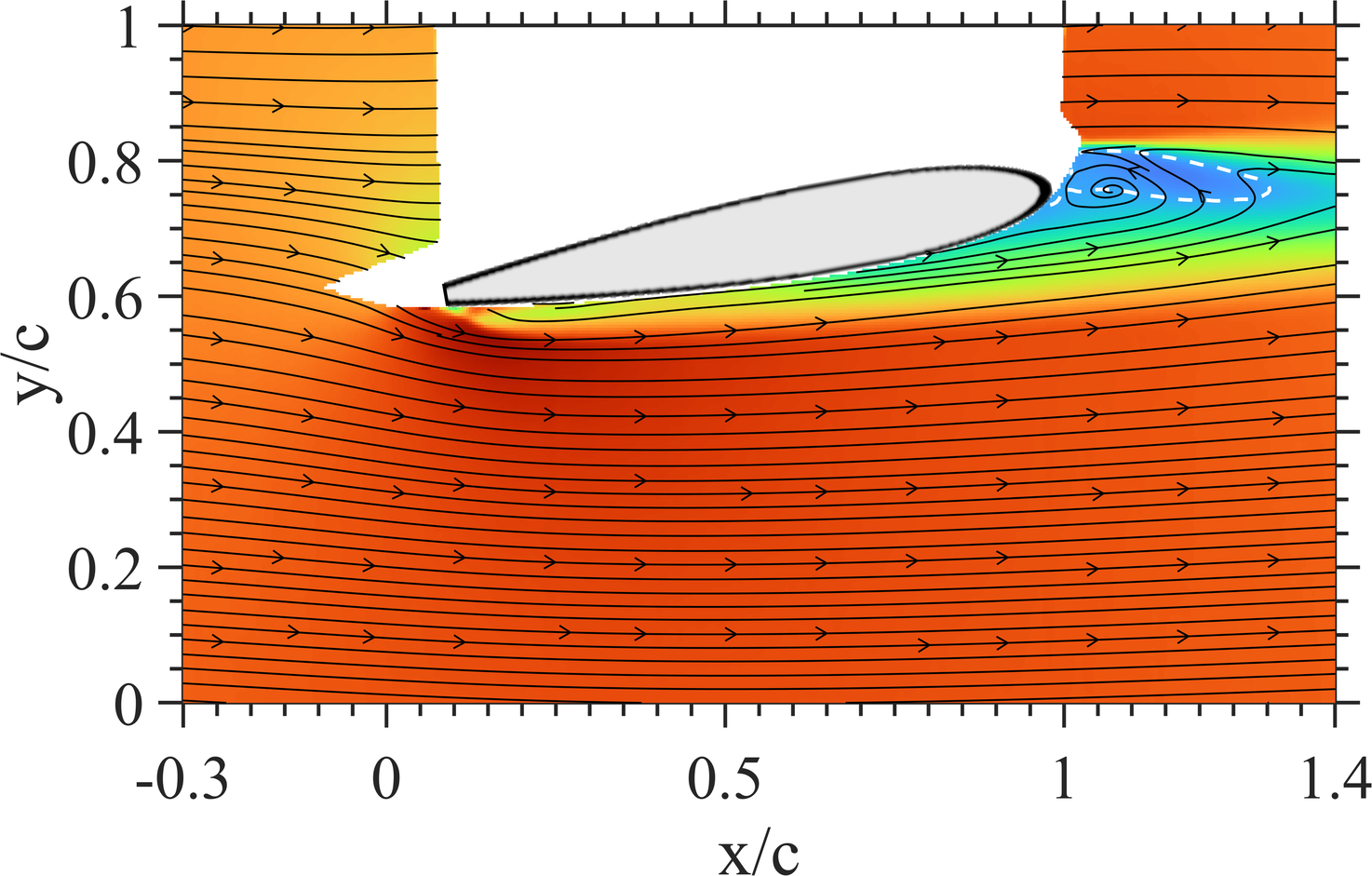}
    \caption{$\alpha$ = 190$^\circ$}
    \label{}
\end{subfigure}
\vskip\baselineskip
\hspace{7em}
\begin{subfigure}[b]{0.351\textwidth}
    \centering 
    \includegraphics[trim=0 0 0 0,clip,width=\textwidth]{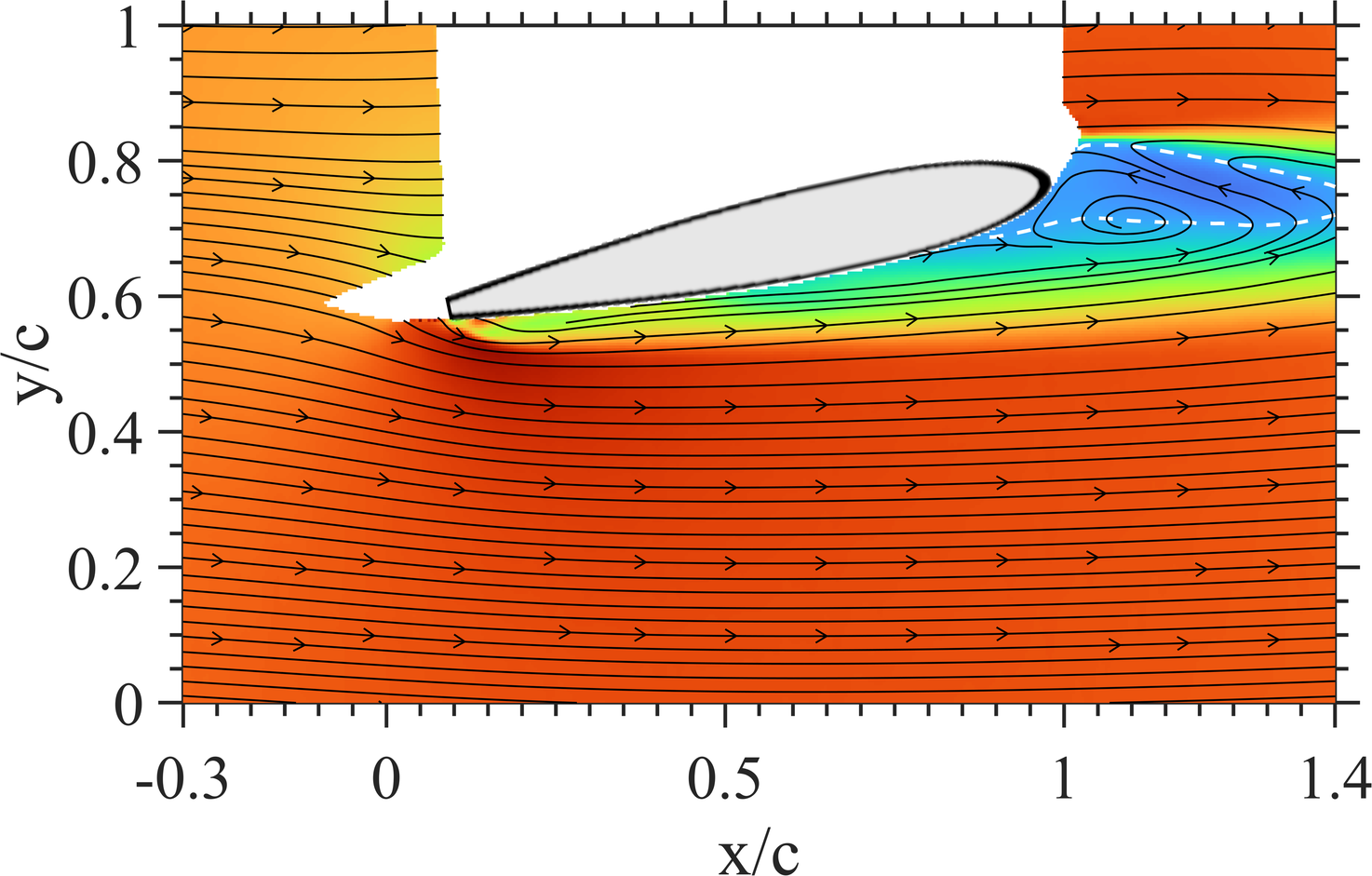}
    \caption{$\alpha$ = 192$^\circ$}
    \label{}
\end{subfigure}
\begin{subfigure}[b]{0.315\textwidth}
    \centering 
    \includegraphics[trim=33 0 0 0,clip,width=\textwidth]{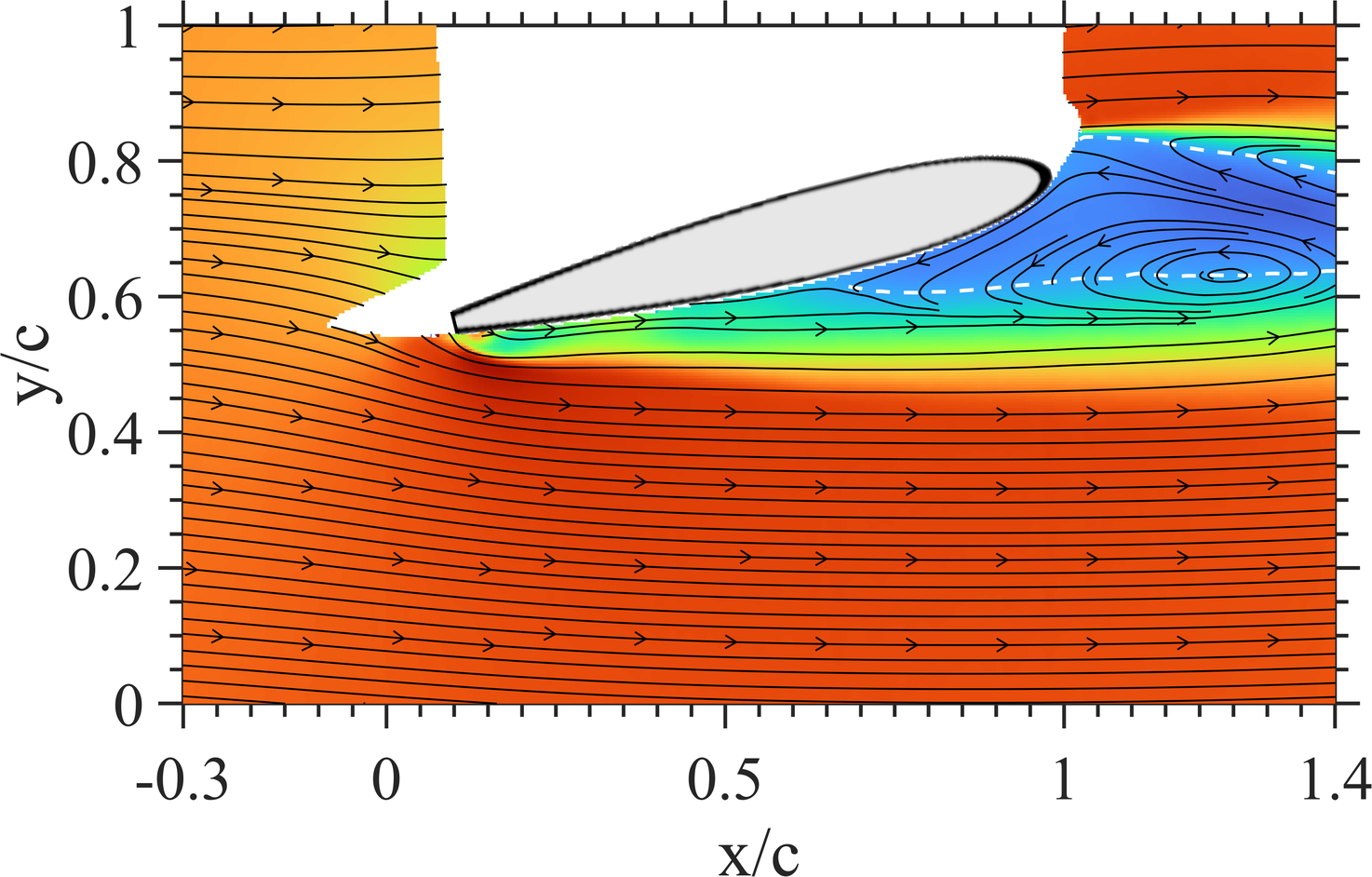}
    \caption{$\alpha$ = 194$^\circ$}
    \label{}
\end{subfigure}
\caption{Contour plots of the time-averaged streamwise velocity component, normalized with freestream velocity for the $\mathbf{A10\lambda05}$ airfoil along the trough of a midspan sinusoid, at static pitch angles from 184$^\circ$ (a) to 194$^\circ$ (f), overlaid with in-plane streamlines. The white dashed contour line represents the location where mean streamwise velocity is zero. Flow direction is from left to right.}
\label{A10P05_T}
\end{figure} 
 Small pockets of recirculation are observed at the leading edge at $\alpha = 192^\circ$ and $194^\circ$ for the $A10\lambda10$ airfoil, which are absent for the latter. The trough region has a bluff geometry as discussed previously. The sharp edge of the bluff region could be forcing a local flow separation. In future designs, a sharp or smooth trailing edge along the entire sinusoid can be investigated to improve the flow characteristics in the trough region. The difference in the cross-stream extent of the shear layer at the leading edge suggests a dependence of the flow features on the effective bluff area of the sinusoid exposed to the incoming flow. The wider-wavelength sinusoid, $A10\lambda10$, presents a larger bluff area, even though both sinusoids have the same thickness at the trough. 
\begin{figure}[H]
\centering
\begin{subfigure}[b]{0.44\textwidth}
    \includegraphics[width=\textwidth]{Images/PIV_plots/Avg_Vx_png/legend_horz.jpg}
\end{subfigure}
\hspace{6em}
\begin{subfigure}[b]{0.09\textwidth}
    \centering 
    \includegraphics[trim=0 12 0 0,clip,width=\textwidth]{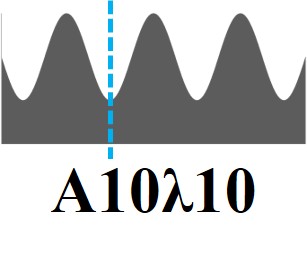}
\end{subfigure}
\vskip\baselineskip
\centering
\begin{subfigure}[b]{0.351\textwidth}
    \centering 
    \includegraphics[trim=0 34 0 0,clip,width=\textwidth]{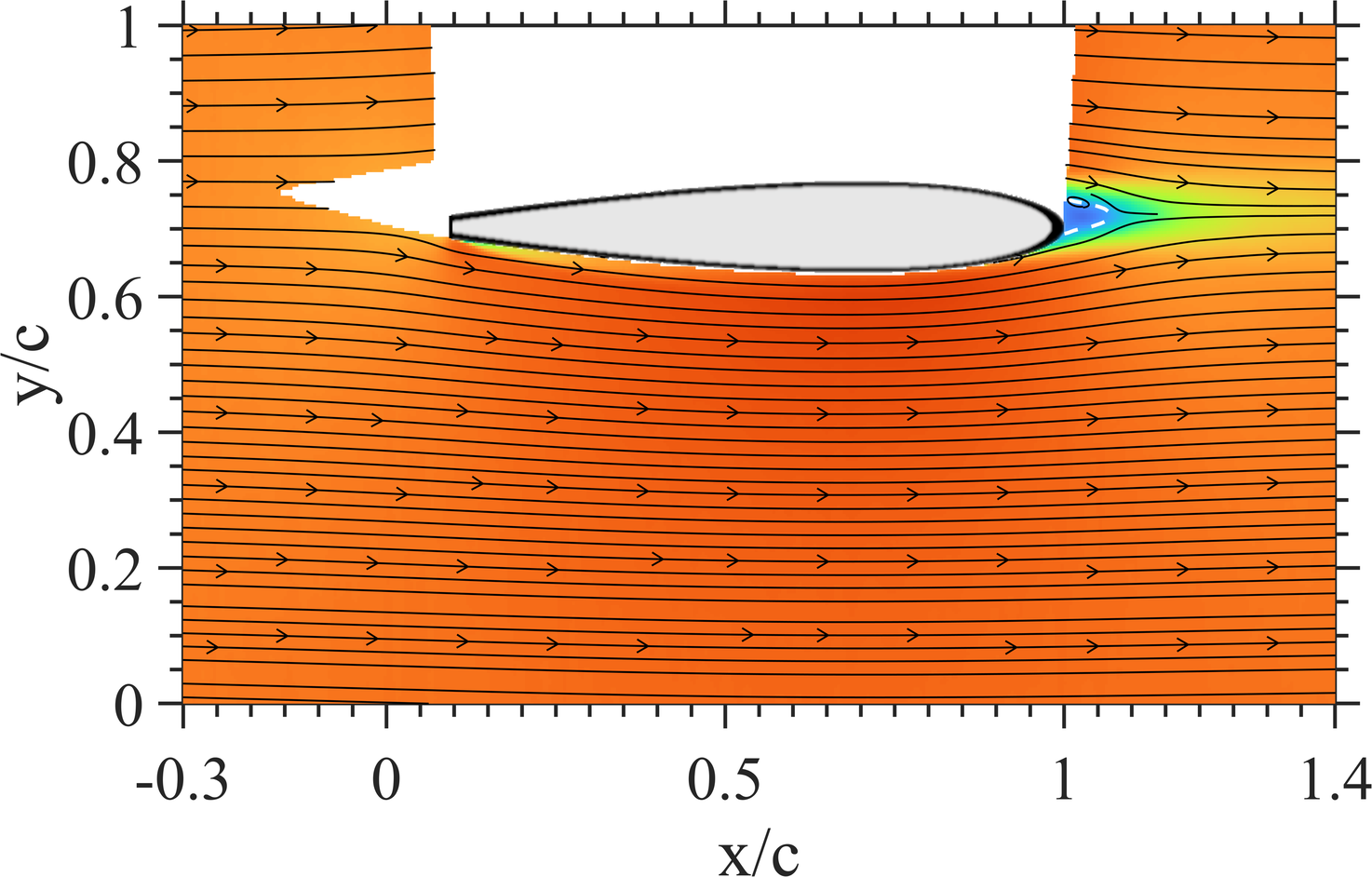}
    \caption{$\alpha$ = 180$^\circ$}
    \label{}
\end{subfigure}
\begin{subfigure}[b]{0.315\textwidth}
    \centering 
    \includegraphics[trim=33 34 0 0,clip,width=\textwidth]{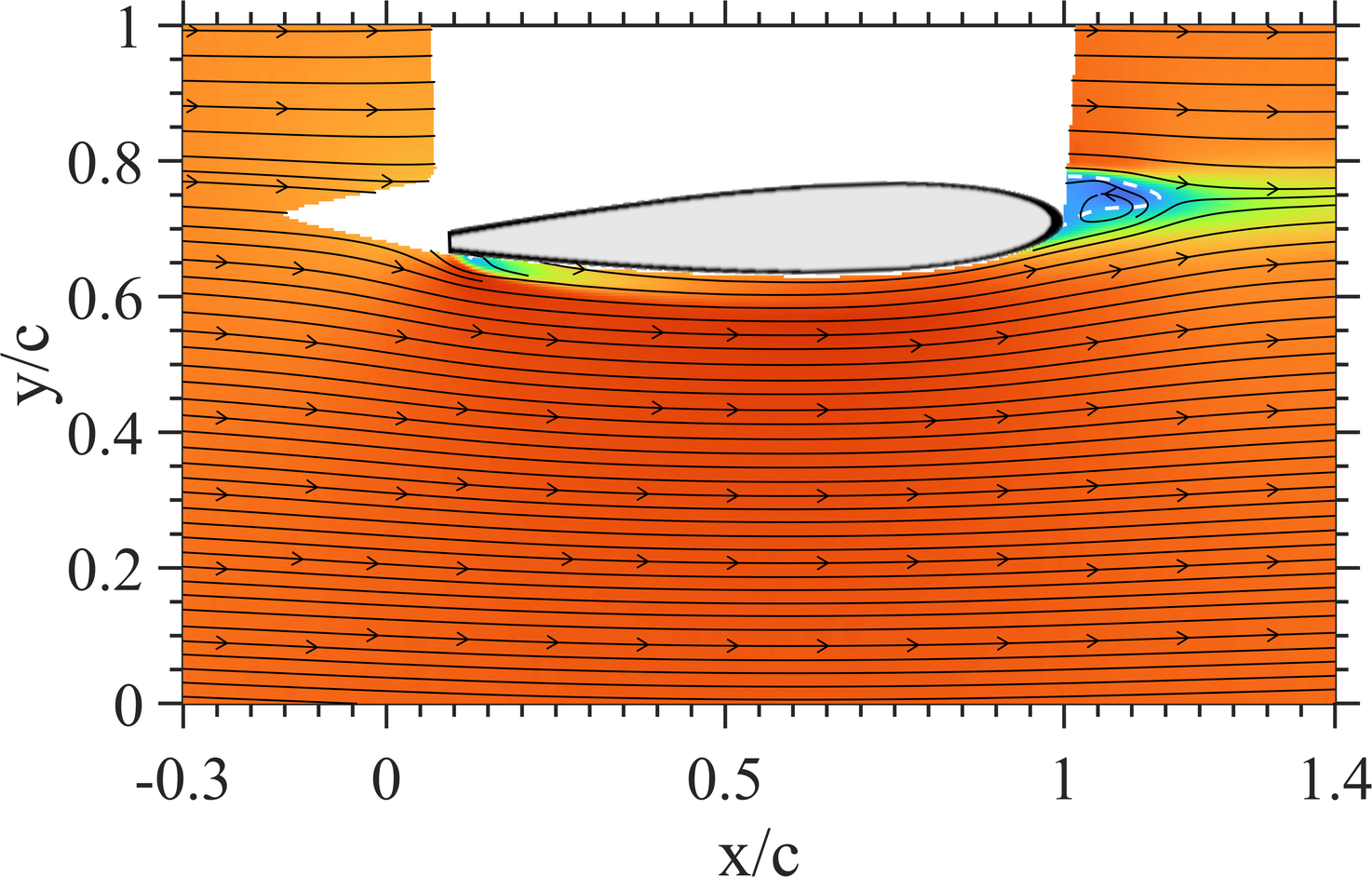}
    \caption{$\alpha$ = 182$^\circ$}
    \label{}
\end{subfigure}
\begin{subfigure}[b]{0.315\textwidth}
    \centering 
    \includegraphics[trim=33 34 0 0,clip,width=\textwidth]{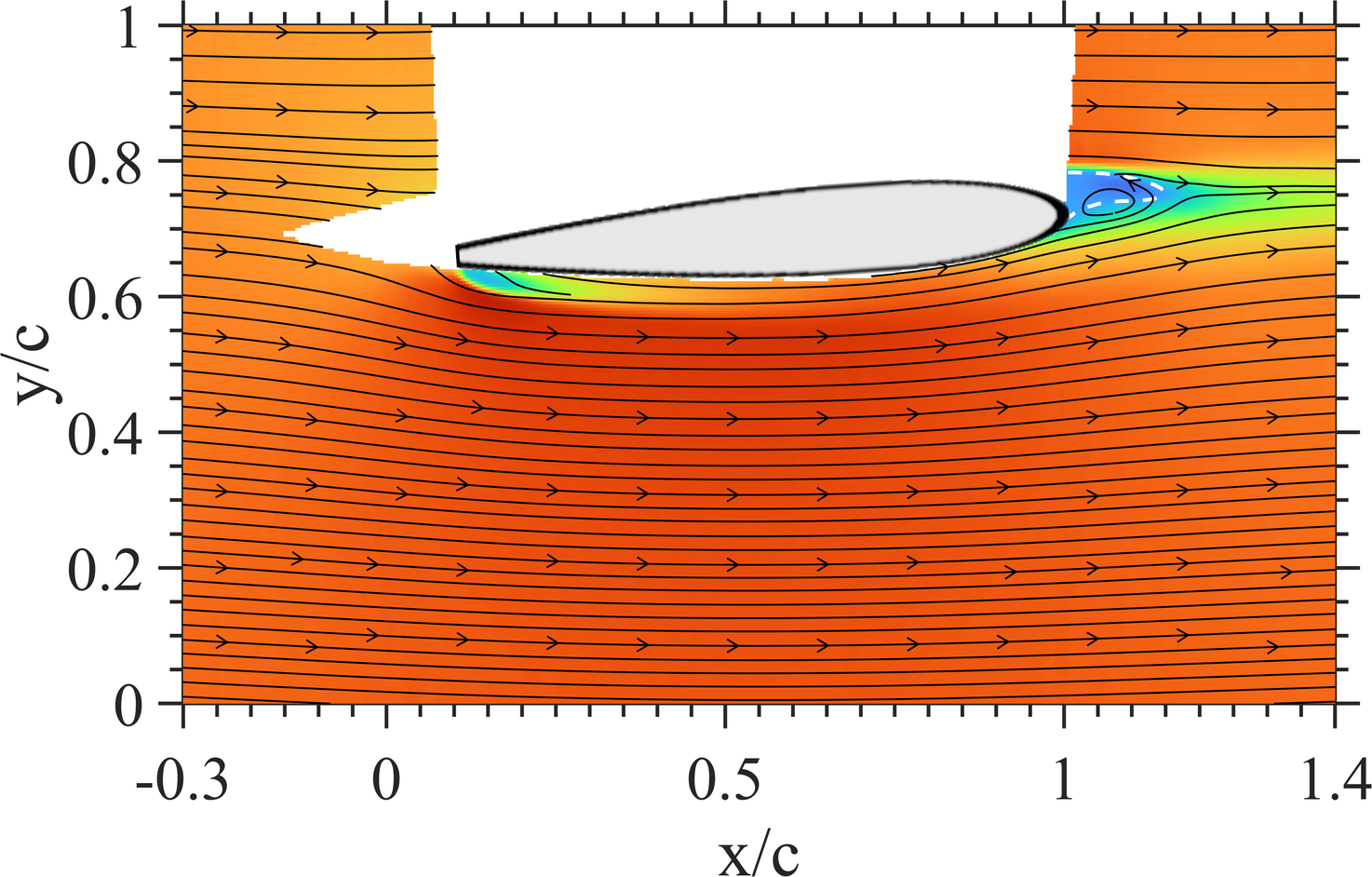}
    \caption{$\alpha$ = 184$^\circ$}
    \label{}
\end{subfigure}
% \vskip\baselineskip
\end{figure}
\begin{figure}[H]\ContinuedFloat
\begin{subfigure}[b]{0.351\textwidth}
    \centering 
    \includegraphics[trim=0 0 0 0,clip,width=\textwidth]{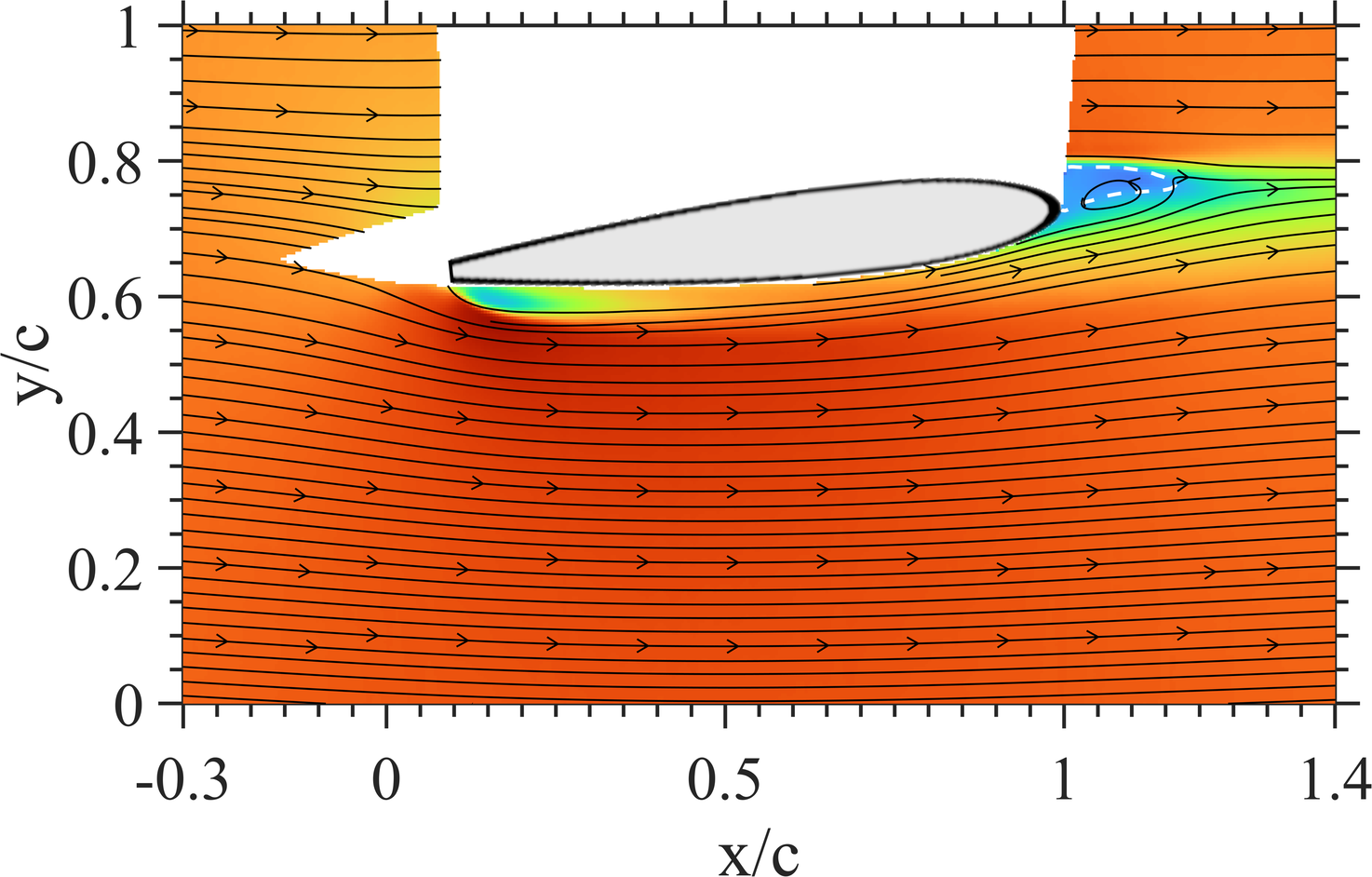}
    \caption{$\alpha$ = 186$^\circ$}
    \label{}
\end{subfigure}
\begin{subfigure}[b]{0.315\textwidth}
    \centering 
    \includegraphics[trim=33 0 0 0,clip,width=\textwidth]{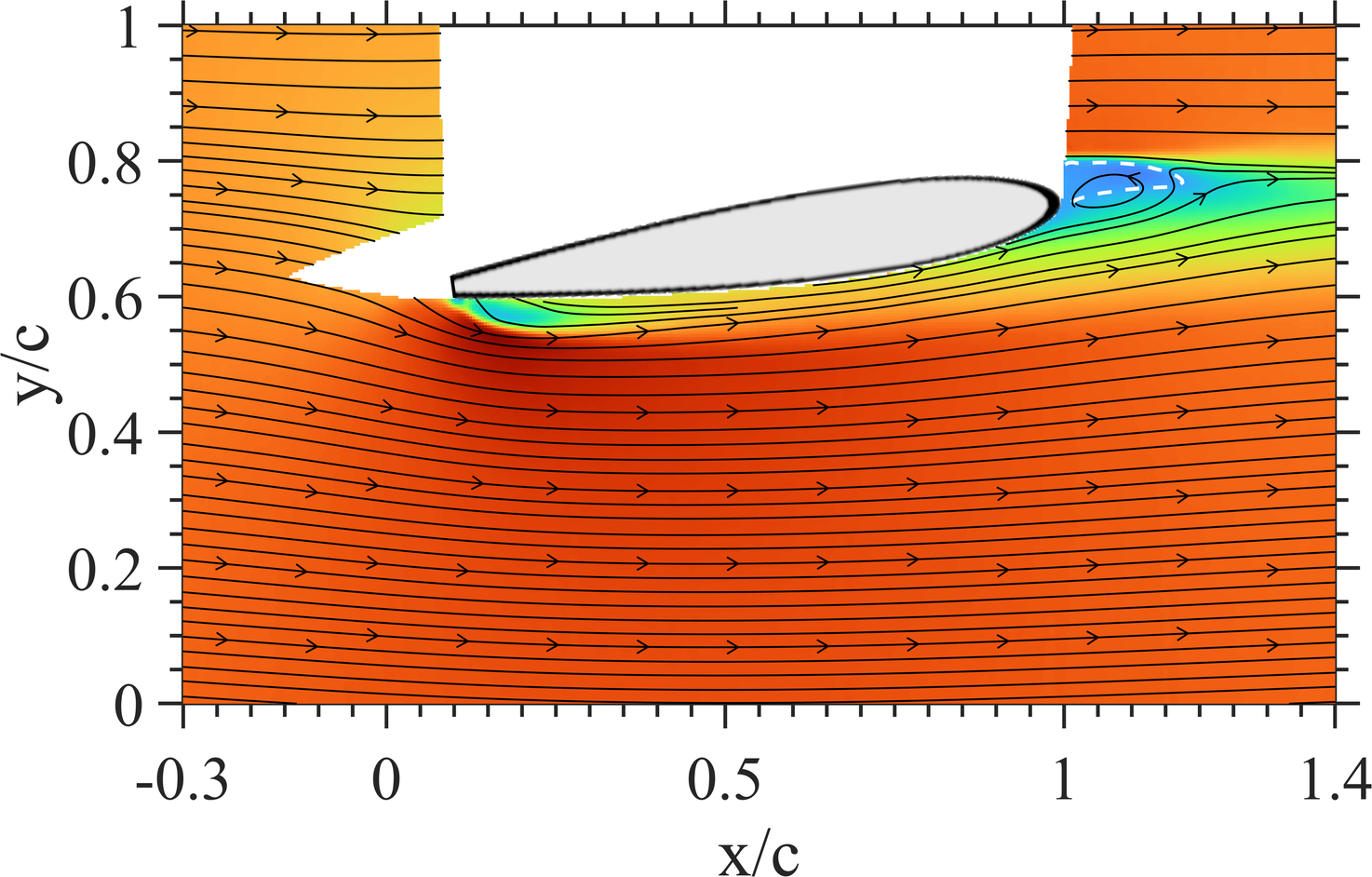}
    \caption{$\alpha$ = 188$^\circ$}
    \label{}
\end{subfigure}
\begin{subfigure}[b]{0.315\textwidth}
    \centering 
    \includegraphics[trim=33 0 0 0,clip,width=\textwidth]{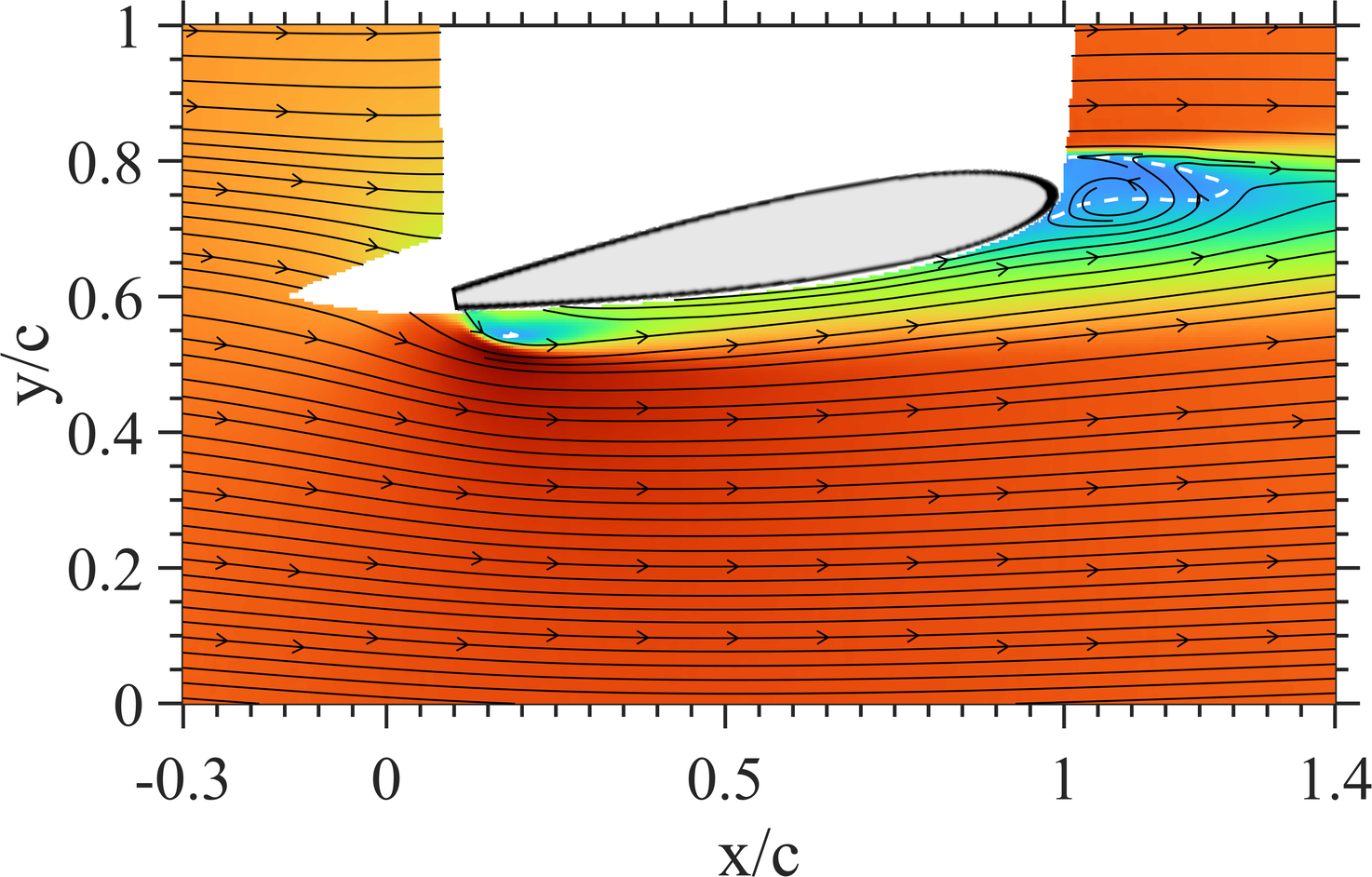}
    \caption{$\alpha$ = 190$^\circ$}
    \label{}
\end{subfigure}
\vskip\baselineskip
\hspace{7em}
\begin{subfigure}[b]{0.351\textwidth}
    \centering 
    \includegraphics[trim=0 0 0 0,clip,width=\textwidth]{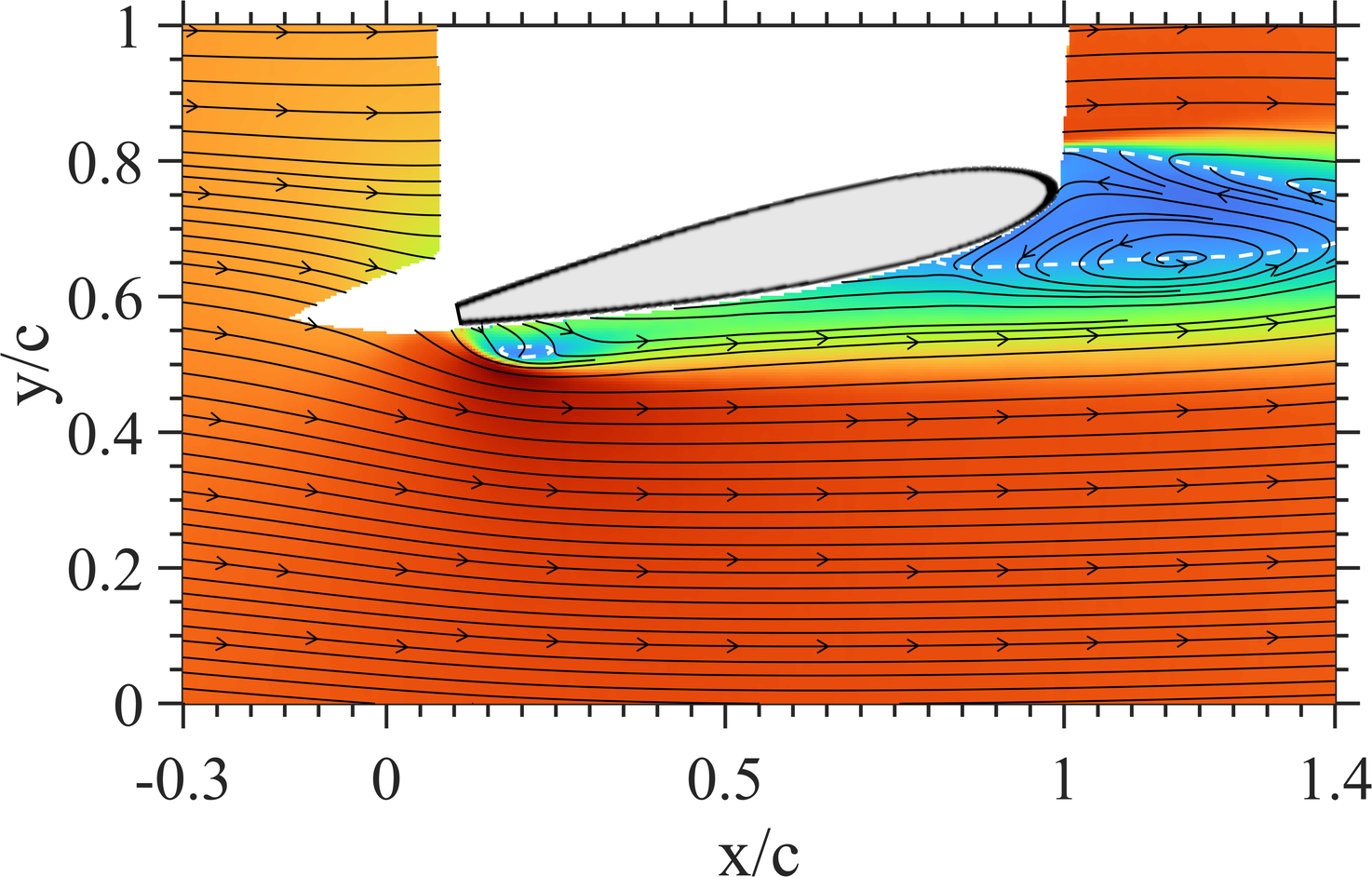}
    \caption{$\alpha$ = 192$^\circ$}
    \label{}
\end{subfigure}
\begin{subfigure}[b]{0.315\textwidth}
    \centering 
    \includegraphics[trim=33 0 0 0,clip,width=\textwidth]{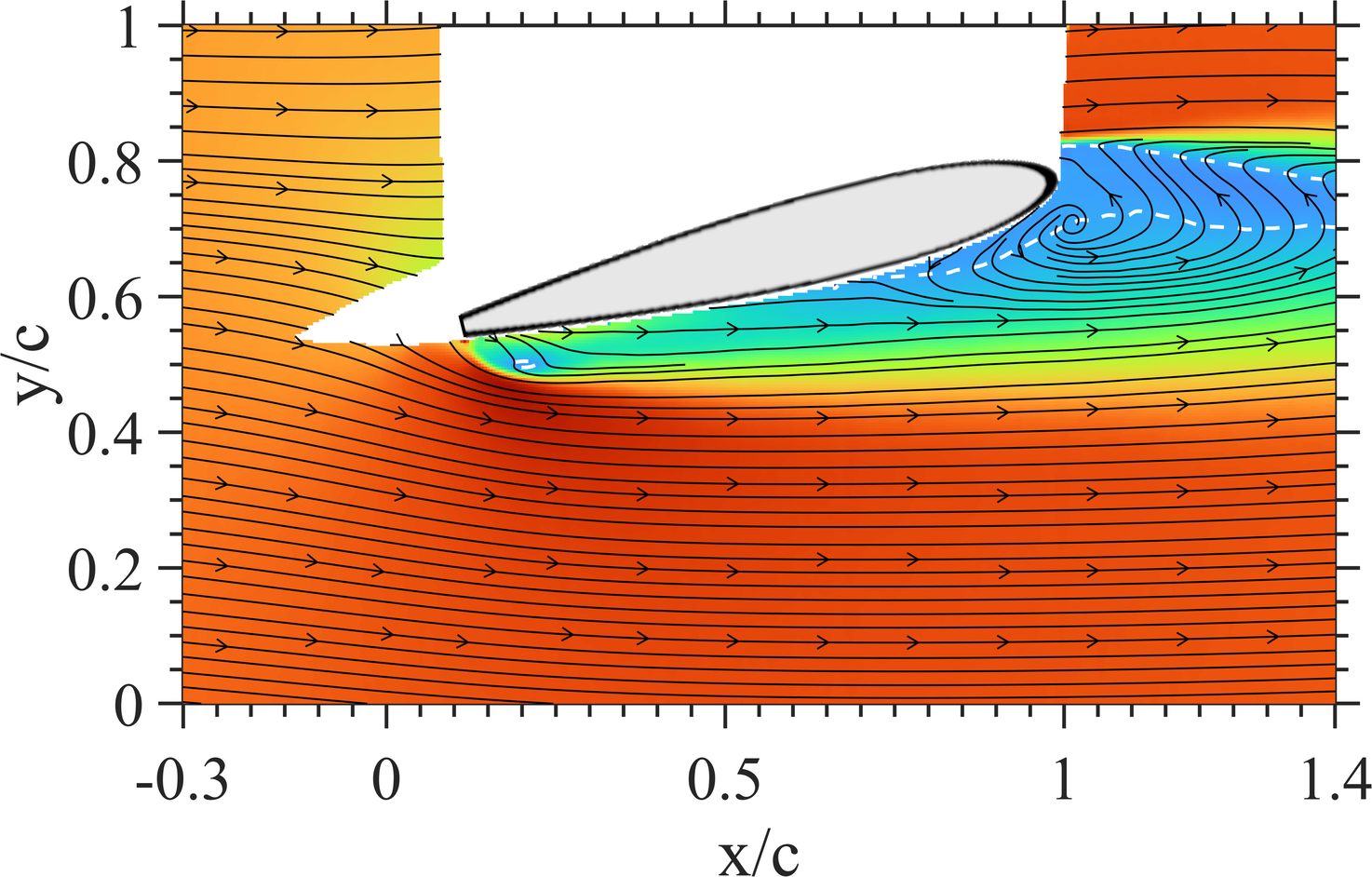}
    \caption{$\alpha$ = 194$^\circ$}
    \label{}
\end{subfigure}
\caption{Contour plots of the time-averaged streamwise velocity, normalized with freestream velocity for the $\mathbf{A10\lambda10}$ airfoil along the trough of a midspan sinusoid, at static pitch angles from 184$^\circ$ (a) to 194$^\circ$ (f), overlaid with in-plane streamlines. White dashed contour lines represent the locations where mean streamwise velocity is zero. Flow direction is from left to right.}
\label{A10P10_T}
\end{figure}
Subsequent acceleration of the thick shear layer as a surface-aligned flow suggests that the streamwise vortices re-energize the shear layer and realign the incoming flow with the airfoil surface. While the effect of actuation is clear, further examination of these vortical flowfields is required to understand the three-dimensional flow physics.

To conclude, the time-averaged streamwise velocity fields show that for the baseline airfoil a clear closed separation region is visible at $\alpha=186^\circ$, which gradually grows into an open separation zone by $\alpha=190^\circ$. In contrast, the flow remains largely attached over the surface for both modified airfoils over the three investigated planes, ultimately separating at the blunt trailing edge. The separation point gradually moves upstream, shifting to $\sim$0.6c over the sinusoidal crest and middle planes, and to $\sim$0.7c over the sinusoidal trough plane, for both airfoils at $\alpha=194^\circ$. Investigations over these three planes show that the largest variation occurs near the leading edge, where the geometry of the sinusoid is observed to play a major role. The flowfield over the airfoil surface and in the wake is three-dimensional and requires further analysis using SPIV.
%%%%%%%%%%%%%%%%%%%%%%%%%%%%%%%%%%%%%%%%%%%%%%%%%%%
\subsection{Static Pitch: Forces and Moments}
\label{Forces}
The section discusses the time-averaged aerodynamic forces and moments experienced by the three airfoils in both forward and reverse flow conditions. While PIV results showed the effect of sinusoidal trailing edges at selected planes, the load results show their overall effect on aerodynamic characteristics. 
\subsubsection{Reverse flow}    
\label{reverse_loads}
\begin{figure}[H]
    \centering
    \begin{subfigure}[b]{0.43\textwidth}
        \includegraphics[width=\textwidth]{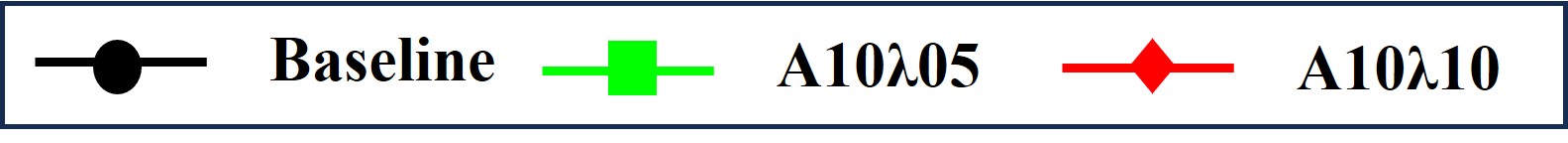}
    \end{subfigure}
    \vskip\baselineskip
    \begin{subfigure}[b]{0.325\textwidth}
        \includegraphics[width=\textwidth]{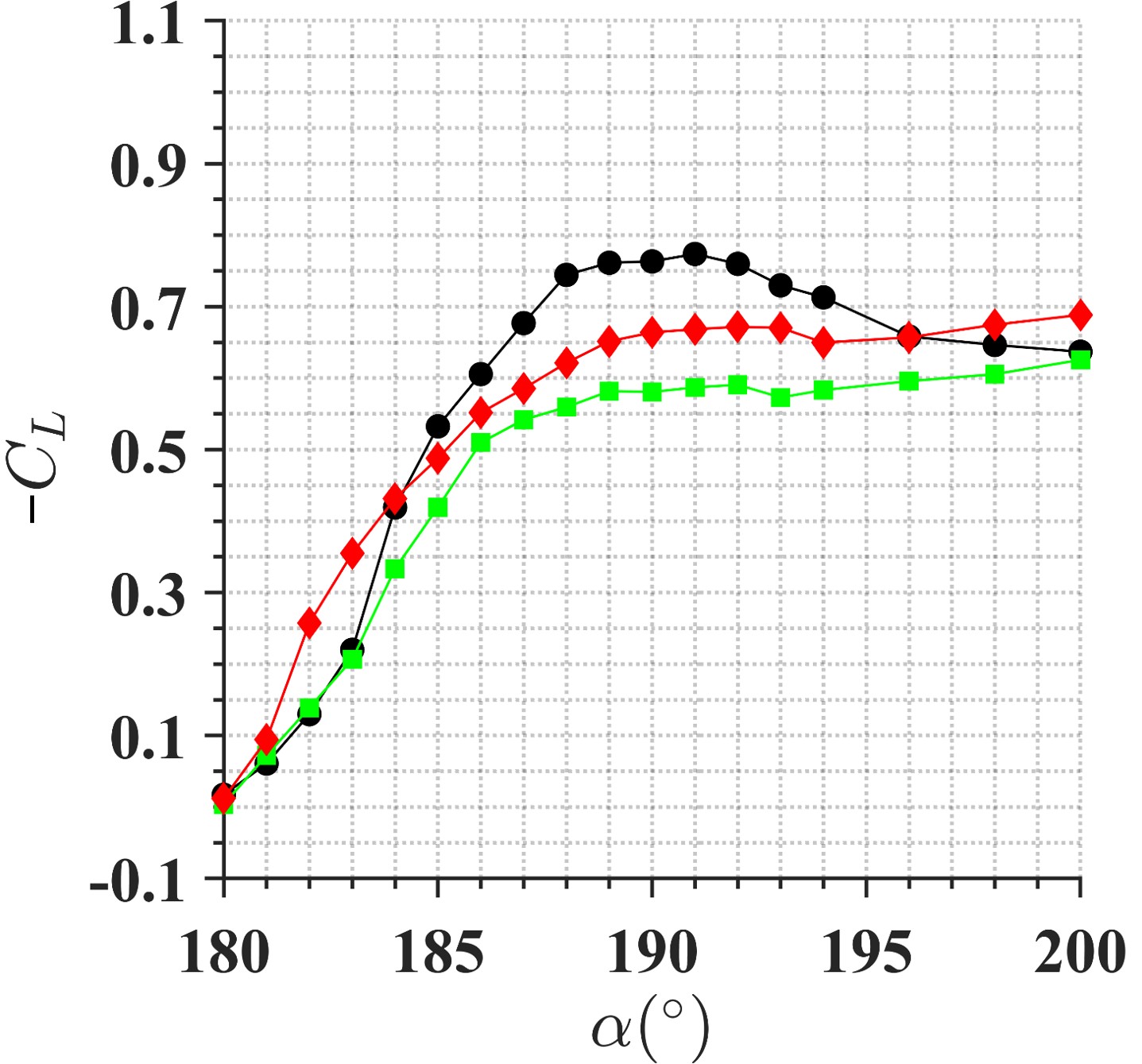}
        \caption{}
        \label{cl_v10}
    \end{subfigure}
    \begin{subfigure}[b]{0.325\textwidth}
        \includegraphics[width=\textwidth]{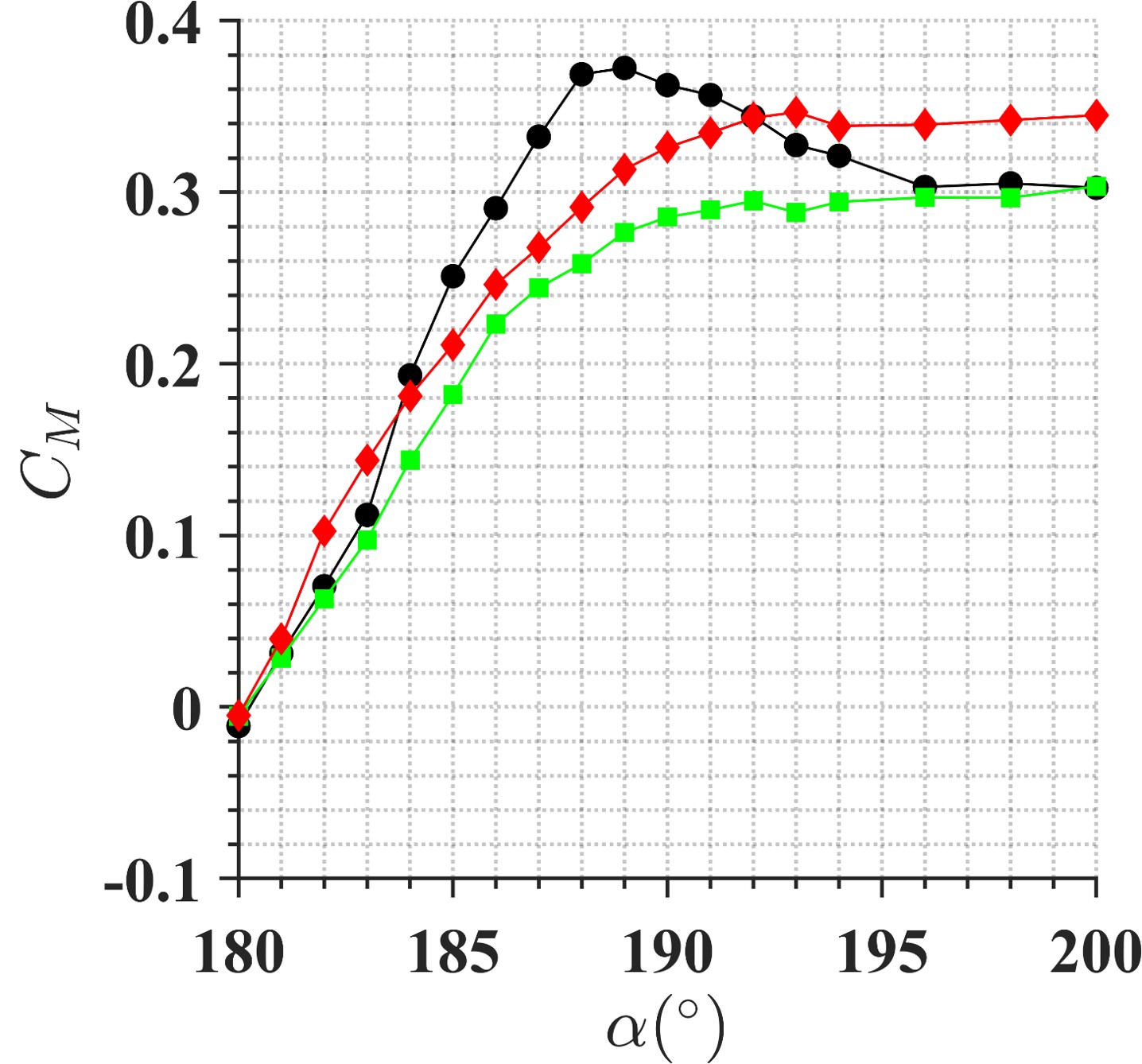}
        \caption{}
        \label{cm_v10}
    \end{subfigure}    
    \begin{subfigure}[b]{0.333\textwidth}
        \includegraphics[width=\textwidth]{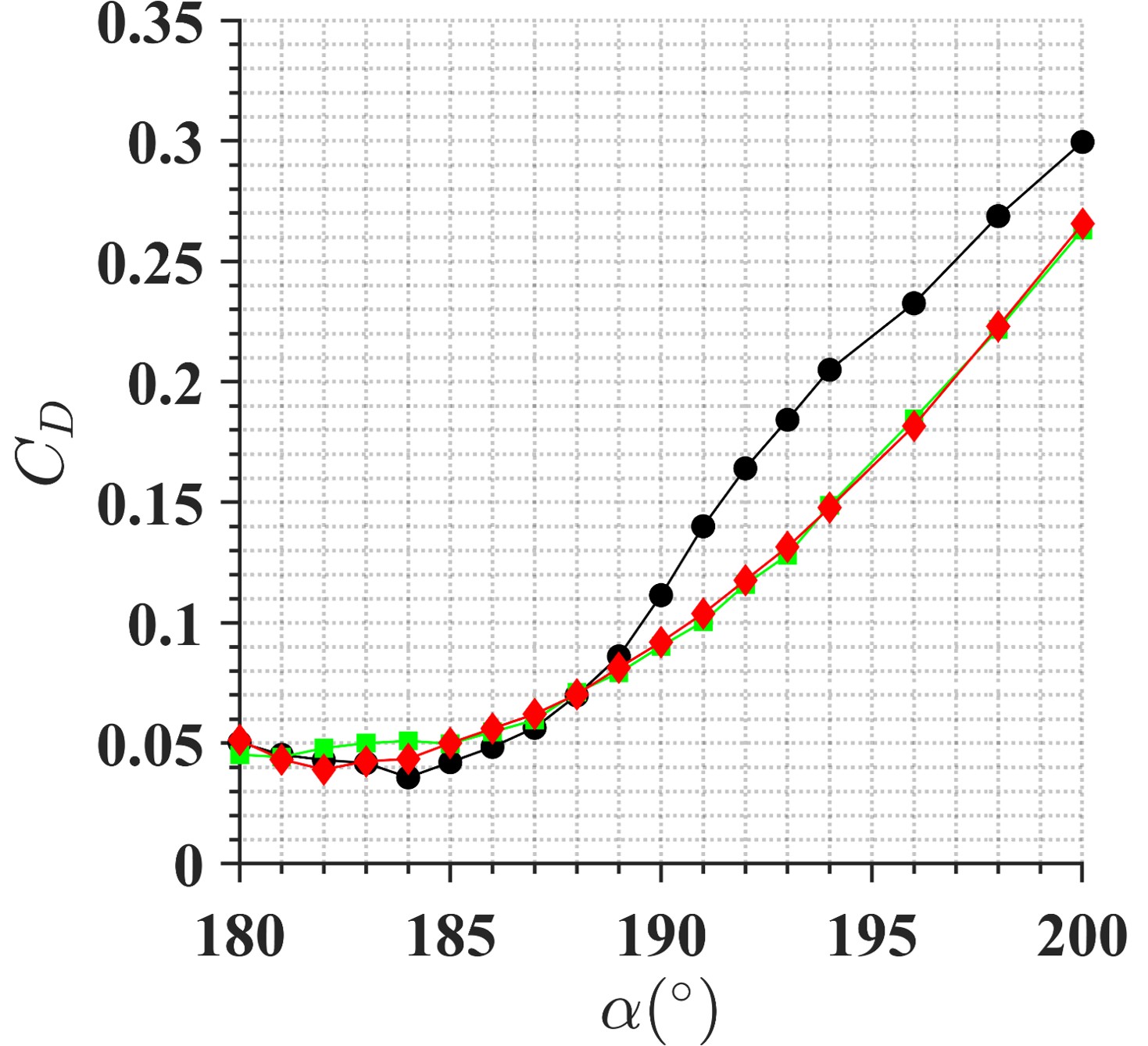}
        \caption{}
        \label{cd_v10}
    \end{subfigure}
    \vskip\baselineskip
    \begin{subfigure}[b]{0.325\textwidth}
        \includegraphics[width=\textwidth]{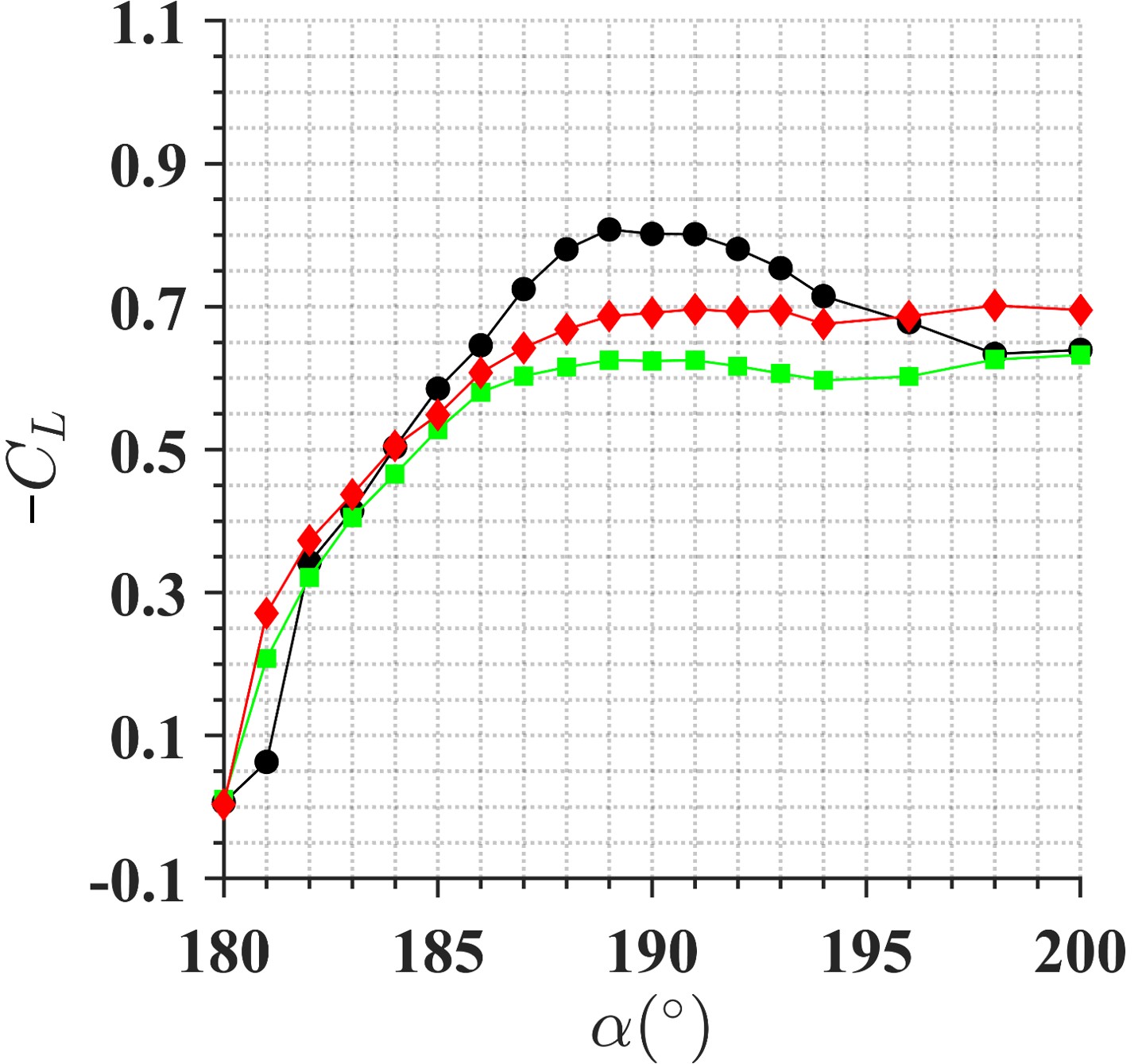}
        \caption{}
        \label{cl_v20}
    \end{subfigure}
    \begin{subfigure}[b]{0.325\textwidth}
        \includegraphics[width=\textwidth]{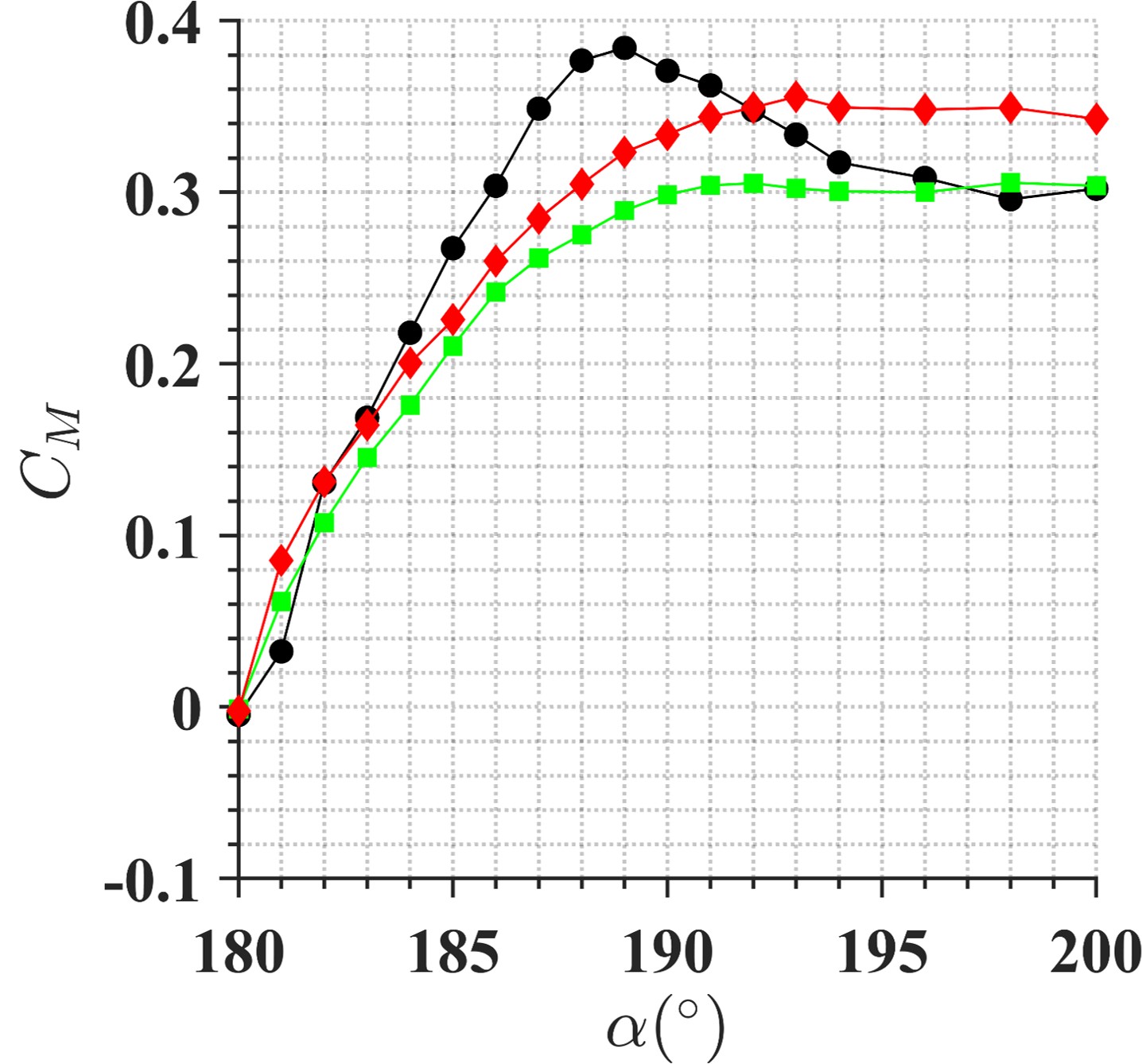}
        \caption{}
        \label{cm_v20}
    \end{subfigure}    
    \begin{subfigure}[b]{0.333\textwidth}
        \includegraphics[width=\textwidth]{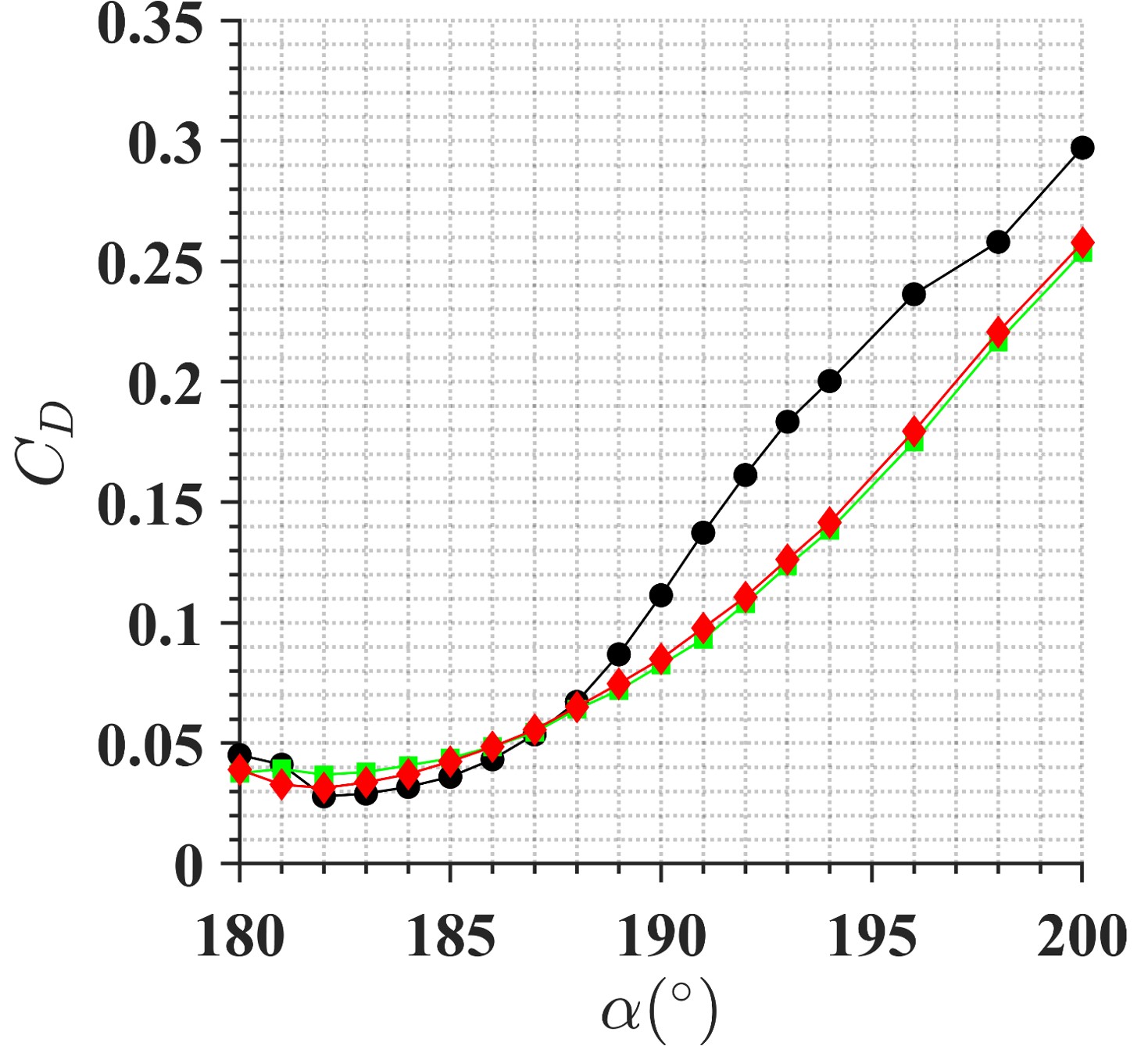}
        \caption{}
        \label{cd_v20}
    \end{subfigure}
    \caption{Coefficients of lift (a, d), pitching moment (b, e), and drag (c, f) for the baseline (black), $\mathbf{A10\lambda05}$ (green), and $\mathbf{A10\lambda10}$ (red) airfoils in reverse flow. Plots are shown for $\mathbf{Re_c = 0.7 \times 10^5}$ (a-c) and $\mathbf{1.4\times10^5}$ (d-f).}
    \label{loads_overall}
\end{figure}

The variations in lift, drag and pitching moment coefficients with angle of attack, for the three airfoils in reverse flow configuration, are shown in  Fig.\ref{loads_overall}. The aerodynamic forces and moments were measured at 0.7c in reverse flow and transferred to 0.75c, which corresponds to the geometric quarter-chord point and the aerodynamic center for a symmetric airfoil under forward flow conditions. This is also the location where the blade is expected to be attached to the hub. 
The coefficients for the three different blades are calculated using their respective planform area, which varies by a small magnitude due to the differences in trailing edge geometry.   
The plots are arranged row-wise, with the figures in the top row (\ref{cl_v10}-\ref{cd_v10}) corresponding to $Re_c = 0.7\times10^5$, and the figures in the bottom row (\ref{cl_v20}-\ref{cd_v20}) corresponding to $Re_c = 1.4\times10^5$.
 The investigated angle of attack ranges from $180^\circ\le \alpha \le 200^\circ$, corresponding to $0^\circ\le \alpha \le 20^\circ$ in reverse flow configuration. Measurements were taken at $1^\circ$ intervals up to $\alpha = 194^\circ$, and thereafter at $2^\circ$ intervals up to $\alpha = 200^\circ$. Observations corresponding to $\alpha \leq 194^\circ$ at $Re_c = 1.4\times10^5$ are compared with the PIV results discussed in the previous section.
 
The lift during reverse flow is directed towards the suction side (Fig.\ref{load_schm}b), and hence is an undesirable effect. Therefore, positive values of $-C_L$ correspond to negative lift on the suction side in Figs. \ref{loads_overall}a and \ref{loads_overall}d. At $\alpha = 180^\circ$, airflow is symmetric on both sides of the baseline airfoil, leading to $-C_L = 0$ at both Reynolds numbers.
The lift increases linearly till $\alpha = 183^\circ$ for $Re_c = 0.7\times10^5$ and till $\alpha = 181^\circ$ for $Re_c = 1.4\times10^5$. In this angle of attack range, the flow is attached on both sides of the airfoil. However, since the bottom side is the suction or low-pressure side, negative lift is generated in reverse flow, even in the absence of flow separation at the leading edge. 
Depending on the angle of attack, the suction from the recirculation region at the blunt trailing edge can also contribute to the negative lift. A sudden increase in the lift slope is observed from $\alpha = 183^\circ$ to $184^\circ$ at $Re_c = 0.7\times10^5$ and from $\alpha = 181^\circ$ to $182^\circ$ at $Re_c = 1.4\times10^5$. PIV results showed a region of accelerated flow near the sharp leading edge of the baseline airfoil at $\alpha = 182^\circ$ for $Re_c = 1.4\times10^5$, which was suggestive of the beginning of flow separation. Comparison of the load and PIV results suggests that the initiation of flow separation leads to a jump in the negative lift. It also suggests that the flow separation starts slightly later at $\alpha = 184^\circ$ for $Re_c = 0.7\times10^5$. The lift coefficient increases almost linearly from $\alpha = 184^\circ$ to $188^\circ$ at $Re_c = 0.7\times10^5$, and from $\alpha = 182^\circ$ to $188^\circ$ at $Re_c = 1.4\times10^5$. PIV results at $Re_c = 1.4\times10^5$ showed this as the range in which the leading-edge separation bubble grows while remaining closed. The low-pressure core of the separation vortex remains close to the surface for a closed separation bubble, generating the suction that produces negative lift. As the separation bubble grows in size and strength, the magnitude of negative lift also increases. The lift coefficient plateaus from $\alpha = 188^\circ$ to $191^\circ$ at both Reynolds numbers. PIV results show this as the range where the leading-edge separation bubble gradually opens and merges with the wake recirculation region. Thereafter, the lift coefficient decays linearly and starts to asymptote near $\alpha = 200^\circ$ at both Reynolds numbers. PIV results show this as the range where the low-pressure vortex core associated with the open-separation region convects away from the surface, leading to the corresponding decay in negative lift. The peak lift coefficients at $Re_c= 0.7\times10^5$ and $Re_c= 1.4\times10^5$ are $-C_{Lmax}\approx 0.75$ and $-C_{Lmax}\approx 0.8$, respectively. They occur in the plateau region where the merger of the two flow separation regions takes place. 

The negative lift generated by the modified airfoils is either higher or almost equal to the baseline, prior to the initiation of flow separation on the baseline airfoil. That is for $\alpha\leq183^\circ$ at $Re_c=0.7\times10^5$ and $\alpha\leq181^\circ$ at $Re_c=1.4\times10^5$, respectively. A probable reason can be that the blunt region of the sinusoidal geometry experiences higher pressure due to stagnation, a component of which contributes to negative lift. Since this blunt region of the $A10\lambda10$ airfoil is more exposed than the $A10\lambda05$ airfoil, higher negative lift is observed for the former. In the range of $184\leq\alpha\leq196^\circ$ at $Re_c=0.7\times10^5$ and $182\leq\alpha\leq196^\circ$ at $Re_c=1.4\times10^5$, both modified airfoils show either similar or lower negative lifts in comparison to the baseline. This is the range where either a closed or an open separation region exists close to the baseline airfoil's surface, which generates a large negative lift. Whereas mitigation of flow separation using the sinusoidal trailing edge leads to a significant reduction. The highest difference is observed in the lift-plateau region of the baseline airfoil. $A10\lambda05$ and $A10\lambda10$ airfoils produce approximate reductions of $20\%$ ($\Delta C_L\approx0.15$) and $13\%$ ($\Delta C_L\approx0.10$), respectively, at $Re_c=0.7\times10^5$, and $19\%$ ($\Delta C_L\approx0.15$) and $12\%$ ($\Delta C_L\approx0.10$), respectively, at $Re_c=1.4\times10^5$. 
PIV results at the sinusoidal crest plane showed that the $A10\lambda10$ airfoil has high near-surface velocities over a larger chordwise extent than the $A10\lambda05$ airfoil. PIV results at the sinusoidal middle and sinusoidal trough planes showed a recirculation region near the leading edge for the $A10\lambda10$ airfoil, which was absent for the $A10\lambda05$ airfoil. Lower pressures associated with higher velocities and suction from the separation-vortex core are probable reasons behind higher negative lift for the $A10\lambda10$ airfoil. Other contributing factors can be the variance in the characteristics of the vortex pairs produced by the sinusoids, and the variance in the suction generated by the wake recirculation region. At higher pitch angles, $\alpha\geq196^\circ$, the negative lift is higher than baseline for the $A10\lambda10$ airfoil, and lower or almost similar for the $A10\lambda05$ airfoil. In this range, the separation vortex core moves farther away from the surface for the baseline airfoil, resulting in a decay of negative lift. However, for both modified airfoils, the wake separation region grows larger and stays close to the surface, leading to higher or similar levels of suction and negative lift. An increase in the lift slope is observed for both modified airfoils at $\alpha=194^\circ$, for both Reynolds numbers. This is where PIV results showed a sudden increase in the size of the wake for both modified airfoils. Moreover, for $\alpha\geq194^\circ$ and at both Reynolds numbers, the $A10\lambda10$ airfoil exhibits higher lift than the $A10\lambda05$ airfoil, with a consistent difference of $\Delta C_L\approx0.05$. The contributing factors are the same as those discussed previously, especially the flow recirculation at the middle and trough regions of the $A10\lambda10$ leading edge, which occurs at higher angles. 

The pitching moment coefficients for the baseline and modified airfoils, at the two Reynolds numbers, are compared in Figs.\ref{cm_v10} and \ref{cm_v20}, respectively. Variations in pitching moment show very similar characteristics to the variations in lift for all three airfoils, showing that the moment from lift is the major contributor. 
A nose-up pitching moment, $+C_M$, is generated during reverse flow. For the baseline airfoil, the pitching moment increases linearly from zero till flow separation initiates at the leading edge. The suction from the flow separation bubble leads to a sudden increase in the moment from $\alpha = 183^\circ$ to $184^\circ$ at $Re_c = 0.7\times10^5$ and from $\alpha = 181^\circ$ to $182^\circ$ at $Re_c = 1.4\times10^5$. Thereafter, the moment increases linearly and peaks at $\alpha=189^\circ$,  with a peak $C_M$ value of approximately $0.38$ at both Reynolds numbers. This is the range where the separation bubble grows but remains closed, and its suction primarily governs the negative lift and pitching moment. The moment reduces as the open separation region develops and moves farther away from the airfoil surface. The plateau region observed for the negative lift plots is absent here. A possible reason for this variation is that the suction generated by the separation bubble and the wake recirculation region both contribute to the negative lift. However, the moment also depends on the distribution of the suction with respect to the geometric quarter-chord. Therefore, while the suction from the separation bubble produces a nose-up pitching moment, the suction from the wake recirculation produces a nose-down pitching moment. The peak observed in the curve possibly arises from the competing effects of these two opposing factors. Another important point to note is that the pitching moment values in reverse flow are significantly higher than in forward flow, although the lift values are comparable. The forward flow results are discussed in the following section. The reason is that the center of pressure during reverse flow is located toward the aerodynamic leading edge \cite{DeannaPassive, Revflowexp, Cambermorphcomputation}, whereas the moment is reported about the aerodynamic three-quarter-chord location, where the blade is generally attached to the hub. The baseline center of pressure is primarily governed by the low-pressure vortex core associated with flow separation. 
The high nose-up pitching moment, together with the pitching moment impulse when flow separation starts, are two of the major drawbacks of reverse flow. Therefore, the portion of the blade undergoing reverse flow experiences a higher moment than the rest of the blade. It also undergoes cycles of very high and low moments, as it cycles between forward and reverse flow during rotor revolution. Both factors can lead to higher torsional loads on the blade and the rotor-hub. High torsional loads together with hysteresis from revolution and cyclic pitching can contribute to structural failure. 

For the modified airfoils, the pitching moments are slightly higher or equal to the baseline, prior to initiation of flow separation on the baseline airfoil. The higher moments come from higher negative lift, which can be due to higher pressures near the trough, as discussed previously. However, the change is gradual, due to the mitigation of leading-edge flow separation. Thus, the sinusoidal trailing edges remove the pitching moment impulse, which is one of the primary problems associated with reverse flow. In the range of $184\leq\alpha<192^\circ$ at $Re_c=0.7\times10^5$ and $182\leq\alpha<192^\circ$ at $Re_c=1.4\times10^5$, both modified airfoils generate either similar or lower pitching moments in comparison to the baseline. The highest reduction is observed at $\alpha = 189^\circ$, where the baseline moment is approximately equal to the peak value of 0.38. The $A10\lambda05$ and $A10\lambda10$ airfoils show reductions of $30\%$ ($\Delta C_M\approx0.11$) and $22\%$ ($\Delta C_M\approx0.08$), respectively, at $Re_c=0.7\times10^5$, and $26\%$ ($\Delta C_M\approx0.10$) and $21\%$ ($\Delta C_M\approx0.08$), respectively, at $Re_c=1.4\times10^5$. 
The nose-up pitching moment for the modified airfoils comes from the lift generated due to attached flow over the suction side. However, the large reduction comes from the mitigation of flow separation and the associated suction from its low-pressure vortex core. At higher pitch angles, $\alpha>192^\circ$, the pitching moment is higher than baseline for the $A10\lambda10$ airfoil, and lower or almost similar for the $A10\lambda05$ airfoil. This suggests that the attached flow over the modified airfoils continues to generate a high moment, while it reduces for the baseline airfoil due to downstream convection of the recirculation region. PIV results showed that the point of flow separation starts to move upstream for both modified airfoils, starting from $\alpha=192^\circ$. Suction from the larger wake recirculation region will lead to an increase in nose-down pitching moment. The asymptotic behavior of the moment curve, observed for both airfoils for $\alpha \geq 192^\circ$, may arise from the competing effects of wake recirculation and attached flow. Additionally, at high pitch angles for the $A10\lambda10$ airfoil, the small recirculation regions developing near the blunt regions of the sinusoids will also contribute towards the nose-up pitching moment. Overall, across the entire range of angle of attack, the $A10\lambda05$ airfoil produces a lower pitching moment than the $A10\lambda10$ airfoil. The supporting reasons are the same as those used to justify the comparatively lower negative lift. 

%$$$$$$$$$$$$$$$$$$$$-------DRAG-------$$$$$$$$
The drag coefficients for the baseline and modified airfoils are shown in Figs.\ref{cd_v10} and \ref{cd_v20}. Stagnation pressure at the aerodynamic leading edge over the pressure side, suction from flow separation at the leading edge on the suction side, and suction from flow separation at the blunt trailing edge are the primary contributors to drag during reverse flow on the baseline airfoil  \cite{Cambermorphcomputation}. The baseline airfoil has a finite drag ($C_D\approx 0.05$) at $\alpha=180^\circ$ at both Reynolds numbers. This is due to skin-friction drag and pressure drag from the bluff-body separation at the trailing edge. The pressure drag makes the total drag considerably higher in comparison to $\alpha=0^\circ$ in forward flow. A small decrease in drag is observed till the beginning of flow separation, which is at $\alpha=184^\circ$ at $Re_c=0.7\times10^5$ and at $\alpha=182^\circ$ at $Re_c=1.4\times10^5$. The corresponding PIV results at $Re_c=1.4\times10^5$ showed a small reduction in wake width in this range. The drag increases beyond this point for the rest of the range of angle of attack tested. The increase occurs at a comparatively lower rate till $\alpha\approx188^\circ$. This is the region where a closed separation bubble exists on the airfoil, and a component of suction contributes towards drag due to the angle of attack. Drag increases at a much higher rate beyond these angles, due to the large pressure drag from the formation and growth of the open separation region.

For the modified airfoils, similar to the baseline airfoil, drag comes from stagnation pressure at the leading edge and suction from flow separation at the trailing edge. Suction from leading-edge flow separation is absent, but suction from local separation near the blunt portions of the sinusoid at high angles of attack will be present. Suction is also generated from the attached flow over the suction side. The component of these suction forces that contributes to drag will vary with angle of attack. Streamwise vortex pairs that develop from the sinusoids can also lead to additional skin friction drag \cite{gad2003flow}. For $\alpha\leq188^\circ$, the modified airfoils exhibit drag values that are nearly identical to or slightly higher than those of the baseline airfoil. In this range, the size of the wake separation region is similar for all three airfoils. However, a reduction can probably be achieved by using a sharp sinusoidal trailing edge geometry. For $\alpha>188^\circ$, drag increases for both modified airfoils, but it is significantly lower than the baseline. For $193^\circ\leq\alpha\leq196^\circ$, the reduction is highest with $\Delta C_D \approx 0.05$, corresponding to a percentage reduction of approximately 22\%-28\% for both airfoils with respect to the baseline. In this range, drag grows for the modified airfoils due to the growth of the wake separation region, among other aforementioned factors. However, it is much smaller than the open separation over the baseline airfoil, leading to the large difference. Though within uncertainty limits, a trend of slightly higher drag is noticed for the $A10\lambda10$ airfoil in comparison to the $A10\lambda05$ airfoil. Contributing factors can be higher regions of accelerated flow at the crest and local recirculation regions at the middle and trough of the sinusoid for the $A10\lambda10$ airfoil.

%$$$-----Summary Loads
In summary,  sinusoidal trailing edge modification is effective in producing a considerable reduction in negative lift, pitching moment, and drag at both Reynolds numbers. It completely removes the problem of pitching moment impulse. The sinusoid with a lower wavelength, $A10\lambda05$, produces a comparatively larger reduction in negative lift and pitching moment. The reduction is primarily observed beyond the angle of attack where the flow separation starts on the baseline airfoil. The highest reductions observed in negative lift and nose-up pitching moment are 20\% and 30\%, respectively. A reduction in drag is observed for $\alpha>188^\circ$. The reduction is similar for both geometries, with the highest reduction nearly 28\% of baseline. Small variations are noticed in the characteristics of all three airfoils with the variation of Reynolds number. 
%%%%%%%%%%%%%%%%%%%%%%%%%%%%%%%%%%%%%%%%%%%%%%%

\subsubsection{Forward flow}
\begin{figure}[H]
    \centering
    \begin{subfigure}[b]{0.43\textwidth}
        \includegraphics[width=\textwidth]{Images/loads_overall/legend.jpg}
    \end{subfigure}
    \vskip\baselineskip
    \begin{subfigure}[b]{0.313\textwidth}
        \includegraphics[width=\textwidth]{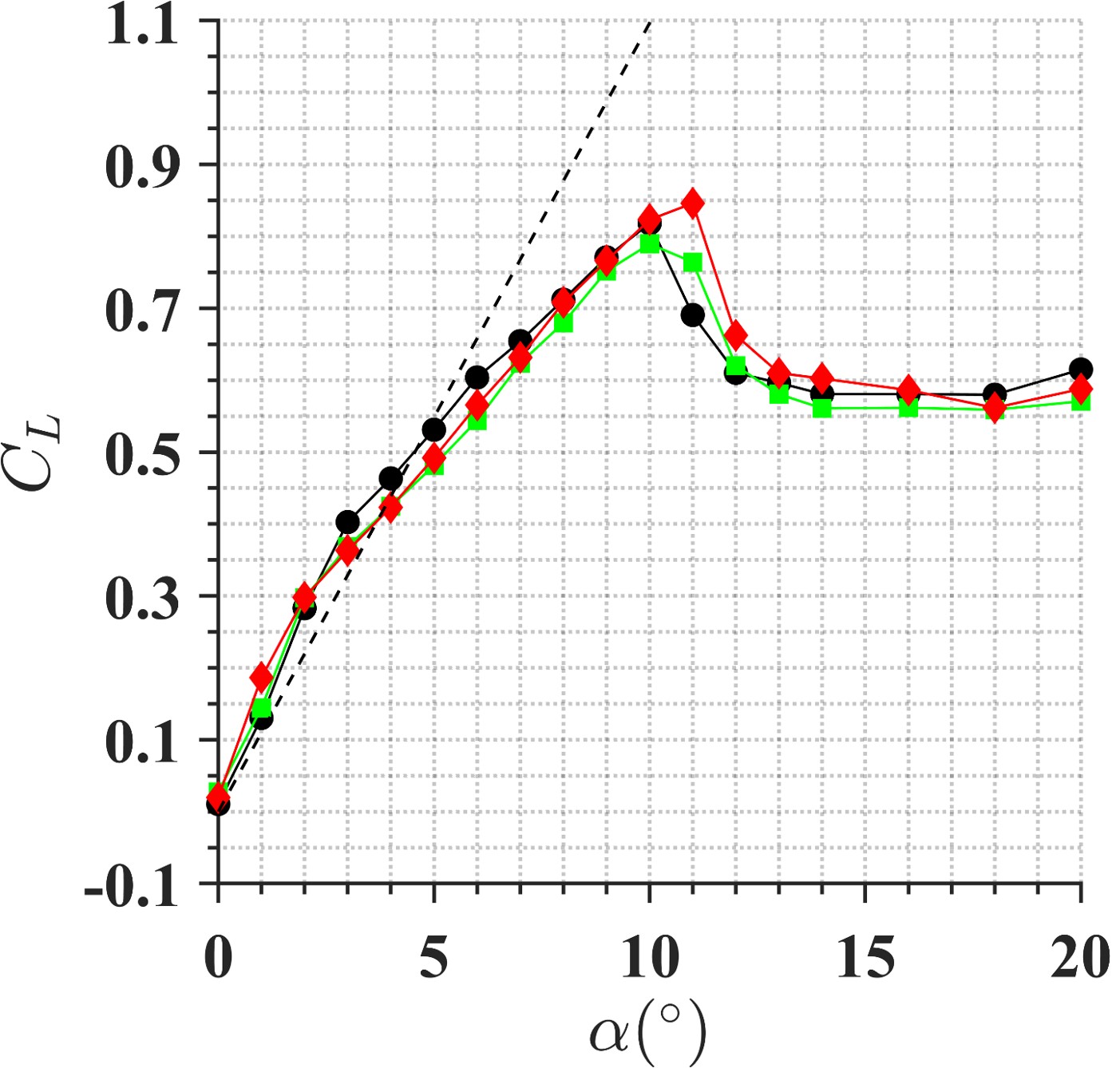}
        \caption{}
        \label{cl_f_v10}
    \end{subfigure}
    \begin{subfigure}[b]{0.313\textwidth}
        \includegraphics[width=\textwidth]{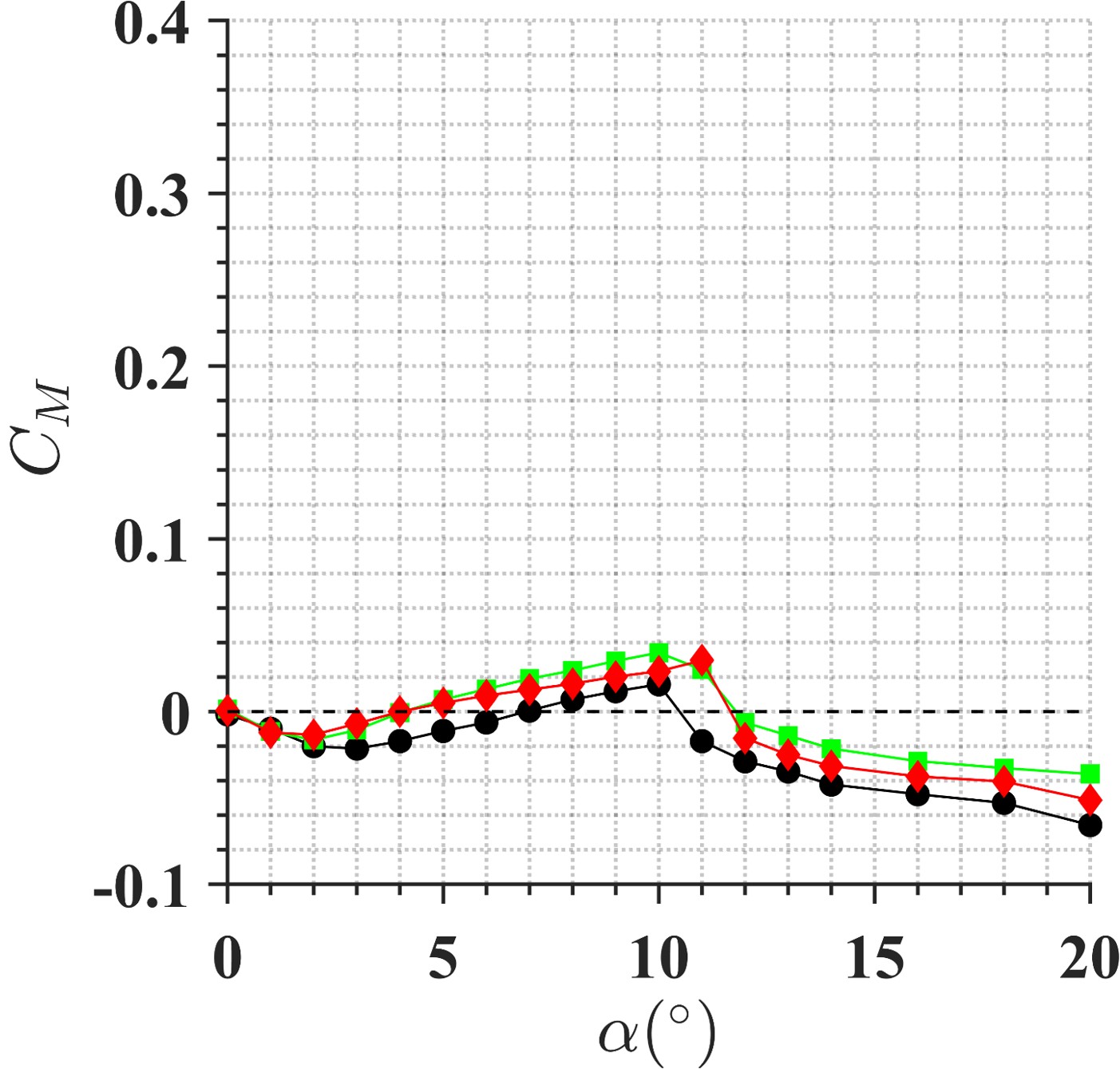}
        \caption{}
        \label{cm_f_v10}
    \end{subfigure}
%    \vskip\baselineskip
    \begin{subfigure}[b]{0.33\textwidth}
        \includegraphics[width=\textwidth]{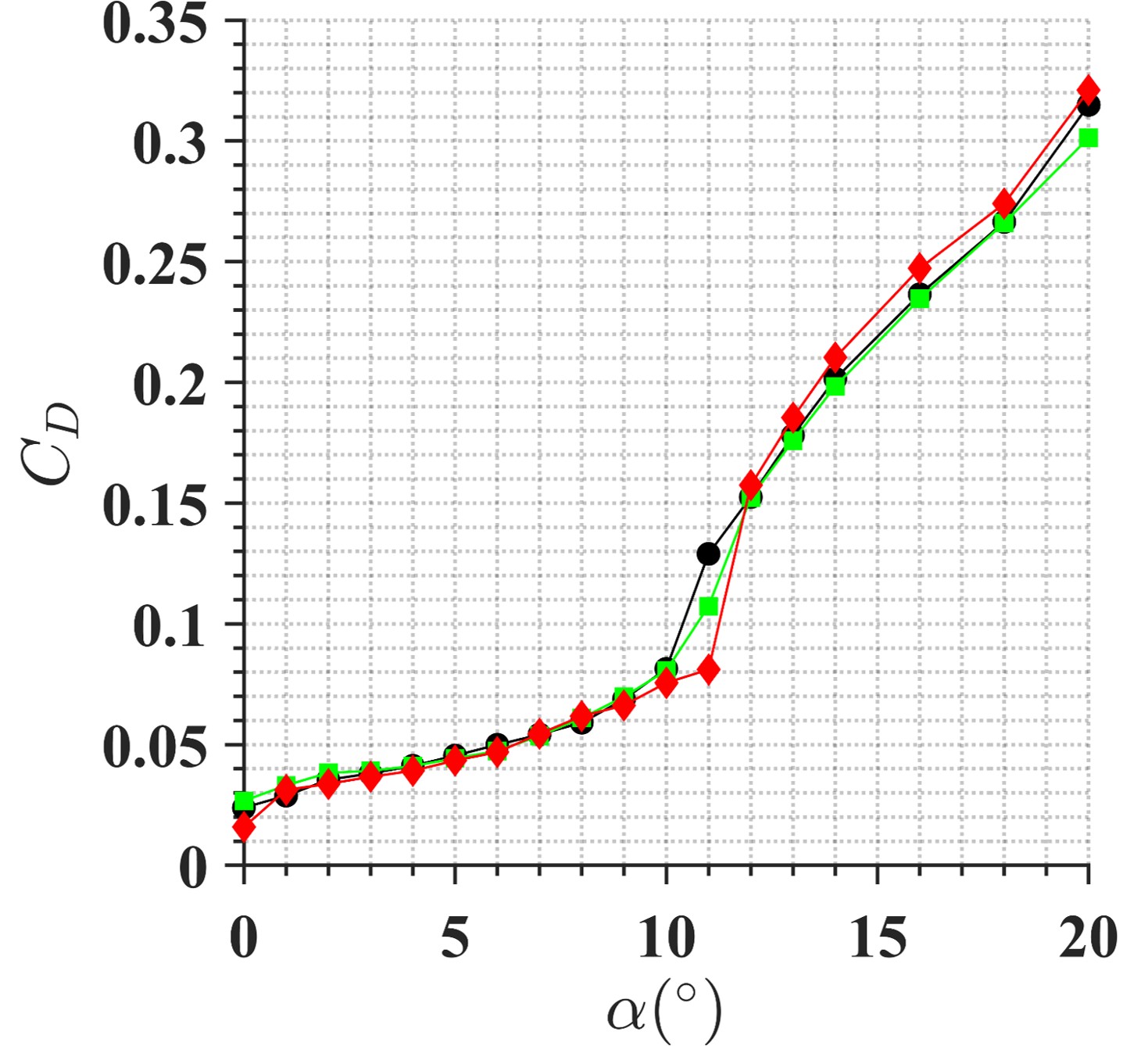}
        \caption{}
        \label{cd_f_v10}
    \end{subfigure}
%    \vskip\baselineskip
    \begin{subfigure}[b]{0.313\textwidth}
        \includegraphics[width=\textwidth]{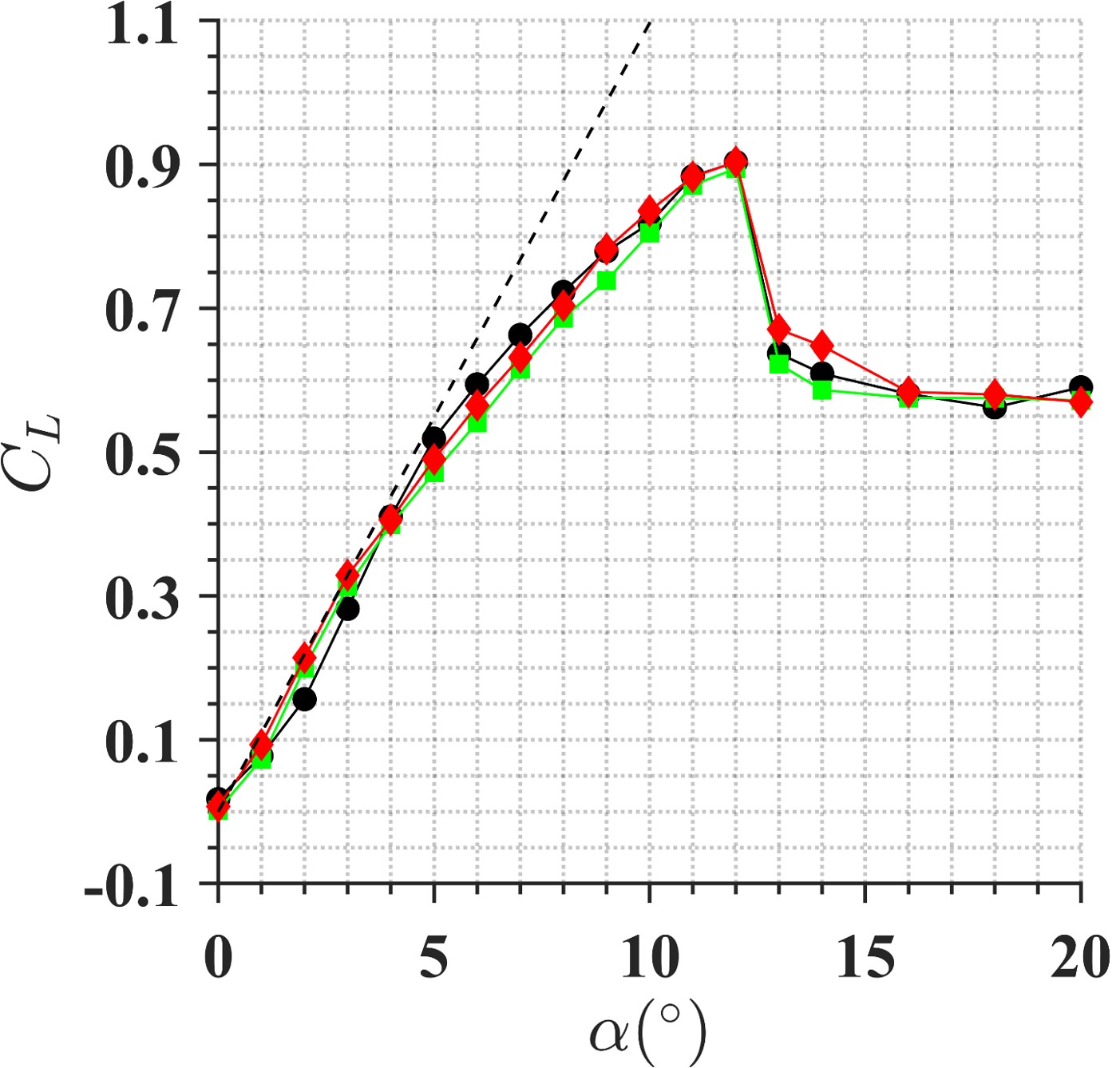}
        \caption{}
        \label{cl_f_v20}
    \end{subfigure}
    \begin{subfigure}[b]{0.313\textwidth}
        \includegraphics[width=\textwidth]{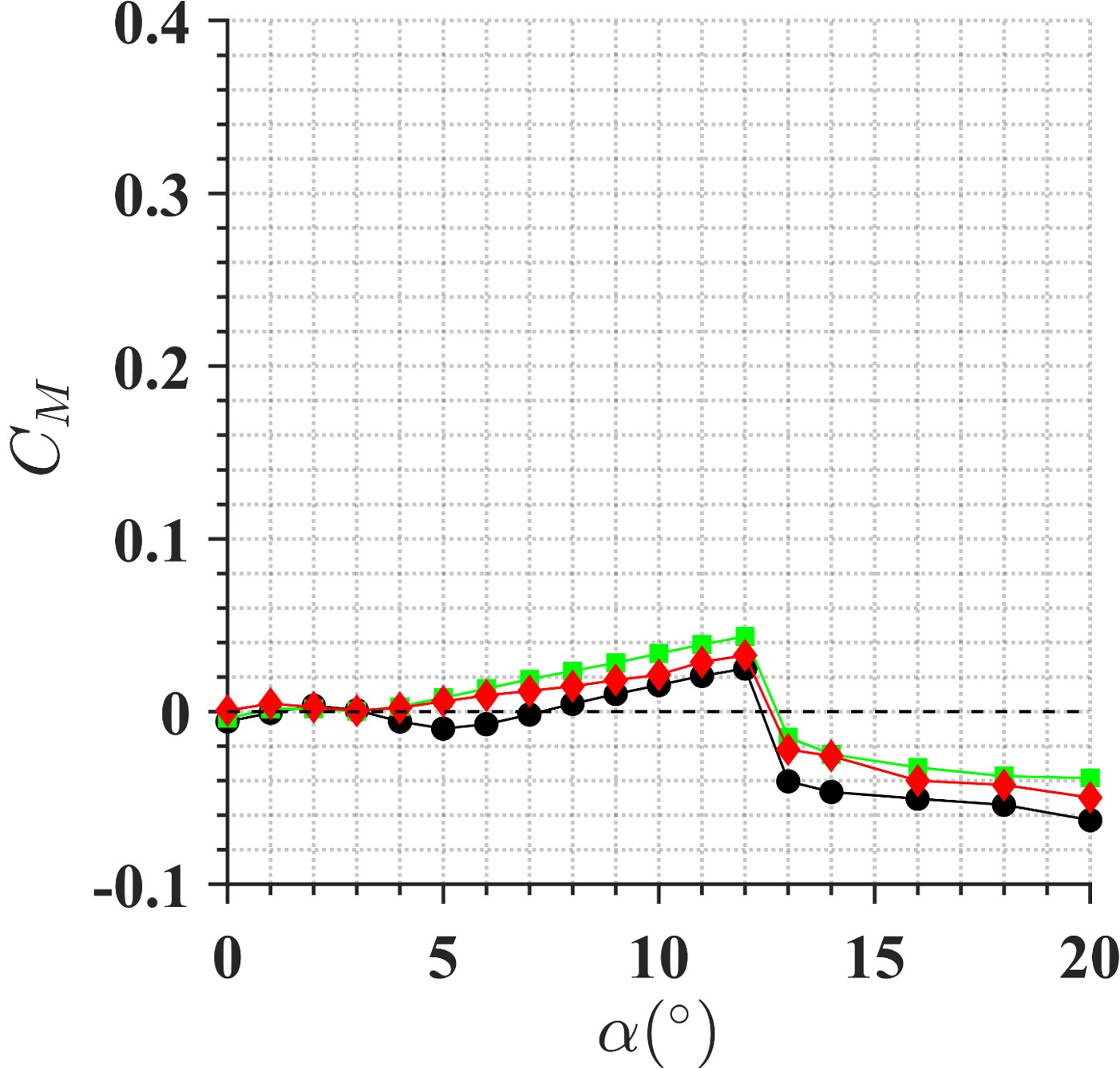}
        \caption{}
        \label{cm_f_v20}
    \end{subfigure}    
    \begin{subfigure}[b]{0.33\textwidth}
        \includegraphics[width=\textwidth]{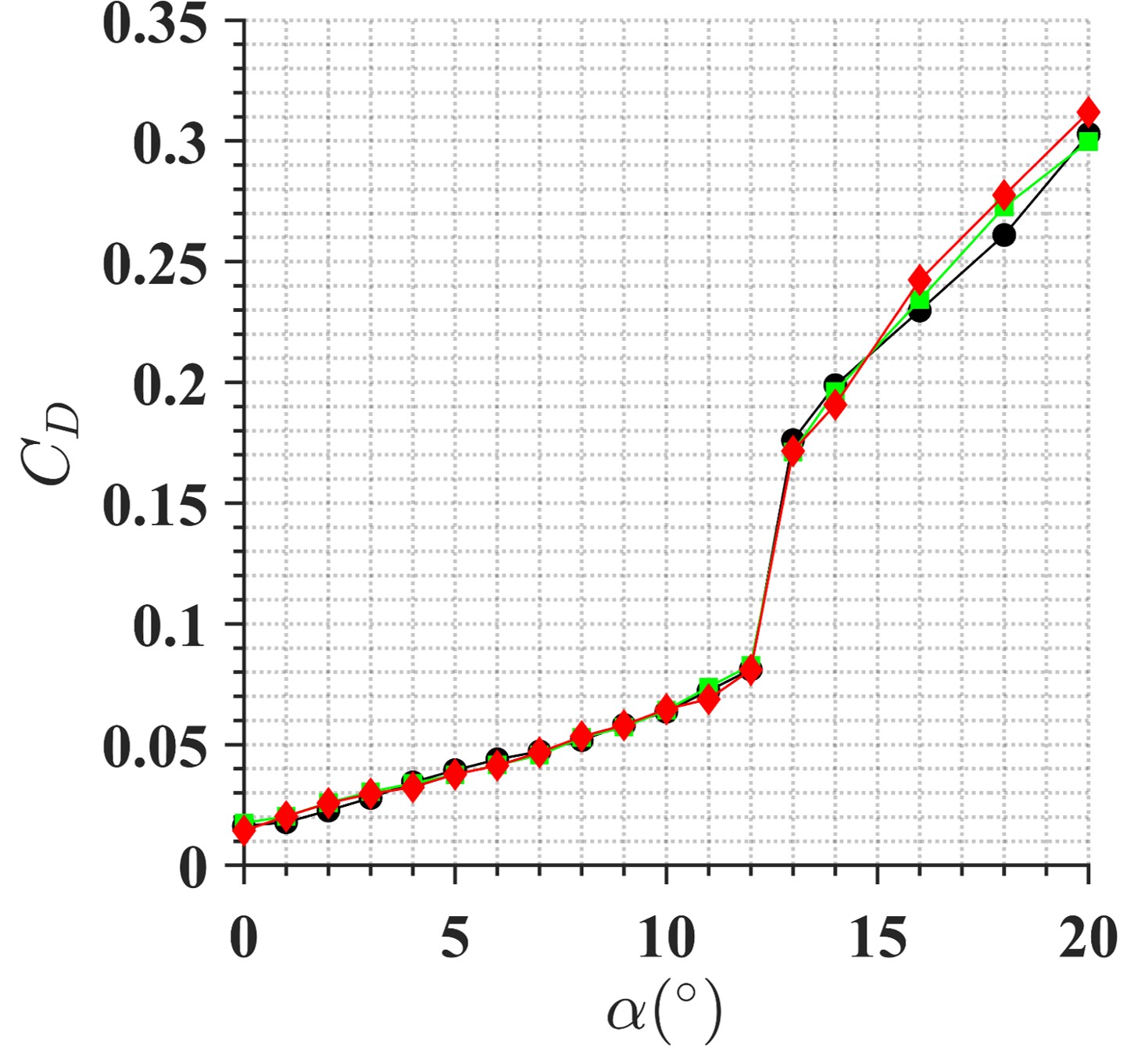}
        \caption{}
        \label{cd_f_v20}
    \end{subfigure}
    \caption{Coefficients of lift (a, d), pitching moment (b, e), and drag (c, f) for the baseline (black), $\mathbf{A10\lambda05}$ (green), and $\mathbf{A10\lambda10}$ (red) airfoils in forward flow. A dashed line with a slope of $\mathbf{2\pi}$ is overlaid on (a) and (d). A dashed line representing $C_M=0$ is overlaid on (b) and (e). Plots are shown for $\mathbf{Re_c = 0.7 \times 10^5}$ (a-c) and $\mathbf{1.4\times10^5}$ (d-f).}
    \label{loads_f_overall}
\end{figure}

Figure \ref{loads_f_overall} shows the variations in the lift, drag, and pitching moment coefficients for the baseline, $A10\lambda05$, and $A10\lambda10$ airfoils under forward flow conditions. The aim of conducting the study was to address the modifications or penalties to forward flow characteristics, as a rotor blade transitions through both forward and reverse flow conditions over its entire revolution. 
The airfoil orientation for these measurements is as shown in Fig. \ref{load_schm}c. The coefficients for the three airfoils are calculated using their respective planform areas. They are reported at the geometric quarter-chord location, 0.25c. The nose-up pitching moment is positive and clockwise in direction.   
 The figure shares several similar characteristics to Fig.\ref{loads_overall}. The sub-figures in the top (\ref{cl_f_v10}-\ref{cm_f_v10}) and bottom (\ref{cl_f_v20}-\ref{cm_f_v20}) rows correspond to $Re_c = 0.7\times10^5$ and $1.4\times10^5$, respectively. The range of angle of attack tested is $\alpha=0^\circ$ to $20^\circ$. Data were acquired for every degree till $\alpha=14^\circ$ and after that every two degrees till $\alpha=20^\circ$. A black dotted line with a slope of $2\pi$, representing the theoretical slope of the lift curve for a thin airfoil in forward flow, is overlaid on the lift curves (Figs.\ref {cl_f_v10} and \ref{cl_f_v20}).

%--------forward lift----------
The lift curve of the baseline airfoil shows good agreement with the thin airfoil solution at low angles of attack, especially at the higher Reynolds number. Deviations can be attributed to viscous effects, wall effects, and the finite thickness of the NACA 0015 airfoil. A noticeable decrease in the slope of the lift curve is observed at $\alpha = 3^\circ$ and $5^\circ$ for the lower and higher Reynolds numbers, respectively. 
The peak lift coefficient occurs at $\alpha = 10^\circ$ for $Re_c = 0.7\times10^5$ and at $\alpha = 12^\circ$ for $Re_c = 1.4\times10^5$, followed by an abrupt drop in magnitude, indicating stall. 
The overall characteristics of the modified airfoils are similar to the baseline airfoil at both Reynolds numbers. At $Re_c=1.4\times10^5$, they attain the same peak lift coefficient and at the same angle of attack as the baseline. However, at $Re_c = 0.7\times10^5$, the $A10\lambda10$ airfoil exhibits a delayed stall, with stall occurring at $\alpha = 11^\circ$, consequently resulting in a higher maximum lift coefficient. At $Re_c = 0.7\times10^5$, for $3^\circ<\alpha<10^\circ$, the modified airfoils produce marginally lower lift than the baseline airfoil. At $Re_c = 1.4\times10^5$, the modified airfoils produce marginally higher lift for $\alpha<4^\circ$, and marginally lower lift for $4^\circ<\alpha\leq 10^\circ$, in comparison to the baseline. Maximum reductions are observed to be approximately 10\% and 5\% for the $A10\lambda05$ and $A10\lambda10$ airfoils, respectively.
%--------forward-pitching moment----
The pitching moment coefficients for all three airfoils are compared in Figs.\ref{cm_f_v10} and \ref{cm_f_v20}. 
The pitching moment coefficient for the baseline airfoil is nose-up and therefore positive, and its magnitude increases almost linearly till stall. After stall, the moment becomes nose-down and therefore negative, and its magnitude continues to increase with angle of attack. Both modified airfoils show similar variation with pitch angle, as the baseline, at both Reynolds numbers. However, the magnitudes are observed to be marginally higher before stall, and marginally lower after stall, especially for the $A10\lambda05$ airfoil. Overall, the pitching moment coefficients of all three airfoils are relatively small, as expected, and are significantly lower than those observed under reverse flow conditions.

%----forward drag------
Figures \ref{cd_f_v10} and \ref{cd_f_v20} compare the variation in drag coefficient with angle of attack for the three airfoils. For all three airfoils, the zero-lift drag, at $\alpha=0^\circ$, is considerably lower than in reverse flow conditions. A similar observation was also made by Lind et al.\cite{Timeaveraged}. Drag for all three airfoils shows an almost linear increase up to stall. A jump in magnitude is observed at stall, followed by a linear increase in the post-stall region. In the pre-stall regime, the modified airfoils have drag values similar to the baseline. In the post-stall regime, marginally higher values are observed for the $A10\lambda10$ airfoil, but similar values are observed for the $A10\lambda05$ airfoil at both Reynolds numbers. In summary, these observations suggest that the penalties in forward flow due to the sinusoidal trailing edge are minimal. However, to understand the exact influence of the sinusoids in forward flow conditions, further pressure and flowfield measurements are required. 
\vspace{3em}
%%%%%%%%%%%%%%%%%%%%%%%%%%%%%%%%%%%%%%%
\section{Conclusions}
Effects of sinusoidal trailing edges on control of reverse flow-induced flow separation were investigated through wind tunnel tests on NACA 0015 airfoil sections at chord-based Reynolds numbers of $0.7\times10^5$ and $1.4\times10^5$. Three NACA 0015 airfoils were tested: a baseline airfoil with a sharp trailing edge and two modified airfoils with sinusoidal trailing edges. The two sinusoidal geometries, $A10\lambda05$ and $A10\lambda10$, both had a spatial amplitude of 10\% of the chord, but spatial wavelengths of 5\% and 10\% of the chord, respectively. Experiments consisted of the measurement of aerodynamic forces and moments using a load cell and analysis of the flowfield on the suction side of the airfoil using planar PIV.

PIV was conducted at the midspan location for the baseline airfoil, and at three planes corresponding to the crest, middle, and trough locations of a midspan sinusoid for the modified airfoils. The range of angle of attack investigated was $\alpha=180^\circ$ to $194^\circ$ in reverse flow conditions, at $Re_c = 1.4\times10^5$.  The time-averaged velocity fields suggest that a closed separation bubble develops near the leading edge of the baseline airfoil as early as $\alpha=182^\circ$. The bubble grows in size with angle of attack, leading to fully separated flow by $\alpha=190^\circ$. In contrast, for both modified airfoils, the flow remains almost completely attached and separates only near the trailing edge till $\alpha=190^\circ$. Depending on the airfoil and interrogation plane, the point of separation moves upstream to 80\%-90\% of the chord length at $\alpha=192^\circ$, and to 60\%-70\% of the chord length at $\alpha=194^\circ$. Overall, the PIV analysis shows that sinusoidal trailing edges are highly effective in mitigating flow separation during reverse flow. Interrogation at the different sinusoidal planes shows that the flowfield is highly three-dimensional, especially near the leading edge and in the wake. The results suggest that the trailing edge sinusoids function as leading edge vortex generators during reverse flow. The downwash from these vortices helps in aligning the flow with the airfoil surface, which in turn mitigates leading edge flow separation and enhances overall flow attachment.

Load cell measurements were conducted from $\alpha=180^\circ$ to $200^\circ$ in reverse flow conditions at both Reynolds numbers. The results showed that the baseline airfoil generates high lift, drag, and nose-up pitching moment coefficients during reverse flow. The primary factor is the suction from the low-pressure vortex core of the separated flow. Therefore, as soon as flow separation starts at the leading edge, a sudden jump is noticed in the lift and moment values, leading to an impulse. Thereafter, the values increase in the range where the separation bubble grows but remains closed, and decrease after the formation of open separation. Both modified airfoils show a large decrease in both negative lift and pitching moment, especially in the angle of attack range where a closed separation bubble exists over the baseline airfoil. They also show a gradual increase in negative lift and moment, thereby eliminating the problem of pitching moment impulse. The sinusoid with the lower wavelength, $A10\lambda05$, shows a larger reduction and also over a larger range of angle of attack, in comparison to $A10\lambda10$. The highest percentage reductions in negative lift and pitching moment are 20\% and 30\%, respectively. However, both modified airfoils are ineffective in the range prior to flow separation over the baseline airfoil. The drag of the baseline airfoil increases at a slower rate during closed separation and at a faster rate during open separation. It is in the second region where both modified airfoils produce significantly lower drag. Both airfoils show similar performance with a maximum reduction of approximately 28\%. The characteristics of all three airfoils show minor variations between the two Reynolds numbers. 
In forward flow conditions, load measurements were conducted from $\alpha=0^\circ$ to $20^\circ$, at both Reynolds numbers. The modified airfoils show similar lift, drag, and pitching moment characteristics as the baseline airfoil, especially at $Re_c = 1.4\times10^5$. At $Re_c = 0.7\times10^5$, the $A10\lambda10$ airfoil shows a delayed stall and a higher maximum lift coefficient. However, in general, the modified airfoils produce marginally lower lift and a higher pitching moment, in comparison to the baseline.  

To conclude, the flowfield and load measurements show that the passive sinusoidal trailing edges are effective in mitigating the adverse effects of reverse flow under static pitch conditions, while minimally affecting forward flow performance. The sinusoidal trailing edge geometry with a lower wavelength, $A10\lambda05$, is found to be more effective in reverse flow while maintaining comparable forward flow characteristics.  

\section{Acknowledgments}
This work was funded by the U.S. Asian Office of Aerospace Research and Development (AOARD), under award no. FA2386-24-1-4020, monitored by Lt Col Jeffrey M. Newcamp (2024-2025) and Dr. James Joo (2025-2026). The authors would like to express their gratitude to Dr. Kamal Poddar at the Indian Institute of Technology Kanpur for providing access to the low-speed wind tunnel facility. 
%\bibliography{bib_thesis}
\bibliography{bib_thesis}

@inproceedings{sikorsky,
  title={Flight dynamics and control modeling with system identification validation of the Sikorsky X2 technology demonstrator},
  author={Fegely, Cody and Juhasz, Ondrej and Xin, Hong and Tischler, Mark B},
  booktitle={American Helicopter Society 72nd Annual Forum, West Palm Beach, FL},
  year={2016}
}

@article{Optimumdesign,
  title={Optimum design of a compound helicopter},
  author={Yeo, Hyeonsoo and Johnson, Wayne},
  journal={Journal of Aircraft},
  volume={46},
  number={4},
  pages={1210--1221},
  year={2009}
}

@article{problemsofdesign,
author = {Tanner, Watson H. and Bergquist, Russel R.},
title = {Some problems of design and operation of a 250- knot compound helicopter rotor},
journal = {Journal of Aircraft},
volume = {1},
number = {5},
pages = {252-259},
year = {1964},
}

@article{slowedrotor,
  title={Experimental investigation and fundamental understanding of a full-scale slowed rotor at high advance ratios},
  author={Datta, Anubhav and Yeo, Hyeonsoo and Norman, Thomas R},
  journal={Journal of the American Helicopter Society},
  volume={58},
  number={2},
  pages={1--17},
  year={2013},
  publisher={Vertical Flight Society}
}

@article{vorticallift,
  title={Vortical lift on retreating rotor blades at high advance ratios: The aerodynamics of the sharp edge vortex},
  author={Hiremath, Nandeesh and Shukla, Dhwanil and Komerath, Narayanan},
  journal={Experiments in Fluids},
  volume={60},
  pages={1--17},
  year={2019},
  publisher={Springer}
}

@book{bramwell,
  title={Bramwell's helicopter dynamics},
  author={Bramwell, Anthony Robert Southey and Balmford, David and Done, George},
  year={2001},
  publisher={Elsevier}
}

@book{principles,
  title={Principles of helicopter aerodynamics with CD extra},
  author={Leishman, Gordon J},
  year={2006},
  publisher={Cambridge university press}
}

@article{corkethomas,
   author = "Corke, Thomas C. and Thomas, Flint O.",
   title = "Dynamic Stall in Pitching Airfoils: Aerodynamic Damping and Compressibility Effects", 
   journal= "Annual Review of Fluid Mechanics",
   year = "2015",
   volume = "47",
   number = "Volume 47, 2015",
   pages = "479-505",
   publisher = "Annual Reviews",
   issn = "1545-4479",
   type = "Journal Article",
  }

@techreport{TorsionalNASA,
  title={Analysis of helicopter rotor blade torsional oscillations due to stall},
  author={Crimi, Peter},
  year={1975},
  institution={NASA}
}

@article{Timeaveraged,
author = {Lind, Andrew H. and Lefebvre, Jonathan N. and Jones, Anya R.},
title = {Time-Averaged Aerodynamics of Sharp and Blunt Trailing-Edge Static Airfoils in Reverse Flow},
journal = {AIAA Journal},
volume = {52},
number = {12},
pages = {2751-2764},
year = {2014},
}

@article{Revflowexp,
  title={Flowfield measurements of reverse flow on a high advance ratio rotor},
  author={Lind, Andrew H and Trollinger, Lauren N and Manar, Field H and Chopra, Inderjit and Jones, Anya R},
  journal={Experiments in Fluids},
  volume={59},
  pages={1--15},
  year={2018},
  publisher={Springer}
}

@article{Reeffects,
author = {Lind, Andrew H. and Smith, Luke R. and Milluzzo, Joseph I. and Jones, Anya R.},
title = {Reynolds Number Effects on Rotor Blade Sections in Reverse Flow},
journal = {Journal of Aircraft},
volume = {53},
number = {5},
pages = {1248-1260},
year = {2016},
doi = {},

URL = { 
    
        
    
    

},
eprint = { 
    
      
    
    

}
}

@article{DeannaPassive,
author = {Ko, Deanna and Guha, Tufan Kumar and Amitay, Michael},
title = {Control of Reverse Flow over a Cantilevered Blade Using Passive Camber Morphing},
journal = {AIAA Journal},
volume = {59},
number = {12},
pages = {5310-5331},
year = {2021},
}

@article{Cambermorphcomputation,
author = {Jacobellis, George and Gandhi, Farhan and Rice, Thomas and Amitay, Michael},
year = {2019},
month = {01},
pages = {},
title = {Computational and Experimental Investigation of Camber-Morphing Airfoils for Reverse Flow Drag Reduction on High-Speed Rotorcraft},
volume = {65},
journal = {Journal of the American Helicopter Society},
}

@inbook{ThomasRF,
author = {Thomas T. Rice and Deanna Ko and Michael Amitay},
title = {Control of Reversed Flow in Static and Dynamic Conditions Using Camber Morphing Airfoils},
booktitle = {AIAA Aviation 2019 Forum},
chapter = {},
pages = {},
year={2020},
month={June},
}

@inproceedings{X2D,
  title={Aerodynamic design of the X2 technology demonstrator™ main rotor blade},
  author={Bagai, Ashish},
  booktitle={Annual forum proceedings-American Helicopter Society},
  volume={64},
  number={1},
  pages={29},
  year={2008},
  organization={American Helicopter Society, INC}
}

@article{Plasmaactuator,
author = {Clifford, Chris and Singhal, Achal and Samimy, Mo},
title = {Flow Control over an Airfoil in Fully Reversed Condition Using Plasma Actuators},
journal = {AIAA Journal},
volume = {54},
number = {1},
pages = {141-149},
year = {2016},
}

@article{LEPHansen, 
    title={Evolution of the streamwise vortices generated between leading edge tubercles},
    volume={788},
    journal={Journal of Fluid Mechanics}, 
    author={Hansen, Kristy L. and Rostamzadeh, Nikan and Kelso, Richard M. and Dally, Bassam B.}, 
    year={2016}, 
    pages={730–766}
}

@article{Whaletubercles,
    author = {Miklosovic, D. S. and Murray, M. M. and Howle, L. E. and Fish, F. E.},
    title = "{Leading-edge tubercles delay stall on humpback whale (Megaptera novaeangliae) flippers}",
    journal = {Physics of Fluids},
    volume = {16},
    number = {5},
    pages = {L39-L42},
    year = {2004},
    month = {05},
}

@article{LEPairfoil,
author = {Johari, H. and Henoch, C. and Custodio, D. and Levshin, A.},
title = {Effects of Leading-Edge Protuberances on Airfoil Performance},
journal = {AIAA Journal},
volume = {45},
number = {11},
pages = {2634-2642},
year = {2007},
}

@article{FlatplateTE,
author = {Moreau, Danielle J. and Doolan, Con J.},
title = {Noise-Reduction Mechanism of a Flat-Plate Serrated Trailing Edge},
journal = {AIAA Journal},
volume = {51},
number = {10},
pages = {2513-2522},
year = {2013},
}

@article{TEextension,
author = {Sundeep, Shivam and Cantos, Sinforiano and Zhong, Siyang and Zhou, Peng},
title = {On Improving the Noise Reduction Performance of Trailing Edge Serrations by Extension},
journal = {AIAA Journal},
volume = {62},
number = {9},
pages = {3354-3364},
year = {2024},
}

@inbook{Revflow_EAG,
author = {Raamesh Balasubramani and Ranga Srinivas Gokul and Tufan K. Guha},
title = {Experimental Study on the Effect of Sinusoidal Trailing Edge for the Control of Reverse Flow on High-Speed Rotorcraft},
booktitle = {AIAA AVIATION FORUM AND ASCEND 2024},
chapter = {},
pages = {},
year = {2024},
month={July},
}

@article{SilentFlight,
  title={Silent owl flight: bird flyover noise measurements},
  author={Sarradj, Ennes and Fritzsche, Christoph and Geyer, Thomas},
  journal={AIAA journal},
  volume={49},
  number={4},
  pages={769--779},
  year={2011}
}

@article{favier2012control,
  title={Control of the separated flow around an airfoil using a wavy leading edge inspired by humpback whale flippers},
  author={Favier, Julien and Pinelli, Alfredo and Piomelli, Ugo},
  journal={Comptes Rendus Mecanique},
  volume={340},
  number={1-2},
  pages={107--114},
  year={2012},
  publisher={Elsevier}
}

@article{jaworski2020aeroacoustics,
  title={Aeroacoustics of silent owl flight},
  author={Jaworski, Justin W and Peake, Nigel},
  journal={Annual Review of Fluid Mechanics},
  volume={52},
  number={1},
  pages={395--420},
  year={2020},
  publisher={Annual Reviews}
}

@book{barlow1999low,
  title={Low-speed wind tunnel testing},
  author={Barlow, Jewel B and Rae, William H and Pope, Alan},
  year={1999},
  publisher={John wiley \& sons}
}

@article{wieneke2015piv,
  title={PIV uncertainty quantification from correlation statistics},
  author={Wieneke, Bernhard},
  journal={Measurement Science and Technology},
  volume={26},
  number={7},
  pages={074002},
  year={2015},
  publisher={IOP Publishing}
}

@article{nelson2024control,
  title={Control of Reverse Flow over Cantilevered Swept Blades Using Passive Camber Morphing},
  author={Nelson, Cooper and Kumar Guha, Tufan and Amitay, Michael},
  journal={AIAA Journal},
  volume={62},
  number={4},
  pages={1503--1516},
  year={2024},
  publisher={American Institute of Aeronautics and Astronautics}
}

@book{gad2003flow,
  title={Flow control: fundamentals and practices},
  author={Gad-el-Hak, Mohamed and Pollard, Andrew and Bonnet, Jean-Paul},
  volume={53},
  year={2003},
  publisher={Springer Science \& Business Media}
}
\end{document}